\documentclass[twocolumn]{aastex6}

\usepackage{enumitem}
\usepackage{color}
\usepackage{multirow}
\usepackage{amsmath}
\usepackage{mathtools}

\newcommand{\msun}{\ensuremath{\rm M_\odot}}

\renewcommand\u[2][]{\,\textrm{#2}^{#1}}
\newcommand{\kms}{\u{km}\u[-1]{s}}

\newcommand{\lya}{Ly$\alpha$}
\newcommand{\ha}{H$\alpha$}
\newcommand{\hb}{H$\beta$}

\newcommand{\oiii}{[\ion{O}{3}]}

\newcommand{\ewlya}{\ensuremath{\rm EW_{\mathrm{Ly}\alpha}}}
\newcommand{\ewlis}{\ensuremath{\rm EW_{\mathrm{LIS}}}}
\newcommand{\vlya}{\ensuremath{v_{\mathrm{Ly}\alpha}}}

\newcommand{\zla}{\ensuremath{z_{\rm Ly\alpha}}}

\newcommand{\sub}[1]{\ensuremath{_\textrm{\scriptsize{#1}}}}

\shorttitle{\lya\ Production and Escape}
\shortauthors{Trainor et al.}

\begin{document}

\title{Predicting \lya\ Emission from
  Galaxies via Empirical Markers of Production and Escape in the KBSS\altaffilmark{1}}

\author{Ryan F. Trainor\altaffilmark{3}, Allison L. Strom\altaffilmark{2,4}, Charles C. Steidel\altaffilmark{5}, Gwen
  C. Rudie\altaffilmark{4}, Yuguang Chen\altaffilmark{5}, Rachel L. Theios\altaffilmark{2}}
\altaffiltext{1}{Based on data obtained at the W.M. Keck Observatory, which is operated as a scientific partnership among the California Institute of Technology, the University of California, and NASA, and was made possible by the generous financial support of the W.M. Keck Foundation.}
\altaffiltext{2}{Carnegie Fellow.}
\altaffiltext{3}{Department of Physics \& Astronomy, Franklin \& Marshall
  College, 415 Harrisburg Pike, Lancaster, PA 17603; ryan.trainor@fandm.edu}
\altaffiltext{4}{Carnegie Observatories, 813 Santa Barbara Street, Pasadena, CA 91101}
\altaffiltext{5}{Cahill Center for Astrophysics, MC 249-17, 1200 E California Blvd, Pasadena, CA 91125}

\begin{abstract}
Ly$\alpha$ emission is widely used to detect and confirm high-redshift
galaxies and characterize the evolution of the intergalactic
medium. However, many galaxies do not display Ly$\alpha$ emission in
typical spectroscopic observations, and intrinsic Ly$\alpha$-emitters
represent a potentially biased set of high-redshift galaxies. In this
work, we analyze a set of 703 galaxies at $2\lesssim z\lesssim3$ with
both Ly$\alpha$ spectroscopy and measurements of other rest-frame
ultraviolet and optical properties in order to develop an empirical
model for Ly$\alpha$ emission from galaxies and understand how the
probability of Ly$\alpha$ emission depends on other observables. We
consider several empirical proxies for the efficiency of Ly$\alpha$
photon production as well as the subsequent escape of these photons
through their local interstellar medium. We find that the equivalent
width of metal-line absorption and the O3 ratio of rest-frame optical
nebular lines are advantageous empirical proxies for Ly$\alpha$ escape
and production, respectively. We develop a new quantity,
$X_\mathrm{LIS}^\mathrm{O3}$, that combines these two properties into
a single predictor of net Ly$\alpha$ emission, which we find describes
$\sim$90\% of the observed variance in Ly$\alpha$ equivalent width
when accounting for our observational uncertainties. We also construct
conditional probability distributions demonstrating that galaxy
selection based on measurements of galaxy properties yield samples of
galaxies with widely varying probabilities of net Ly$\alpha$
emission. The application of the empirical models and probability
distributions described here may be used to infer the selection biases
of current galaxy surveys and evaluate the significance of
high-redshift Ly$\alpha$ (non-)detections in studies of reionization
and the intergalactic medium.
\end{abstract} 


\section{Introduction} \label{sec:intro}

The \lya\ line of hydrogen is a powerful tool for detecting and characterizing
high-redshift galaxies. Large samples of galaxies
have been selected through \lya\ emission via narrow-band surveys
(e.g.,
\citealt{cowie1998,steidel2000,rhoads2000,trainor2015,ouchi2018}) or
IFU spectroscopy (e.g., \citealt{bacon2015}). Likewise, its strength in emission or
absorption makes \lya\ extremely efficient for spectroscopically
confirming the redshifts of galaxies selected by other means,
including broad-band imaging surveys. 

\lya\ also provides valuable information about the properties of
galaxies and their surrounding gas, both in emission and absorption
(e.g., \citealt{rakic2012,rudie2012a}). \lya\ emission clearly signifies
the presence of embedded star formation and/or AGN
activity\footnote{While an external ionizing field can in principle
  illuminate ``dark galaxies'' unpolluted by star formation, current
  detection limits prohibit the detection of these pristine halos in
  all but the most extreme environments (see e.g.,
  \citealt{can05,kol10,can12,trainor2013}). In general, \lya\ is therefore
  an effective tracer of \textit{local} ionizing photon production.}  in a galaxy, and resonant scattering of
\lya\ photons can cause this light to trace the gas distribution on
scales comparable to the virial radius of the galaxy halo (e.g.,
\citealt{steidel2011,momose14,wisotzki2016}) or even beyond the halo
radius for very luminous quasars (e.g.,
\citealt{cantalupo2014,martin2015}). In addition, the apparent \lya\
emission from galaxies at the highest 
redshifts is a useful diagnostic of the intergalactic medium (IGM):
the apparent drop-off in the fraction of galaxies exhibiting strong
\lya\ emission at $z\gtrsim 6-7$ (e.g., \citealt{pentericci2011,schenker2012}) likely points to the
increasing neutral fraction at this epoch, and this evolution thus constrains the tail
end of cosmic reionization \citep{robertson2015}.

However, the same physical processes of emission and scattering that
make \lya\ such a promising tool also introduce significant
challenges to its utility. Because strong \lya\ emission facilitates efficient galaxy
detection and redshift confirmation -- but not all galaxies exhibit
strong \lya\ emission -- there are potential selection
biases both in \lya-selected galaxy samples and
in samples of broad-band-selected galaxies that are vetted
through rest-UV spectra. In particular, star-formation is a necessary
but insufficient condition for detectable \lya\ emission; only
$\sim$50\% of L$_*$ galaxies at $z\sim3$ show \lya\ in net emission in
slit spectroscopy \citep{sha03, steidel2011}, although this fraction appears
to increase toward lower galaxy masses and continuum luminosities
(e.g., \citealt{stark2013,oyarzun16}). As such, \lya-selected (or
\lya-confirmed) galaxy samples will lack non-star-forming galaxies as
well as a large fraction of star-forming galaxies. Resonant scattering, as well
as the ``fluorescent'' generation of 
\lya\ recombination emission, can be used to map out extended \ion{H}{1}
illuminated by an external or internal engine, but this same
scattering often serves to impede the identification of the size,
location, and intrinsic properties of the energizing source.

Finally, our incomplete knowledge of the galaxy-scale determinants of
strong \lya\ emission impedes our ability to use it as an IGM
tracer. Given that many star-forming galaxies do not show strong \lya\
emission at $z\sim3$ where the neutral fraction of the IGM is minimal,
it is not always clear which galaxies at $z\gtrsim 6-7$ are intrinsic
emitters of \lya\ photons and whether their lack of apparent emission
indicates suppression by the IGM. Recent studies of \lya\ as a
tracer of IGM opacity have been careful to compare galaxies at similar
redshift epochs, where intrinsic galaxy evolution is likely to be
minimal (e.g., from $z\sim7$ to $z\sim6$;
\citealt{mason2018,pentericci2018,hoag2019}). However, both the
overall trend in increasing intrinsic \lya\ emission as a function of
redshift (e.g., \citealt{stark2010}) as well as individual
measurements of strong \lya\ emission even at $z>7$
\citep{roberts-borsani2016,stark2017} indicate that the variation of intrinsic
\lya\ emission among galaxies is important to understand for a full
accounting of the reionzation process.
Similarly, the uncertain connections
between \lya\ emission and physical properties of galaxies prevents
the clear identification of the selection biases intrinsic to \lya\
selection and redshift confirmation, even as we expect some such
biases to be present.

Advancing the utility of \lya\ emission as a tool for detecting and
characterizing galaxies therefore requires a model for understanding
-- and potentially predicting -- when this
emission is expected based on a galaxy's other properties.
Any such model of galaxy-scale \lya\ emission must include two
disparate sets of processes: (1) the 
production of \lya\ photons in \ion{H}{2} regions, and (2) the subsequent
transmission (or absorption) of these photons through the surrounding \ion{H}{1} gas.

The latter set of processes -- those pertaining to \lya\ scattering,
transmission, and escape from galaxies -- have been subject to
detailed study for two decades, eased in part by the
availability of ISM diagnostics near \lya\ in the rest-UV spectra of
galaxies. In particular, much work has demonstrated that \lya\
emission (parameterized by the \lya\ equivalent width, \ewlya)
correlates strongly with the optical depth or covering fraction of Lyman-series
lines or low-ionization metal lines (hereafter LIS lines,
parameterized by \ewlis), in the
sense that galaxies with stronger \lya\ emission exhibit weaker
Lyman-series or LIS absorption 
\citep{kunth1998,sha03,jones2012,steidel2010,trainor2015,du2018}.
Weak absorption lines are likely an indicator that the covering
fraction or optical depth of \ion{H}{1} gas is relatively small, and in some
cases the covering fraction of LIS and/or Lyman-series lines have been
shown to be much less than unity, particularly for \lya-emitting
or Lyman-continuum-emitting galaxies \citep{trainor2015,steidel2018}\footnote{Note that 
the \ion{H}{1} covering fraction inferred from Lyman-series lines may not
be the same as the LIS covering fraction (see e.g., \citealt{henry2015}),
but both are found to decrease on average with increasing \ewlya.}.

In a related phenomenon, strong \lya\ emission lines are found
to be narrow in velocity space and close to the systemic redshift of
the emitting galaxy (e.g., \citealt{erb2014,trainor2015}) as well as
spatially compact (e.g., \citealt{steidel2011,momose2016}, but c.f. \citealt{wisotzki2016}). These relationships suggest that net \lya\
emission is maximized when these photons can escape their parent galaxy with
minimal scattering both in velocity and in physical space.

Common among each of the above observables -- LIS and Lyman-series
absorption, spatial 
and spectral scattering -- is that they are associated with modulation
of the \lya\ photons that occurs after these photons have left their
original star-forming regions. Conversely, more recent studies have
identified trends linking \ewlya\ to signatures of the star-forming
regions themselves. Much of this recent work has been enabled by the
development of efficient near-infrared spectrometers such as MOSFIRE
\citep{mclean2012} and the HST/WFC3-IR grisms (e.g., \citealt{atek2010,brammer2012}) that can
detect the faint rest-frame-optical emission lines used to
characterize the gas around young stars in high-$z$ galaxies. 

\citet{mclinden2011}, \citet{finkelstein2011}, and
\citet{nakajima2013} found strong \oiii\ lines in a total of 8
\lya-selected galaxies at $z\sim2-3$, while \citet{song14} localized
10 \lya-selected galaxies in the
N2-BPT\footnote{The N2-BPT compares
  log([\ion{N}{2}] $\lambda6583/$\ha) [N2] 
  vs. log([\ion{O}{3}] $\lambda5007/$\hb) [O3], an emission line
  diagnostic similar to those presented by \citet{baldwin1981}, but
  introduced by \citet{veilleux1987} in its modern form.} plane,
suggesting that high-redshift galaxies with strong \lya\ emission have
low metallicities and high nebular excitation. 
\citet{trainor2016} found similarly extreme N2-BPT lines ratios (i.e.,
high O3 and low N2)
for a stacked sample of 60 $L\sim0.1L_*$ galaxies with strong \lya\
emission at $z\sim2.5$, while \citet{erb2016} and \citet{hagen2016}
demonstrated that galaxies selected for strong \oiii\ emission and/or
weak N2 ratios are strong \lya\ emitters and share many physical
properties with \lya-selected galaxies.

In addition to presenting results for faint \lya-emitting galaxies,
\citet{trainor2016} also show that galaxies ranging from
strong \lya-emitters to \lya-absorbers can be described as a sequence
in the N2-BPT plane, a phenomenon that appears to be primarily linked
to the variation in the excitation state of gas in star-forming \ion{H}{2}
regions, rather than to extreme variation in gas-phase metallicity. In
that paper, we also argue that \lya\ emission and high
nebular excitation are linked by their association with strong sources
of ionizing emission within galaxies, including massive stars with low
Fe abundances as discussed at length by \citet{steidel2016} and
\citet{strom2017}. Taken together, the results of these rest-frame optical studies are
consistent with the expectation that \lya\ production is accompanied
by numerous other forms of recombination emission and
collisionally-excited emission that originate in the same star-forming
regions as \lya, although the subsequent transmission of these
non-\lya\ photons is much less
sensitive to the surrounding \ion{H}{1} distribution. 

It is therefore clear that the net \lya\ emission on galaxy scales
depends on both the properties of star-forming regions (the sites of
\lya\ production) and the distribution of the surrounding \ion{H}{1}
gas that modulates \lya\ escape. Here we propose a holistic, empirical framework for accounting 
for both of these processes. Using the largest sample of galaxies with
simultaneous spectroscopic measurements of \lya, the rest-UV
continuum, and a series of rest-frame optical transitions, we identify
empirical discriminants of \lya\ production and escape, and we 
demonstrate that the combination of these observable markers can
predict the net \lya\ emission of galaxies more reliably than individual galaxy properties.

The paper is organized as follows: Sec. 2
presents details of our galaxy observations and the assembly of our
sample; Sec. 3 describes the methods used to quantify our empirical
markers of the efficiency of \lya\ production and escape in a given
galaxy and their correlations with \ewlya; Sec. 4 presents our combined model for
predicting \lya\ based on multiple markers; Sec. 5 presents the
conditional probability distribution of detecting net \lya\ emission
as a function of other galaxy properties; Sec. 6 provides
discussion comparing our results to previous work; and Sec. 7
summarizes our conclusions.

\section{Observations} \label{sec:obs}

\subsection{Galaxy sample} \label{sec:sample}

The galaxies presented here are selected from the Keck Baryonic
Structure Survey (KBSS; \citealt{rudie2012a}) and KBSS-MOSFIRE
\citep{steidel2014,strom2017}, which together comprise a set of
rest-UV and and rest-optical spectra of more than 1000 galaxies
across 15 fields at $1.5\lesssim z\lesssim3.5$. The galaxies are
selected using optical colors in the $U_n$,
$G$, and $\mathcal{R}$ bands to identify Lyman-break galaxy
analogs (LBGs) over the target range of redshifts; a more detailed
description of the photometric selection is given by
\citet{steidel2004}. The full distributions of KBSS-MOSFIRE galaxies'
redshifts, masses, and star-formation rates are given in
\citet{strom2017}, but they occupy the ranges $10^9\lesssim
M_*/\msun\lesssim10^{11}$ and $3\lesssim$
SFR\footnote{SFR = star formation
  rate}$/(\msun\,\mathrm{yr}^{-1})\lesssim300$, as inferred from
reddened stellar population synthesis models\footnote{The SED fitting
  is performed using \citet{bruzual2003} solar metallicity models, a
  \citet{chabrier2003} initial mass function, and \citet{calzetti2000}
attenuation curve. For a full description of the fitting methodology,
see \citet{reddy2012}.} fit to Keck LRIS and
MOSFIRE broadband photometry. The galaxies have
typical dark-matter halo masses $M_h\approx 8\times10^{11}\,\msun$
estimated through galaxy-galaxy clustering measurements of the
spectroscopic KBSS sample
\citep{trainor2012}. The median apparent magnitude of the KBSS-MOSFIRE
sample is $\langle\mathcal{R}\rangle=24.4$, which corresponds to an
absolute magnitude $\langle M_\mathrm{UV}\rangle=-20.7$ ($\langle
L\rangle\approx L_*$; \citealt{reddy2008}) at the median
redshift of the sample, $\langle z\rangle=2.3$.

For this work, we select the subset of the full KBSS-MOSFIRE sample
that have rest-UV spectroscopic coverage of the \lya\ line and surrounding
region ($1208\u{\AA}<\lambda_\mathrm{rest}<1227\u{\AA}$) from Keck/LRIS \citep{oke1995,steidel2004} as well as a galaxy
redshift measured from rest-optical MOSFIRE \citep{mclean2012} spectroscopy (typically
from the \oiii\ $\lambda\lambda4959,5008$ doublet and/or \ha). Details
of the rest-UV and rest-optical spectroscopic data are given in the
sections below. The total subset of the KBSS-MOSFIRE sample used in
this paper comprises 703 galaxies, although most of the individual
empirical parameters described in Sec. \ref{sec:measurements} are
not measured robustly for every galaxy. The requirements for making
each measurement and the number of galaxies for which each 
is made are given explicitly in Sec. \ref{sec:measurements} with
numbers also given in Table~\ref{table:correlations}.

\subsection{LRIS observations} \label{sec:lris}

The rest-UV galaxy spectra were obtained with Keck I/LRIS-B
\citep{steidel2004} in multislit mode over a series of observing runs
between August 2002 and August 2016. Approximately 2/3 of spectra were taken using the
4000/3400 grism, which produces a resolution $R\sim 800$ in the
typical seeing conditions of 0\farcs6-0\farcs8\footnote{LRIS slitmasks
for the observations were made with 1\farcs2 slits, so the spectral
resolution of the (typically spatially unresolved) galaxies in our
spectroscopic sample is limited by the smaller seeing disk diameter.}. The remaining
spectra were taken using the 600/4000 grism, which produces a
resolution $R\sim 1300$ in the same seeing conditions. The blue edge of
the LRIS-B observations is typically determined by the atmospheric
limit at $\lambda \sim 3200\u{\AA}$, while the red edge is determined by
a dichroic that splits the light between LRIS-B and LRIS-R; dichroics
with transition wavelengths
$\lambda_\mathrm{dich}=5000\u{\AA},\,5600\u{\AA}$, or $6800\u{\AA}$ were
used to collect the spectra in this sample. These constraints, combined
with the redshift distribution of our sample, allow us to sample the
rest-frame spectrum of most objects over the range $1000\u{\AA}\lesssim
\lambda_\mathrm{rest}\lesssim1700\u{\AA}$. Each object was typically
observed for 1.5 hours of $1800\u{s}$ integrations, but a 
subsample of $\sim$150 galaxies included in the KBSS-LM1 project
\citep{steidel2016} were each observed for $\sim$10$-$14 hours.

LRIS-B spectra were reduced using a custom suite of IDL and IRAF
routines. Raw two dimensional spectrograms were rectified using a
slit-edge tracing algorithm, and the resulting rectilinear
spectrograms were flat-fielded, background-subtracted, and subjected
to cosmic-ray rejection. Individual two-dimensional spectrograms were
stacked after accounting for shifts in the spatial and spectral
directions due to instrument flexure between exposures.\footnote{Note
  that MOSFIRE corrects the optical path internally to account for instrument
  flexure, so analogous shifts are not necessary for
  our MOSFIRE spectra (Sec.~\ref{sec:mosfire}).} A
one-dimensional spectrogram was extracted for each object, and
wavelength calibration was performed using arc-line spectra
observed through the slitmask during daytime telescope operations,
which were then individually shifted to match the wavelength of the
$5577\u{\AA}$ sky-line in each science spectrum in order to account for
instrument flexure. Finally, spectra are corrected to vacuum,
heliocentric wavelengths and rebinned to a common wavelength
scale. Further details regarding the software routines 
used in the KBSS data-reduction process are given by
\citet{steidel2003,steidel2010}.

\begin{figure*}[ht]
\center
\includegraphics[width=\linewidth]{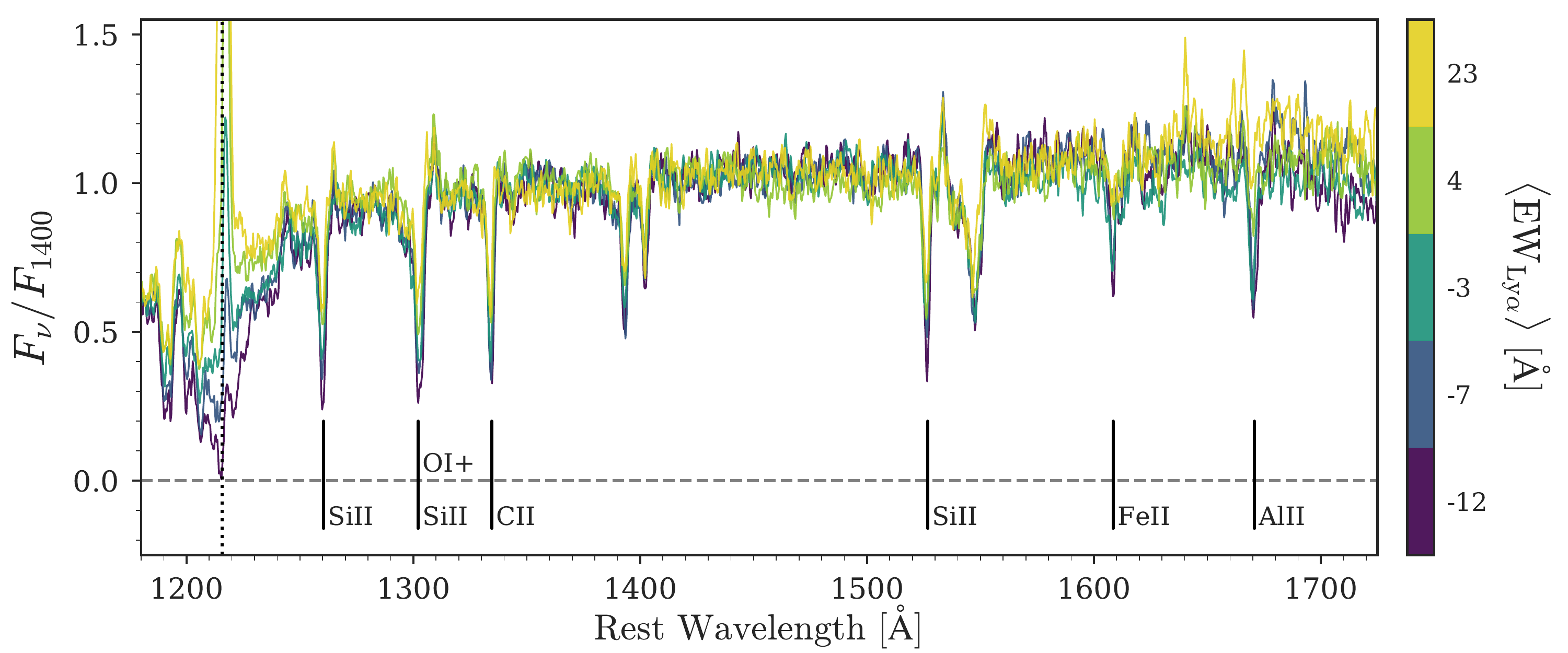}
\caption{Stacked spectra of 703 galaxies sorted in quintiles of
  \ewlya; the median \ewlya\ for each sample is given by the colorbar
  on the right. The dashed vertical line indicates \lya\ while the solid
  vertical lines indicate the LIS transitions described in
  Table~\ref{table:lislines} and Sec.~\ref{sec:ewlis}. The LIS
  line strengths vary similarly across all 6 distinct absorption
  features, such that strong LIS absorption is associated with weak
  \lya\ emission and vice versa.}
\label{fig:spec_eqw_norm_wide}
\end{figure*}

\begin{figure*}[ht]
\center
\includegraphics[width=\linewidth]{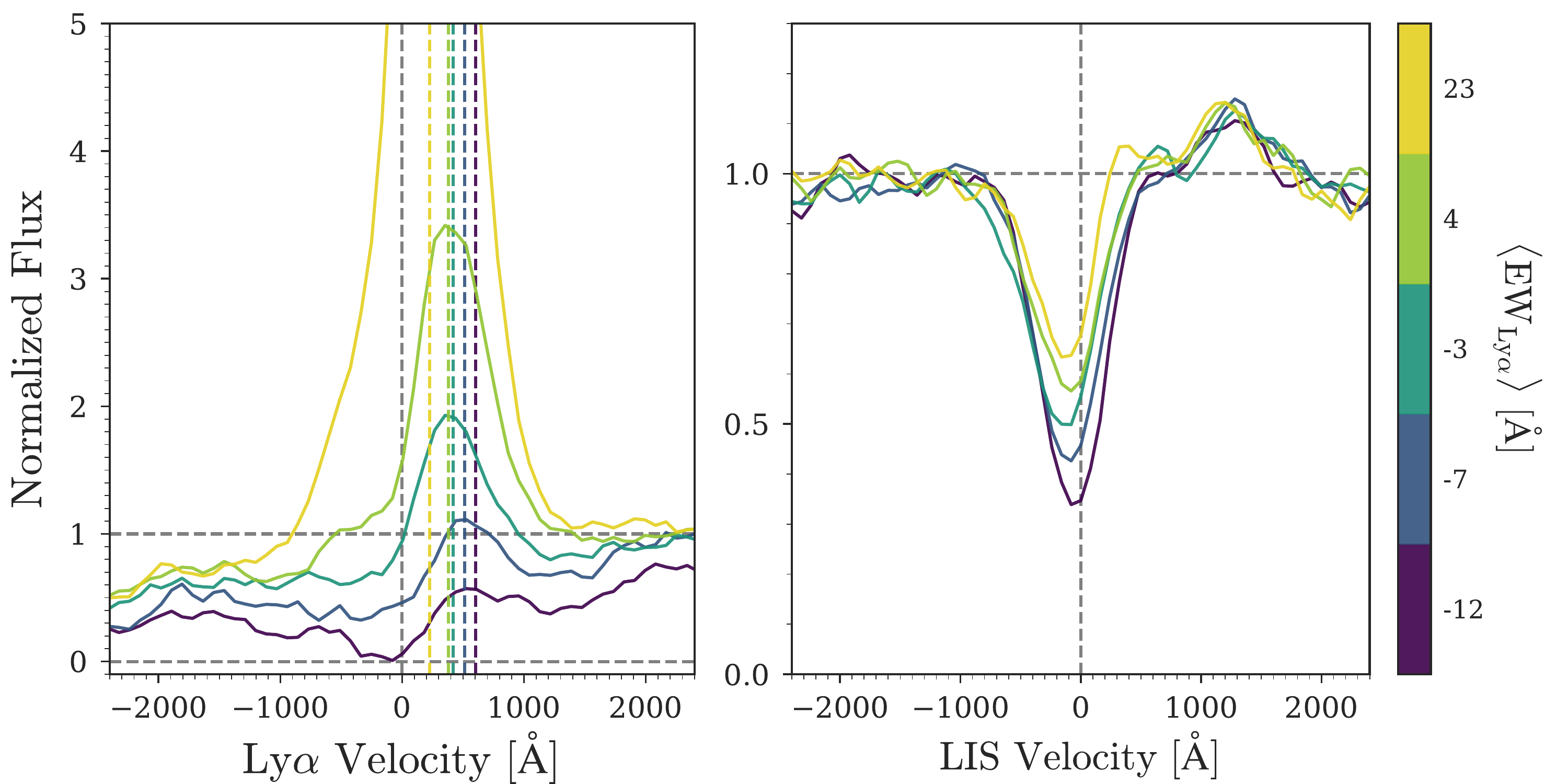}
\caption{Stacked spectra in quintiles of \ewlya, as in
  Fig.~\ref{fig:spec_eqw_norm_wide}. The left panel is zoomed in on the
  \lya\ line, and the dashed colored lines indicate the velocity of
  the \lya\ centroid for each stacked subsample. The continuum for each
  stack is normalized at $\lambda_\mathrm{rest}=1400\u{\AA}$ as in
  Fig.~\ref{fig:spec_eqw_norm_wide}. The right panel shows the
combined LIS absorption profile for each stack, constructed by
averaging the profiles of each of the LIS absorption features labeled
in Fig.~\ref{fig:spec_eqw_norm_wide} according to the process
described in Sec.~\ref{sec:ewlis} including normalization to the
local continuum. Note that the stacked profiles
show the same trends described in the text for individual objects:
increased \ewlya\ is strongly associated with decreased (i.e., less
redshifted) \vlya\ and increased (i.e., weaker and less negative) \ewlis.}
\label{fig:spec_eqw_norm_zoom}
\end{figure*}

Because many objects were observed multiple times over the period
since observations of the KBSS galaxy sample began, all extant
spectra for a given galaxy were averaged -- weighted according to their
exposure time and a visual inspection of each spectrum -- to produce a
final spectrum for each 
galaxy in our sample. Stacked LRIS spectra are displayed in
Figs.~\ref{fig:spec_eqw_norm_wide}$-$\ref{fig:spec_eqw_norm_zoom}. 

\subsection{MOSFIRE observations} \label{sec:mosfire}

The MOSFIRE observations for the KBSS-MOSFIRE survey are described in
detail elsewhere \citep{steidel2014,strom2017}. Briefly, galaxies are
observed in the $J$, $H$, and/or $K$ bands using 0\farcs7 slits, and
the resulting two-dimensional spectrograms are reduced using the
MOSFIRE data-reduction pipeline\footnote{https://github.com/Keck-DataReductionPipelines/MosfireDRP} provided by the instrument
team. Wavelength calibration is performed by identifying OH sky-lines in
all spectral regions except in the red end of the $K$ band
($\lambda_\mathrm{obs}\gtrsim2\u{$\mu$m}$), where arc lamp spectra are
used due to the paucity of sky lines. The absolute flux
scaling, slit-loss corrections, and  cross-band calibration are
performed through observations of a slit star on each mask, as
discussed in detail by \citet{strom2017}. As described in that paper,
the spatial extent of the KBSS galaxies (which are marginally resolved
in typical atmospheric conditions) causes the slit losses for KBSS
galaxies to exceed that measured directly from the slit star.

Line measurements are performed using the IDL program MOSPEC
\citep{strom2017}. Typically, galaxy redshifts and line fluxes are fit
simultaneously in a single band ($J$, $H$, or $K$), with all nebular
lines in the band constrained to have the same redshift and velocity
width. For the \oiii\ $\lambda\lambda4960,5008$ doublet, the known
line flux ratio $f_{5008}/f_{4960}=3$ is also enforced. Line
widths and redshifts are not forced to match between bands (e.g., for
\oiii\ in the $H$ band and \ha\ in the $K$ band, as would be observed at
$z\approx 2-2.6$), but galaxies with redshift measurements in multiple bands
are checked for consistency. Each galaxy is then assigned a nebular redshift
$z_\mathrm{neb}$ based on the rest-frame optical redshift with the
smallest uncertainty. For galaxies with redshift measurements in
multiple bands, the typical agreement is less than $\Delta z=0.0002$.

\subsection{SED models and SFRs} \label{sec:seds}

Photometry of the KBSS fields and spectral-energy distribution (SED)
modeling of the KBSS galaxy sample is described by \citet{steidel2014}
and \citet{strom2017} using SED-fitting methodology described by
\citet{reddy2012}. Models are from the \citet{bruzual2003}
library and assume solar metallicity, a \citet{chabrier2003} initial
mass function (IMF), a \citet{calzetti2000} attenuation relation, and a
constant star-formation history with a minimum age of 50 Myr. Stellar
masses ($M_*$), star formation rates (SFR$_\text{SED}$), and continuum-based
reddening (E$(B-V)_{\rm{SED}}$) estimates are obtained from the SED fitting.

In addition to SED-based SFRs, \ha\ SFRs
are calculated for the majority of the KBSS-MOSFIRE sample as described by
\citet{strom2017} using the MOSFIRE measurements described
above. These SFRs and sSFRs (sSFR = 
SFR$_{\rm{H}\alpha}/M_*$) are calcuated assuming a
\citet{kennicutt1998} H$\alpha$-SFR relation adjusted for a
\citet{chabrier2003} IMF. Dust-corrections for \ha\ SFRs are
calculated as described in Sec.~\ref{sec:ebmv}.

\section{Empirical Galaxy Measurements} \label{sec:measurements}

As described in Sec.~\ref{sec:intro}, the primary goals of this paper
are (1) to characterize the empirical relationships between \lya\ emission
and various other galaxy properties; and (2) to interpret these
relationships in terms of the production and escape of \lya\
photons. Below, we define the various observables used throughout this
paper, and we categorize them in terms of whether they are likely to
primarily relate to the production or escape of \lya\ photons.

\subsection{\ewlya} \label{sec:ewlya}

First, we must define a metric of the \textit{efficiency} of
\lya\ emission: the \lya\ equivalent width (\ewlya). We focus on
\ewlya\ rather than the total \lya\ luminosity for multiple reasons. Firstly,
we find that the \lya\ luminosity is primarily correlated with other
descriptors of the total luminosity of a galaxy, so measuring the
\lya\ luminosity per unit UV continuum luminosity is a more
interesting descriptor of the (in)ability of \lya\ photons to escape
relative to non-\lya\ photons at similar wavelengths. Secondly,
characterization of \ewlya\ is possible in slit spectra even
without precise flux calibration, so our predictions for \ewlya\ can
be more easily applied to actual observations of galaxies with
uncertain slit losses.\footnote{However, note that the spatial
  scattering of \lya\ photons (as described in Sec.~\ref{sec:intro})
  causes \ewlya\ to be sensitive to the differential slit losses in
  \lya\ vs. the continuum (although \ewlya\ is still less
  sensitive to slit losses than the total \lya\ luminosity).}

The value of \ewlya\ is measured for each object in our
sample directly from the one-dimensional object spectrum in a manner
similar to that described in \citet{trainor2015,trainor2016}. Each
rest-UV spectrum is first shifted to the rest frame based on its
nebular redshift $z_\mathrm{neb}$. In order
to account for the wide variety of \lya\ line profiles in a systematic
manner, the \lya\ line flux, $F_{\mathrm{Ly}\alpha}$, was directly integrated by
summing the continuum-subtracted line flux over 
the range $1208\u{\AA}<\lambda_\mathrm{rest}<1227\u{\AA}$, roughly the
maximum range of wavelengths found to encompass the \lya\ line in our
spectra.\footnote{This velocity range is also chosen to minimize
  contaminating absorption due to the \ion{Si}{3} $\lambda$1206 transition.} In velocity space, this corresponds to a range
$-1900\kms<v<2800\kms$ with respect to \lya. The local continuum flux $f_{\lambda,\mathrm{cont}}$ used for
subtraction is estimated as the median flux in the range 
$1225\u{\AA}<\lambda_\mathrm{rest}<1250\u{\AA}$. The \lya\ line flux
$f_{\mathrm{Ly}\alpha}$ is therefore positive for net \lya\ emission lines
and negative for net \lya\ absorption. The line flux and equivalent width are thus
defined by the following expressions:

\begin{align}
F_{\rm Ly\alpha}&=\int_{1208\u{\AA}}^{1227\u{\AA}} \left(f_\lambda-f_{\lambda,\mathrm{cont}}\right) d\lambda\label{eq:flya}\\[10pt]
\ewlya&=\frac{F_{\mathrm{Ly}\alpha}}{f_{\lambda,\mathrm{cont}}}\frac{1}{1+z_\mathrm{neb}}\,\,.
\end{align}

Again, this procedure causes galaxies with net \lya\ emission in their
one-dimensional slit spectra to have $\ewlya>0$, and galaxies
exhibiting net \lya\ absorption to be assigned $\ewlya<0$. Note
that, in each case, the assigned value of \ewlya\ is likely to be an
underestimate of the the intrinsic value owing to the spatial scattering of
\lya\ photons, which preferentially lowers the observed ratio of \lya\ photons
to continuum photons in a centralized aperture. Typical relative slit
losses of \lya\ 
respect to the nearby continuum in similar samples are found to be $2-3\times$ (see,
e.g., \citealt{steidel2011,trainor2015}). However, we have no way to
determine the relative \lya\ slit loss for the majority of individual
objects in our sample, so we do not attempt to do so
here. Furthermore, in this work we are primarily concerned with the
factors that determine the net \lya\ emission on galaxy scales;
variation in the total \lya\ emission of the galaxy-plus-halo system with
physical and environmental properties of the galaxies will be
discussed in future work.

Uncertainty in \ewlya\ is determined based on the uncertainty in the
\lya\ flux (usually a small factor) as well as the uncertainty in the
local continuum (usually the dominant factor, particularly for
high-\ewlya\ sources). In some cases, correlated noise in the
continuum spectrum causes the formal uncertainty in the local continuum
to be unrealistically low based on a visual inspection of the
spectrum. To account for this fact, a separate estimate of the
continuum flux $f_{\lambda,\mathrm{cont}}$ is measured for each
galaxy over the range $1220\u{\AA}<\lambda_\mathrm{rest}<1300\u{\AA}$
(i.e., over a $\sim$3$\times$ larger range of wavelengths than the
first estimate), and this
second continuum estimate is used to recalculate
$F_{\mathrm{Ly}\alpha}$ and \ewlya. If the new \ewlya\ value differs
from the original value by more than the uncertainty on \ewlya\
calculated originally, then the uncertainty on \ewlya\ is replaced by
the absolute value of the difference between the two estimates of
\ewlya. This procedure increases the uncertainty on \ewlya\ for 59
objects (8\% of the total sample), and the total variance in \ewlya\
associated with our estimated measurement error among our entire
sample increases by 5\%.

Furthermore, if the measured continuum value is smaller than
$2\times$ the estimated uncertainty on the continuum, then \ewlya\ is
defined to be a lower limit:
\begin{equation}
\ewlya>\frac{F_{\mathrm{Ly}\alpha}}{2\sigma_{\lambda,\mathrm{cont}}}\frac{1}{1+z_\mathrm{neb}}
\end{equation}

\noindent where $\sigma_{\lambda,\mathrm{cont}}$ is the formal
uncertainty in the local \lya\ continuum. This correction applies to
13 objects (2\% of our total sample), and \ewlya\ for these objects is assumed to
take the value of their 2$\sigma$ error in the analysis that
follows. In this manner, a value and 
uncertainty for \ewlya\ is estimated for each of the 703 galaxies in
our sample.

Spearman rank correlation statistics between \ewlya\ and a series of other
empirical quantities measured among the galaxies in our sample are
given in Table~\ref{table:correlations} below. Definitions and
measurement methodologies for each of these quantities are given in the
sections that follow.

\subsection{Proxies for \lya\ escape} \label{sec:escape}

\subsubsection{\vlya} \label{sec:vlya}

As discussed in Sec. 1 above, increasing shifts of the \lya\
emission line with respect to the systemic redshift are associated with
decreasing \ewlya, and this trend likely corresponds to the fact that
\lya\ photons must scatter significantly in redshift and/or physical
space to escape regions of high optical depth. For this reason, the
difference of the \lya\ redshift (\zla)  from systemic ($z_\mathrm{neb}$) may be
regarded as a proxy for the optical depth experienced by \lya\
photons transiting the galaxy ISM and CGM and is thus related to the
probability of \lya\ photon escape.

We define the \lya\ redshift and offset velocity based on the centroid
of the \lya\ line flux:

\begin{deluxetable}{lllcrr}
\tablecaption{Correlations \label{table:correlations}}
\tablewidth{0pt}
\tablehead{
\multicolumn{3}{c}{Quantities} & \colhead{$N_\mathrm{gal}$\tablenotemark{a}} & \colhead{$r_\mathrm{Sp}$\tablenotemark{b}} & \colhead{$\log_{10}(p)$}
}

\startdata
\multicolumn{6}{l}{\it Escape-related quantities}\\
EW$_{\mathrm{Ly}\alpha}$ & vs. & $v_{\mathrm{Ly}\alpha}$ & 496 & $-$0.56 & $-$41.8\\ 
& vs. &EW$_\mathrm{LIS}$ & 669 &  0.35 & $-$20.5\\
& vs. & E($B-V$)$_{\rm{SED}}$& 637 & $-$0.23 & $-$8.2\\
& vs. & E($B-V$)$_{\rm{neb}}$& 208 & $-$0.14 & $-$1.3\\[5pt] 
EW$_\mathrm{LIS}$ & vs. & $v_{\mathrm{Ly}\alpha}$ & 479 &  0.28 & $-$9.5\\ 
& vs. & $f_\mathrm{esc,rel}$ & 368 & $-$0.42 & $-$16.5\\ 
& vs. & $f_\mathrm{esc,abs}$ & 188 & $-$0.50 & $-$12.4\\[5pt]\hline
\multicolumn{6}{l}{\it Production-related quantities}\\
EW$_{\mathrm{Ly}\alpha}$ & vs. & $M_{*,\rm{SED}}$& 637 & $-$0.15 & $-$3.9\\
& vs. & SFR$_{\rm{SED}}$& 637 & $-$0.17 & $-$5.0\\
& vs. & SFR$_{\rm{H}\alpha}$& 208 & $-$0.05 & $-$0.4\\
& vs. & sSFR$_{\rm{H}\alpha}$ & 199 & 0.23 & $-$3.0\\
& vs. & $M_{\rm{UV}}$& 637 & $-$0.01 & $-$0.1 \\
& vs. & O32$_{\rm{raw}}$  & 316 &  0.43 & $-$15.0\\ 
& vs. & O32$_{\rm{corr}}$ & 174 &  0.47 & $-$10.3\\ 
& vs. & O3 & 395 &  0.40 & $-$15.3\\[5pt]\hline 
\multicolumn{6}{l}{\it Production + escape}\\
EW$_\mathrm{LIS}$ & vs. & O3 & 377 &  0.21 & $-$4.9\\ 
& vs. & O32$_{\rm{corr}}$ & 174 &  0.47 & $-$10.3\\[5pt] 
EW$_{\mathrm{Ly}\alpha}$ & vs. & $X_\mathrm{LIS}^\mathrm{O3}$ \tablenotemark{c} & 377 &0.49 & $-$23.8\\
\enddata
\tablenotetext{\rm a}{Number of galaxies for which correlation is calculated}
\tablenotetext{\rm b}{Spearman correlation coefficient}
\tablenotetext{\rm c}{$X_\mathrm{LIS}^\mathrm{O3}$ is defined by Eq.~\ref{eq:xlya}}
\end{deluxetable}

\begin{align}
\lambda_{\rm Ly\alpha}^{\rm obs}&=\frac{\int\lambda f_\lambda d\lambda}{\int f_\lambda d\lambda}\label{eq:lamlya}\\
\zla&=\frac{\lambda_{\rm Ly\alpha}^{\rm obs}-\lambda_{\rm
      Ly\alpha}^{\rm rest}}{\lambda_{\rm Ly\alpha}^{\rm rest}}\\
\vlya&=\left(\frac{\zla-z_{\rm neb}}{1+z_{\rm neb}}\right)\rm{c}
\end{align}
\noindent where c is the speed of light and the integrals in
Eq.~\ref{eq:lamlya} are evaluated 
over the range $1208\u{\AA}<\lambda_\mathrm{rest}<1227\u{\AA}$, as is
done to estimate the total \lya\ flux (Eq.~\ref{eq:flya}). Note,
however, that in this case we integrate the raw flux over the \lya\
region rather than integrating the continuum-subtracted flux. This
choice does not appreciably change the assigned value of \vlya\ when
\ewlya\ is large, but it significantly reduces the noise on \vlya\
when $\ewlya\to 0$, in which case the denominator would approach zero
for an analogous equation to Eq.~\ref{eq:lamlya} weighted by
continuum-subtracted flux.

Note also that the flux in the \lya\ transition need not exceed
$f_{\lambda,\mathrm{cont}}$ in order to measure a \lya\ velocity; if
a galaxy exhibits \lya\ \textit{absorption} that is preferentially blueshifted,
the assigned \lya\ velocity will be positive and thus similar to a galaxy with
redshifted \lya\ \textit{emission}. 

Velocity uncertainties are determined by a Monte Carlo analysis in which
a randomly-generated error array consistent with the per-pixel
uncertainty is added to the \lya\ region of the
spectrum and the velocity is measured as above. This process is repeated 1000
times per spectrum, and the estimated velocity uncertainty is
1.5$\times$ the median absolute deviation\footnote{Note that
  $\sigma \approx 1.5\times$MAD is a simple estimator of scale that is insensitive
  to outliers and recovers the usual standard deviation when applied
  to a gaussian distribution (see, e.g., \citealt{rousseeuw1993}).} of the Monte Carlo velocity
values. While velocity measurements can be made in this way for every
spectrum in our sample, we include in our analysis of 
\vlya\ only those spectra with a
\lya\ velocity uncertainty $\sigma_{\rm
  Ly\alpha}<750\,\kms$.\footnote{Objects with larger velocity
  uncertainties typically have relatively low signal-to-noise ratios in the
  UV continua as well as minimal \lya\ absorption and emission, making
  the ``velocity'' of the \lya\ line an ill-defined quantity.} Because
the \lya\ velocity also depends on the accuracy of $z_{\rm neb}$, we
also require that \ha\ and/or \oiii\ $\lambda 5008$ are detected with
at least 5$\sigma$ significance. When these cuts are made, 496
galaxies in our sample have reliably measured values of \vlya. The
Spearman rank correlation statistic between \vlya\ and \ewlya\ is $r_{\rm
  Sp}=-0.56$ ($p=1.6\times10^{-42}$), indicating a strong,
highly-significant correlation. This relationship is consistent with
previous work (e.g., \citealt{erb2014}) as well as our results from
stacked spectra shown in Fig.~\ref{fig:spec_eqw_norm_zoom} (left panel).

As described in Sec.~\ref{sec:intro}, our eventual goal is to predict
the value of \ewlya\ for a galaxy in 
the absence of its direct measurement (since the \lya\ flux is not always
directly observable). Unfortunately, \vlya\ may be ineffective as such
a predictor for two reasons. Firstly, \vlya\ cannot be measured in
cases where \lya\ is not directly measurable (e.g., when the
transition is censored by the IGM or contaminated by other
emission). Secondly, even in cases
where \vlya\ is measurable (e.g., among some galaxies at high
redshift), any intergalactic 
absorption that suppresses \ewlya\ is also likely to change the
observed value of \vlya. Because scattering of \lya\ by both the ISM and the
surrounding IGM will produce degenerate shifts on \vlya, \vlya\
itself cannot be expected to separate between these two effects. With this in mind, we caution against the use
of \vlya\ to predict the intrinsic value of \ewlya.

\subsubsection{\ewlis} \label{sec:ewlis}

The second proxy we use for the ease of \lya\ photon escape is the
strength of absorption lines corresponding to
low-ionization interstellar gas. Unfortunately, even the strongest
interstellar absorption features are difficult to measure reliably in
individual spectra. In order to increase the significance of these
detections, we construct a ``mean'' LIS absorption profile for each galaxy
spectrum as follows.

\begin{deluxetable}{lllcr}
\tablecaption{LIS Transitions \label{table:lislines}
}
\tablewidth{0pt}
\tablehead{
Ion & $\lambda\sub{vac}$\tablenotemark{a} (\AA) & $f\sub{osc}$\tablenotemark{b} &
\phantom{x}EW$\sub{ion}$\tablenotemark{c} (\AA)
}

\startdata
\ion{Si}{2} & 1260.418 & 1.22 & $1.74\pm0.06$ \\
\ion{O}{1} & 1302.169 & 0.0520 & \phantom{\tablenotemark{d}}$2.37\pm0.08$\tablenotemark{d} \\
\ion{Si}{2} & 1304.370 & 0.0928 & \phantom{\tablenotemark{d}}$2.37\pm0.08$\tablenotemark{d} \\
\ion{C}{2} & 1334.532 & 0.129 & $1.54\pm0.08$ \\
\ion{Si}{2} & 1526.707 & 0.133 & $1.40\pm0.10$ \\
\ion{Fe}{2} & 1608.451 & 0.0591 & $1.11\pm0.14$ \\
\ion{Al}{2} & 1670.787 & 1.77 & $1.13\pm0.21$\\
\enddata
\tablenotetext{\rm a}{Vacuum wavelength of transition}
\tablenotetext{\rm b}{Oscillator strength from the NIST Atomic Spectra
Database (www.nist.gov/pml/data/asd.cfm)}
\tablenotetext{\rm c}{Equivalent width of absorption in a stacked spectrum
 of all 703 galaxies in sample
 (Fig.~\ref{fig:spec_eqw_norm_wide})}
\tablenotetext{\rm d}{The \ion{O}{1}
$\lambda$1302 and \ion{Si}{2} $\lambda$1304 absorption lines 
are blended, so they are measured as a single absorption feature with
the given (combined) EW.}
\end{deluxetable}
 
\begin{figure*}
\center
\includegraphics[width=0.8\linewidth]{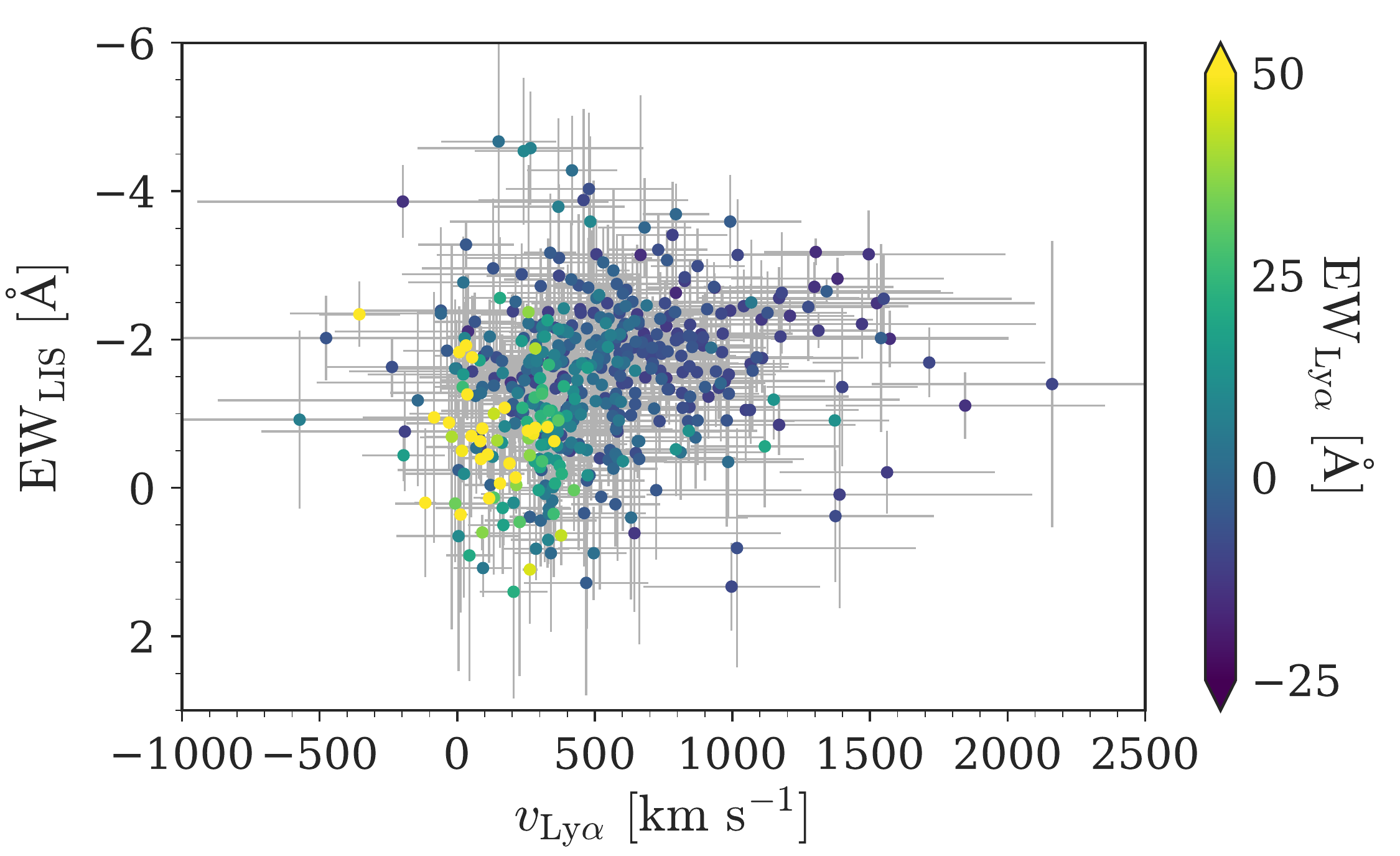}
\caption{Centroid \lya\ velocity vs. rest-frame equivalent
  width of LIS absorption for 452 galaxies, with colors denoting \lya\
  equivalent width. Two clear trends are visible: (1) stronger LIS
  absorption (EW$_\mathrm{LIS}<$0) is associated with increasing \lya\
  redshift ($v_{\mathrm{Ly}\alpha}>0$); and (2) strong \lya\ emission
  (EW$_{\mathrm{Ly}\alpha}>0$, yellow points) is associated with both weak LIS
  absorption and small \lya\ redshifts.}
\label{fig:lis_vs_vlya}
\end{figure*}

Seven LIS transitions covered by the majority of our rest-UV spectra
are identified in Table~\ref{table:lislines}. 
\ion{O}{1} $\lambda1302$ and \ion{Si}{2} $\lambda1304$ are blended at
the typical spectral resolution of our observations, so six distinct
absorption features can be individually measured. For
each transition, the spectral region within $\pm5000\kms$ of the
rest-frame wavelength is interpolated onto a grid in velocity space
and normalized to its local continuum (defined as the median flux
in the region $>$$1000\kms$ and $<$$5000\kms$ from the transition
wavelength in either direction). The six LIS absorption profiles are
then averaged with equal weighting, and an
effective rest-frame equivalent width in absorption is measured for the stacked
profile via direct integration according to the following expression, which we define as
\ewlis:

\begin{align}
\ewlis&=\int_{\lambda_1}^{\lambda_2} \left(1-\frac{f_\lambda}{f_\mathrm{cont}}\right) d\lambda\label{eq:ewlis}
\end{align}

\noindent where $\lambda_1$ and $\lambda_2$ correspond to
$\pm1000\kms$ from the rest-frame line center. The uncertainty on \ewlis\ is 
defined to be the standard deviation of absorption equivalent widths
calculated as above for random $2000\kms$ intervals in nearby regions of the
rest-UV spectrum. Because the reliability of our \ewlis\ measurement
is extremely sensitive to the strength of the FUV continuum, we only
consider measurements of \ewlis\ for which the local continuum is
detected with S/N $> 20$ in the stacked profile; this sample includes 625 objects.

For 162 objects in this sample, one or more LIS transitions fall above
the red edge of the LRIS-B spectrum. For 19 objects,
one or more LIS transitions are flagged as discrepant: they either
correspond to a EW more than $15\sigma$ away from the median EW of the
other transitions, or they otherwise lie in a region of the spectrum
that appears significantly noisier than average based on a visual
inspection. In any of these cases, the missing or flagged transitions
are omitted, and the mean \ewlis\ value and uncertainty are calcuated
from the remaining transitions. In total, the number of objects for
which 6 (5, 4, 3, 2, 1) transitions contribute to \ewlis\ is 456 (75, 56,
42, 8, 0).

As described above, the vast majority of cases where one or more
transitions are omitted occur because of a lack of red spectral
coverage, such that the redder transitions in
Table~\ref{table:lislines} are preferentially omitted. Given that the
two strongest transitions are also the two bluest (and thus, least
likely to be omitted), there is potential for our spectral coverage to
introduce a trend between \ewlis\ and the number of included
transitions. Separating galaxies by the number of included LIS
transitions ($N_\mathrm{LIS}$), the median value of \ewlis\ for each subset is
(1.43\AA, 1.49\AA, 1.27\AA, 1.40\AA, 1.55\AA) for $N_\mathrm{LIS}=($6,
5,4, 3, 2$)$. The lack of a systematic trend between \ewlis\ and
$N_\mathrm{LIS}$ suggests that our \ewlis\ values are not particularly
sensitive to the precise subset of LIS transitions included.

The distribution of \ewlis\ vs. \vlya\ is shown in
Fig.~\ref{fig:lis_vs_vlya} for the 479 objects for which both
quantities are measured robustly according to the criteria described
above. The Spearman correlation coefficient for these two parameters
is $r_{\rm Sp}=0.28$ ($p=3.2\times10^{-10}$; see
Table~\ref{table:correlations}), indicating a moderate (although 
highly statistically significant) correlation between these two
proxies for ISM optical depth (or porosity) and the likely ease
of \lya\ photon escape. The color-coding by \ewlya\ in
Fig.~\ref{fig:lis_vs_vlya} demonstrates that strong \lya\ emission is
associated with weak LIS absorption and a small shift of
the \lya\ line with respect to systemic, in agreement with the
expectations outlined above. The correlation between \ewlis\ and
\ewlya\ is moderately strong and highly significant ($r_{\rm
  Sp}=-0.35$, $p=3\times10^{-21}$).

We note that \ewlis\ has multiple practical advantages over
  \vlya\ as a predictor of \ewlya. Unlike \vlya,
  \ewlis\ is likely to be unaffected by IGM absorption, since the
  metallicity of intergalactic gas will be negligible compared to
  that of the enriched galactic outflows traced by metal-line
  absorption. In addition, in the event that the \lya\ transition is
  censored by the IGM, local \ion{H}{1} or contaminating emission, the
  longer-wavelength 
LIS transitions may still be measurable in many realistic cases at
both low and high redshifts. For these
reasons, our analysis that follows utilizes
\ewlis\ as our primary proxy for \lya\ escape.

\subsubsection{\rm E($B-V$)} \label{sec:ebmv}

Because the escape of \lya\ photons depends on the distribution of dust
in galaxies as well as \ion{H}{1}, we also consider the relationship
between \ewlya\ and E($B-V$) (see also discussion in
\citealt{trainor2016} and \citealt{theios2019}).

E($B-V$)$_{\rm{SED}}$ is measured via SED-fitting as described in
Sec.~\ref{sec:seds} for 637 galaxies. Comparing these values to
\ewlya\ yields a moderate, highly-signficant correlation ($r_{\rm
  Sp}=-0.23$, $p=3\times10^{-8}$).

We also measure E($B-V$)$_{\rm{neb}}$ based on the Balmer decrement
(\ha/\hb) as described by \citet{strom2017}. Briefly, the
slit-loss-corrected \ha\ and \hb\ fluxes are compared to the canonical
ratio \ha/\hb\ = 2.86 for Case-B recombination at $T=10^4$ K
\citep{osterbrock1989}. Galaxies with \ha/\hb\ $< 2.86$ are assigned
E($B-V$)$_{\rm{neb}}$ = 0, while galaxies with \ha/\hb\ $> 2.86$ are
assigned a value of E($B-V$)$_{\rm{neb}}$ based on a
\citet{cardelli1989} Galatic attenuation relation. The median value of
E($B-V$)$_{\rm{neb}}$ for KBSS galaxies is 0.26, and the interquartile
range is 0.06$-$0.47\citep{strom2017}.

As discussed by \citet{trainor2016} and \citet{strom2017}, the \hb\
and \ha\ emission lines are measured in separate exposures in
KBSS-MOSFIRE observations; at typical redshifts $2.0<z<2.6$, the lines
fall in the $H$ and $K$ NIR atmospheric bands, respectively. For this
reason, we present values of E($B-V$)$_{\rm{neb}}$ only for those
galaxies with $>$5$\sigma$ measurements of \ha/\hb\ including the
uncertainties in the individual line fluxes as well as the cross-band
calibration. This cut limits our sample of secure
E($B-V$)$_{\rm{neb}}$ measurements to 208 galaxies, which display a
weak correlation with \ewlya\ ($r_{\rm
  Sp}=-0.14$, $p=0.05$). 

\subsubsection{$f_\mathrm{esc}$} \label{sec:fesc}

The most direct measure of the efficiency of \lya\ photon escape is
the actual escape fraction of \lya, hereafter
$f_\mathrm{esc}$.\footnote{$f_\mathrm{esc}$ here should not be
  confused with the escape fraction of Lyman-\textit{continuum} (i.e.,
  ionizing) photons. The $f_\mathrm{esc}$ defined here is described
  elsewhere in the literature as $f_{\mathrm{esc,Ly}\alpha}$, but we
  will omit the \lya\ subscript for simplicity in this paper.} Any
determination of this escape fraction relies on an estimation of the
true number of \lya\ photons produced in galaxies, which can then be
compared to the observed \lya\ flux. In practice, the observed \lya\ flux is
compared to the observed \ha\ flux, with the latter value scaled by
the expected intrinsic flux ratio
$\left(F_{\mathrm{Ly}\alpha}/F_{\mathrm{H}\alpha}\right)_\mathrm{int}\approx 8.7$ for case-B
recombination.\footnote{Note that various values are assumed for
  $\left(F_{\mathrm{Ly}\alpha}/F_{\mathrm{H}\alpha}\right)_\mathrm{int}$
in the literature, but the uncertainty on the aperture correction for
\lya\ in our data dwarfs the uncertainty on the intrinsic flux ratio,
and our measured trends between $f_\mathrm{esc}$ and other parameters
are insensitive to the chosen value regardless. The value 8.7 is
motivated by the calculations of \citet{dop03} and is consistent with
previous studies \citep{atek2009,hayes2010,henry2015,trainor2015}.}

\begin{figure*}
\center
\includegraphics[height=2.2in]{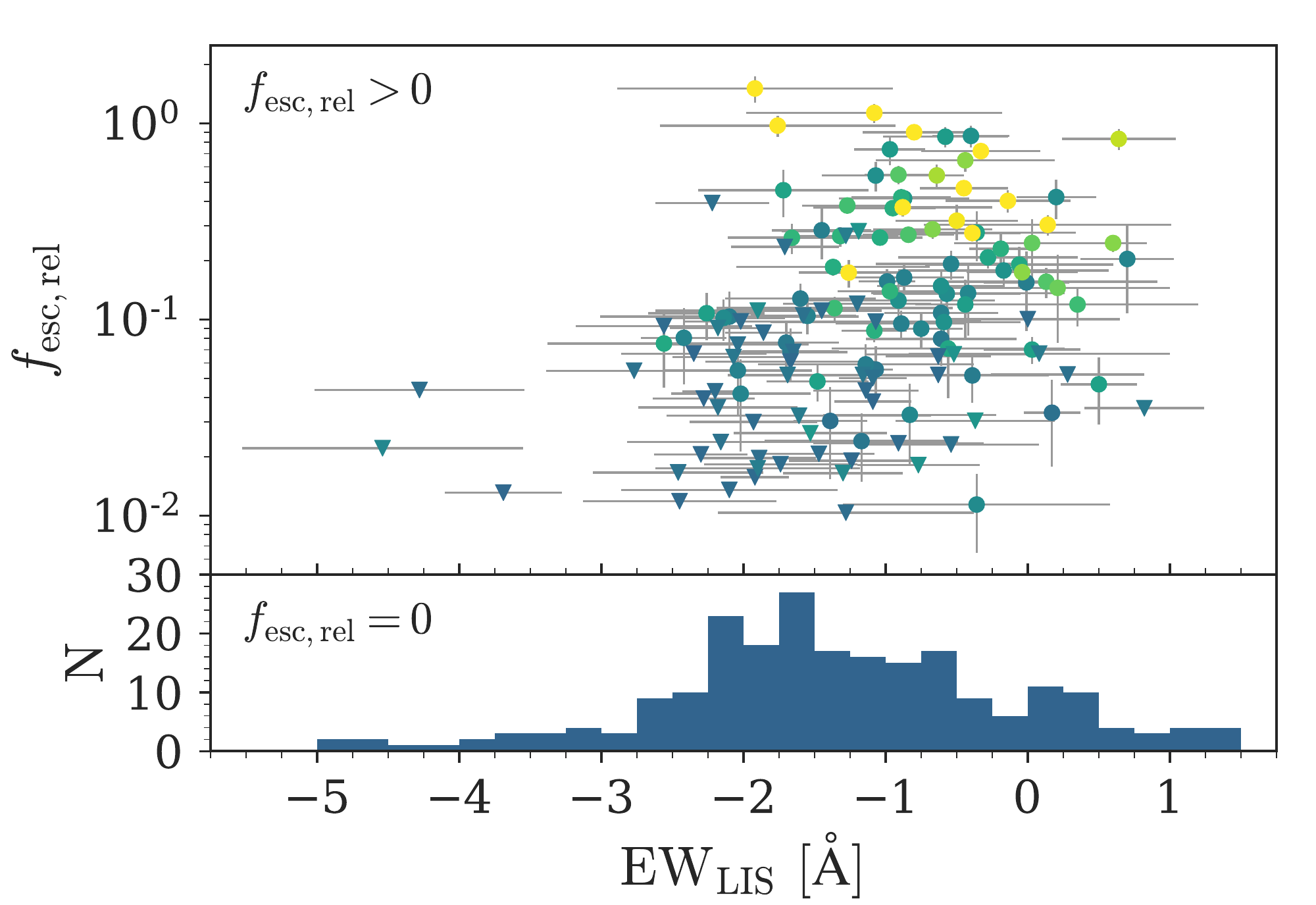}\includegraphics[height=2.2in]{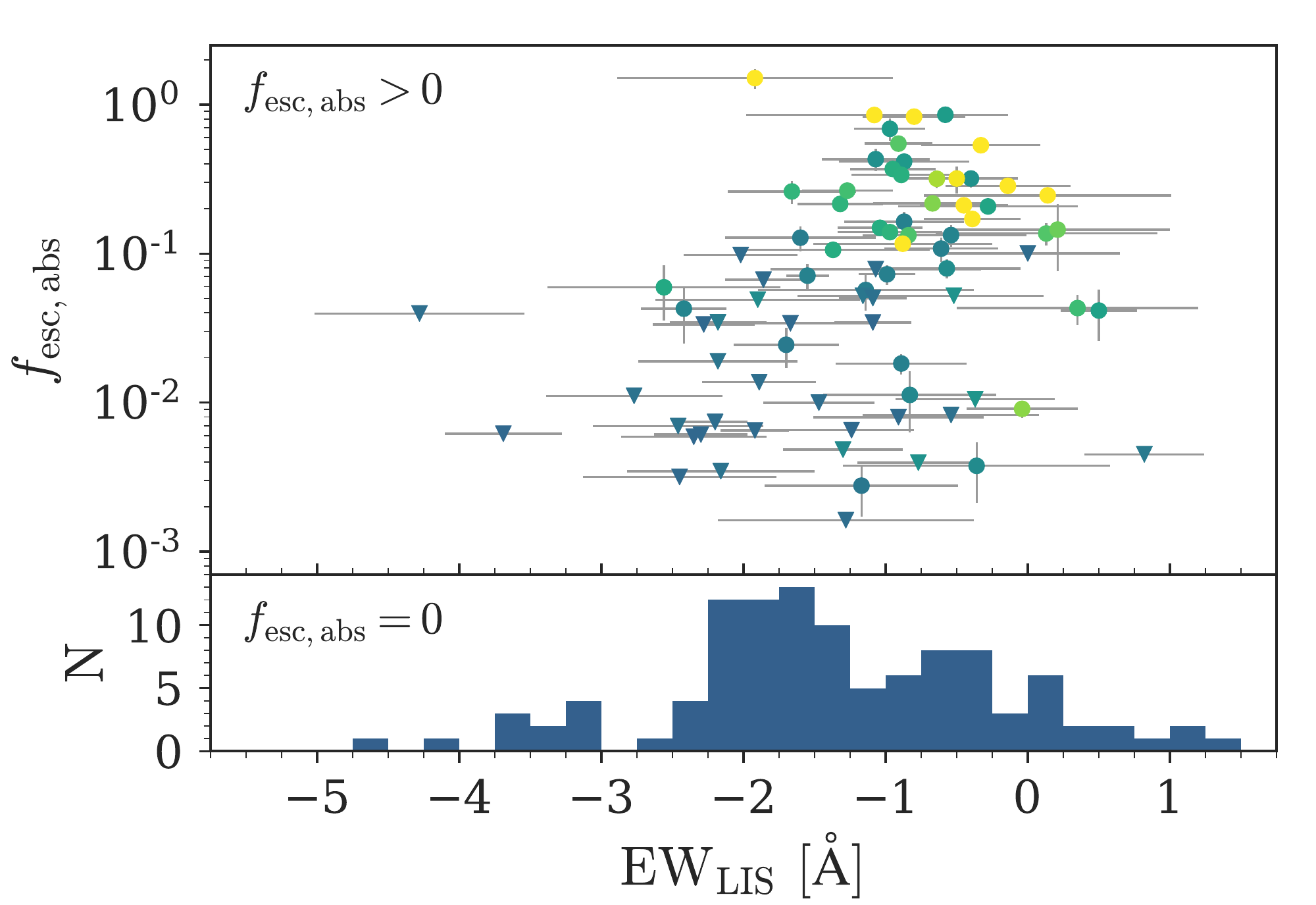}\includegraphics[height=2.2in]{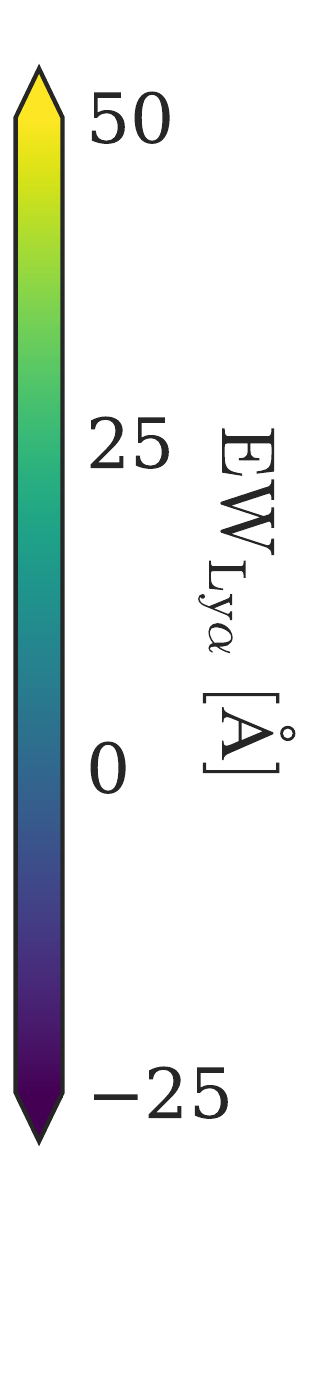}
\caption{The \lya\ photon escape fraction vs. the rest-frame equivalent width of
  LIS absorption, with color coding by \lya\ equivalent width. Circles
  indicate formal detections, while triangles indicate 2$\sigma$ upper
  limits on $f_\mathrm{esc}$. The left
  panel gives the escape fraction of \lya\ photons relative to \ha\
  (Eq.~\ref{eq:fescrel}) for 368 galaxies with
  detected \lya\ and \ha\ emission. The right panel gives the
  absolute escape fraction of \lya\ photons   (Eq.~\ref{eq:fescabs}) for 188 galaxies that also
  have robust estimates of the Balmer decrement (used to dust-correct the
  \ha\ flux). Bottom panels show the EW$_\mathrm{LIS}$ distribution for
  galaxies with $F_{\mathrm{Ly}\alpha}<0$.}
\label{fig:fesc_vs_lis_lya}
\end{figure*}
 
\ha\ provides an effective proxy for the intrinsic \lya\ luminosity
because the former is
not significantly scattered by \ion{H}{1}; however, it nonetheless suffers
extinction by interstellar dust. The intrinsic \ha\ flux (and thus,
the instrinsic \lya\ flux) can therefore only be determined using 
the absolute attenuation $A_{\mathrm{H}\alpha}$, which is typically estimated
from the inferred nebular reddening (i.e., E($B-V$) as defined in Sec.~\ref{sec:ebmv}) and
the application of an attenuation relation that is appropriate to the
galaxy at hand. As discussed by \citet{theios2019}, no single attenuation relation is
able to self-consistently describe the host of photometric and
spectroscopic properties inferred for KBSS galaxies. Given these
ambiguities in the attenuation correction, we present both the relative
(i.e., dust-uncorrected) and absolute (i.e., dust-corrected) \lya\
escape fractions based on the following definitions:
\begin{align}
f_{\rm esc,rel}&=\frac{F_{\rm Ly\alpha}}{8.7\times F_{\rm H\alpha,obs}}\label{eq:fescrel}\\[5pt]
f_{\rm esc,abs}&=\frac{F_{\rm Ly\alpha}}{8.7\times F_{\rm H\alpha,corr}}\label{eq:fescabs}
\end{align}

\noindent where $F_{\rm H\alpha,obs}$ is the observed \ha\ flux and
$F_{\rm H\alpha,corr}$ is that corrected based on the observed
E($B-V$)$_{\rm{neb}}$ and the application of a \citet{cardelli1989}
attenuation relation. In both cases, $F_{\rm Ly\alpha}$ is the
\emph{observed} \lya\ flux as defined in Eq.~\ref{eq:flya}, which is
not dust-corrected.

We calculate $f_{\rm esc,rel}$ via Eq.~\ref{eq:fescrel} for 368
galaxies for which we have a 
5$\sigma$ detection of \ha\ (we
take $f_{\rm esc,rel}=0$ where $F_{\rm Ly\alpha}\le 0$) as well as a
measurement of \ewlis.  As described
in Sec.~\ref{sec:ebmv}, only 208 galaxies have a robust
measurement of E($B-V$)$_{\rm{neb}}$; when combined with the
S/N cuts on \ha\ and \ewlis, this leaves 188 galaxies for 
which $f_{\rm esc,abs}$ may be calculated with confidence via
Eq.~\ref{eq:fescabs} (although 
subject to the remaining uncertainty in the attenuation law as well as
differential slit losses in \lya\ vs. \ha). 

Fig.~\ref{fig:fesc_vs_lis_lya} displays the relationship between
\ewlis\ and $f_{\rm esc}$ in both its absolute and relative forms. In
either case, the
two quantities have a relatively strong correlation given the
measurement uncertainties ($r_{\rm Sp}=-0.50$,
$p=6\times10^{-13}$ for 188 galaxies for $f_{\rm esc,abs}$;
Table~\ref{table:correlations}). Again, our analysis that follows is
restricted to using \ewlis\ as a proxy for \lya\ escape because $f_{\rm
esc}$, like \vlya, is not typically measurable in cases where we would
like to predict \ewlya.

\subsection{Proxies for \lya\ production} \label{sec:production}

\subsubsection{\rm SFR, Mass, and Luminosity} \label{sec:photparams}
 
We now consider parameters that may be associated with \lya\
production. As noted above, \ha\ 
luminosity should be a fairly direct proxy for the intrinsic luminosity of a galaxy
in \lya. However, it is less related to the \textit{efficiency} of
\lya\ production as described by \ewlya: dust-corrected $L_{\rm
  H\alpha}$ is uncorrelated with
\ewlya\  in our sample ($r_{\rm Sp}=-0.05$, $p=0.4$) for the 208
galaxies with robust estimations of E($B-V$) and $A_{\rm{H}\alpha}$
(see Sec.~\ref{sec:ebmv}); the same
correlation holds between \ewlya\ and SFR$_{\rm{H}\alpha}$ since
SFR$_{\rm{H}\alpha}$ is linearly related to
$L_{\rm{H}\alpha}$. \footnote{This calculation 
  includes 208 galaxies with
robust dust corrections as described in Sec.~\ref{sec:ebmv}; the
correlation with dust-uncorrected $L_{\rm H\alpha}$ is similarly weak
despite the much larger sample.}

Our photometry-based estimates of SFR$_{\rm{SED}}$ display slightly
stronger relationships with \ewlya\ ($r_{\rm Sp}=-0.17$,
$p=10^{-5}$), and the 
correlation for stellar mass is very similar ($r_{\rm
  Sp}=-0.15$, $p=10^{-4}$) for the 637 galaxies with SED
fits and \ewlya\ measurements. sSFR
(SFR$_{\rm{H}\alpha}$/$M_*$) displays a slightly stronger correlation
with \ewlya\ ($r_{\rm
  Sp}=0.23$, $p=10^{-3}$) with lower significance due
to the smaller sample size of objects with the necessary measurements
of both SFR$_{\rm{H}\alpha}$ and $M_*$ (199 galaxies).

Rest-UV absolute magnitudes M$_{\rm{UV}}$ are measured from the
$G$ and $\mathcal{R}$ band magnitudes. The apparent
magnitude corresponding to a rest-frame wavelength
$\lambda_\text{rest}=1450$\AA\ is estimated by taking a weighted
average of the $G$ and $\mathcal{R}$ based on the redshift
of the galaxy.\footnote{Note that the \lya\ line falls within the $G$
  band for $z\gtrsim 2.45$, which includes roughly one quarter of our
  galaxy sample. For the galaxies in this redshift interval, we
  correct the inferred value of $M_\text{UV}$ based on the
  spectroscopic measurement of \ewlya. This correction produces a
  median change $\Delta 
  M_\text{UV}\approx0.03$, although $\Delta
  M_\text{UV}\approx0.5$ for the few most extreme \lya-emitters in our
  sample (\ewlya $\gtrsim 100$\AA).} This apparent
magnitude $m_\text{UV}$ is then converted 
to the absolute magnitude $M_\text{UV}$ based on the redshift of the
source and the luminosity distance calculated assuming a $\Lambda$CDM cosmological
model with $H_0=70\,\mathrm{km}\,\mathrm{s}^{-1}\,\mathrm{Mpc}^{-1}$,
$\Omega_m=0.3$, and $\Omega_\Lambda=0.7$. In this manner, M$_{\rm{UV}}$
is measured for each of the 637 galaxies in our SED-fit sample. This
parameter shows the weakest relationship with \ewlya\ of any quantity
we measure, with $r_{\rm
  Sp}=-1.5\times10^{-2}$ and $p=0.71$. This lack of association
between UV luminosity and \ewlya\ is remarkable given the high \ewlya\
values associated with faint, high-$z$ galaxies in other recent work;
these trends are discussed further in Sec.~\ref{sec:discussion}.

\subsubsection{\rm O32} \label{sec:o32}

By definition, the intrinsic \ewlya\ of a galaxy is the ratio of \lya\
photons to UV continuum photons, where the latter are generated
directly by OB stars and the former are generated by the gas excited and
ionized by these same stars. It is therefore sensible that \ewlya\ would be
strongly associated with the excitation and ionization states of the
gas in star-forming regions.

The O32 line ratio is
one commonly-used indicator of nebular ionization
\citep{sanders2016,steidel2016,strom2017,strom2018}:

\begin{align}
\text{O32} \equiv \log\left(\frac{[\text{O III}]\, \lambda\lambda
  4960,5008}{[\text{O II}]\, \lambda\lambda 3727,3729}\right)
\label{eq:o32}
\end{align}

For the
ionization-bounded \ion{H}{2} regions typically assumed in
photoionization models of star-forming galaxies, O32 is approximately
proportional to log($U$), where $U$ denotes the ``ionization
parameter'', the local number of hydrogen-ionizing photons per
hydrogen atom. (see discussion by \citealt{steidel2016}). Furthermore, O32 has
previously been found to 
correlate strongly with \lya\ emission (e.g.,
\citealt{trainor2016,nakajima2016}).

Notably, however, recent work has
suggested that elevated O32 values may correspond in some cases to density-bounded
\ion{H}{2} regions, in which the ionized region is not entirely
surrounded by neutral gas\footnote{Essentially, the local ratio of
  \ion{O}{2} to \ion{O}{3} increases toward the edge of the
  Str{\"o}mgren sphere for an ionization-bounded nebula. For a nebula
  that is optically thin to ionizing photons, this ionization front
  (and its associated region of stronger \ion{O}{2} emission) is
  not present. See e.g., \citet{pellegrini2012}.} \citep{nakajima2013,trainor2016,izotov2016}. In particular,
several recent detections of escaping Ly-continuum (rest-frame
H ionizing) photons from galaxies at low and high redshift have been
accompanied by elevated O32 ratios (e.g.,
\citealt{debarros2016,izotov2016,izotov2018, fletcher2018}, but
c.f. \citealt{borthakur2014} and \citealt{shapley2016} who find
Ly-continuum escape in the absense of extreme O32). In these
scenarios, an association of large O32 with high \ewlya\ may reflect a
combination of both increased ionizing photon production \emph{and}
increased probability of photon escape due to the lack of surrounding
neutral gas. 

O32 measurements for the KBSS sample are calculated as described in
\citet{strom2017}. Briefly, the [\ion{O}{3}]
$\lambda\lambda$4960,5008 and [\ion{O}{2}] $\lambda\lambda$3727,3729
line fluxes are measured as described in
Sec.~\ref{sec:mosfire}. We require
that both [\ion{O}{3}] and [\ion{O}{2}] be detected with S/N $> 3$,
resulting in a sample of 316 galaxies with a measured raw O32
value. Dust-corrected O32 values are measured for a smaller sample of
174 objects that meet both the requirements described above as well as the
cut on the S/N of the Balmer decrement described in
Sec.~\ref{sec:fesc}. For these measurements, each of the
[\ion{O}{3}] and [\ion{O}{2}] emission lines are corrected for
extinction using the measured Balmer decrement and a \citet{cardelli1989}
attenuation curve before calculating the line ratio. These two
O32 estimators have the highest individual correlations with \ewlya\
of any ``production''-related parameter: $r_{\rm
  Sp}=0.43$ ($p=10^{-15}$) for the raw O32 measurements with a
slightly higher correlation strength and lower significance for the
smaller sample of dust-corrected O32 values
(Table~\ref{table:correlations}). However, O32 also displays a
strong correlation with \ewlis\ ($r_{\rm
  Sp}=0.47$), perhaps reinforcing the idea that O32 is not wholly
a measure of \lya\ production. 

\subsubsection{\rm O3} \label{sec:o3}

The O3 ratio is another indicator of nebular ionization and
excitation:
\begin{align}
\text{O3} \equiv \log\left(\frac{[\text{O III}]\, \lambda 5008}{\text{H}\beta}\right)\quad.
\label{eq:o3}
\end{align}

As discussed by \citet{trainor2016}, the O3 ratio is strongly
associated with O32 for the high-excitation galaxies typical at
$z\gtrsim 2$: the two quantities are correlated with $r_{\rm
  Sp}=0.74$ in the KBSS sample. Likewise, \citet{strom2018}
demonstrate that O3 is an effective indicator of log($U$) through extensive
photoionization modeling of the KBSS galaxy sample. O3 therefore has
many of the same advantages as O32 for indicating \lya\ production.

However, O3 has two significant
advantages over O32. Firstly, O3 relies on two emission lines at similar
wavelengths, which makes the 
ratio insensitive to both dust extinction and cross-band
calibration. Secondly, O3 is insensitive to the differences between
density-bounded and ionization-bounded \ion{H}{2} regions, so it may indicate
nebular excitation (and \lya\ production) in a manner more decoupled
from the physics of \lya\ escape. Based on these advantages, we rely
on O3 as our primary metric of \lya\ production efficiency for the
remainder of this work.

Using the same line-fitting process described above to
estimate the line fluxes and uncertainties, we calculate the O3 ratio
for every galaxy that has S/N $> 3$ for both  [\ion{O}{3}] and \hb,
a total of 395 objects. O3 has a correlation with
\ewlya\ that is only marginally weaker than the corresponding
correlation for O32 ($r_{\rm  Sp}=0.40$, $p=5\times10^{-16}$).

\subsection{Summary of correlations with \lya} \label{sec:correlations}

Again, Spearman rank correlation statistics for \ewlya\ and \ewlis\
with other measured quantities are presented in
Table~\ref{table:correlations}. Each quantity is calculated
for a different number of objects according to the cuts described
above, and the $p$ values of every measured correlation (which depend both on
the measured $r$ and the number of objects in the sample) are highly
significant ($p\ll 1$ in nearly all cases). However, there is a wide range of
$r$ values, indicating that certain parameters explain only a small
fraction in the total variation in \ewlya\ despite the statistical
significance of their correlation.

Note
that \ewlis\ is strongly correlated with the escape fraction of \lya\,
as expected based on the arguments that both of these quantities are
related to the ability of \lya\ photons to escape galaxies (see Sec.~\ref{sec:ewlis} and
Fig.~\ref{fig:fesc_vs_lis_lya}). Conversely, \ewlis\ has a
much weaker\footnote{While the correlation is highly significant at
  $p<10^{-6}$, the low rank correlation coefficient $r_{\rm Sp}=0.21$ indicates
  that most of the variation in \ewlis\ is not associated with variation
  in O3.} correlation with O3 despite the fact that both quantities
show relatively strong correlations with \ewlya. We interpret this relationship in
the sections that follow, but it is suggestive of the fact that these
two quantities capture different processes (escape and production)
related to the observed \ewlya.

\begin{figure*}
\center
\includegraphics[width=0.8\linewidth]{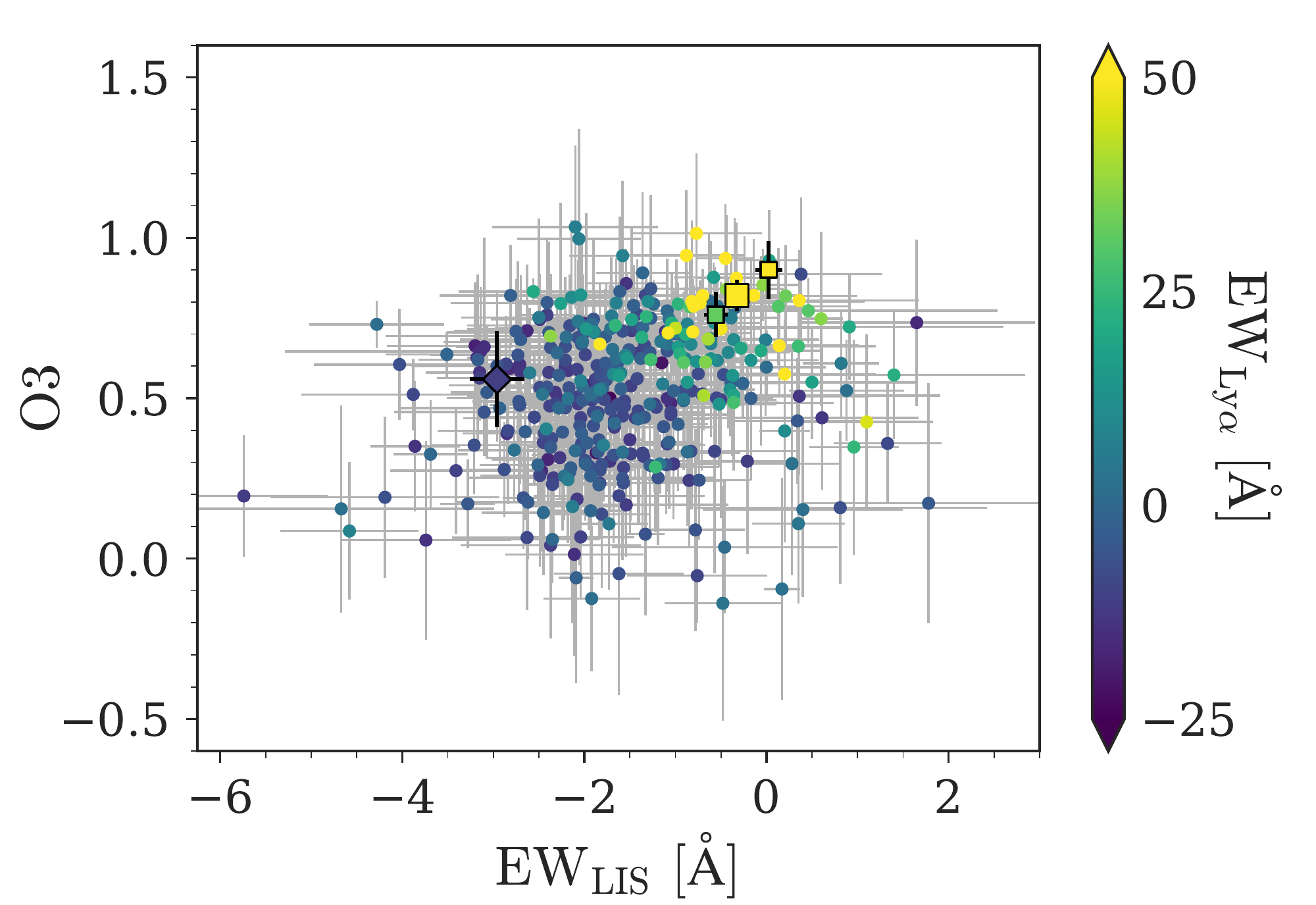}
\caption{O3 ratio ($\equiv\log($[\ion{O}{3}]/\hb$)$) vs. rest-frame equivalent
  width of LIS absorption for 377 galaxies, with colors denoting \lya\
  equivalent width. Note that O3 and EW$_\mathrm{LIS}$ are not
  strongly correlated with each other, but EW$_{\mathrm{Ly}\alpha} >0$ is strongly
  associated with both weak LIS absorption (EW$_\mathrm{LIS}\approx
  0$) and strong [\ion{O}{3}] emission (O3 $\gtrsim 0.5$). Square
  boxes with black borders correspond to stacked measurements of 
  faint $L\sim0.1L_*$ \lya-selected galaxies from
  \citet{trainor2015,trainor2016}. The large square denotes the full
  sample, and the smaller squares denote measurements based on
  splitting about the median value of \ewlya. Diamond with black
  border represents MS 1512-cB58 based on measurements from
  \citet{pettini2002} and \citet{teplitz2000}.}
\label{fig:o3_vs_lis}
\end{figure*}

 Fig.~\ref{fig:o3_vs_lis} visually demonstrates this same
result. While O3 and \ewlis\ are themselves not closely
correlated, there is a clear trend toward high \ewlya\ in the
upper-right corner of Fig.~\ref{fig:o3_vs_lis} (i.e., the region of
high O3 and/or \ewlis\ $\sim$ 0) and low \ewlya\ in the
lower-left corner (i.e., the region of low O3 and/or
strongly-negative \ewlis).

Fig.~\ref{fig:o3_vs_lis} also includes
measurements from stacked spectra of 
  faint $L\sim0.1L_*$ \lya-selected galaxies from
the KBSS-\lya\ survey; these measurements are shown as boxed
points. The \ewlis\ measurements are described by 
\citet{trainor2015}, while the O3 measurements are described by
\citet{trainor2016}. The faint galaxy measurements follow the same
trend as the individual measurements from the brighter KBSS galaxies,
in that high \ewlya\ is associated with the upper-right corner of the
parameter space.

For comparison, we also include measurements of MS 1523-cB58, a
gravitationally lensed galaxy with $M_*\approx 10^9$~\msun, SFR
$\approx 50-100$~\msun\ yr$^{-1}$, and a young age $\sim$9~Myr
\citep{siana2008}. The reported O3 value is based on
spectroscopic measurements reported by \citet{teplitz2000}, while the
\lya\ and LIS equivalent widths are new measurements from the Keck/ESI
spectrum presented by \citet{pettini2002}. Despite its young age and
large star-formation rate -- both of which would predict a high rate
of \lya\ production -- the spectrum of MS 1512-cB58 (hereafter cB58)
displays net \lya\ absorption,\footnote{Note that 
the detailed cB58 \lya\ profile displays weak \lya\ emission
superimposed on a much stronger damped \lya\ absorption profile, as
described by \citet{pettini2000, pettini2002}. Due to the lower S/N of
our KBSS \lya\ measurements, we simply describe each galaxy as a net
absorber or emitter for the purposes of this paper.}
consistent with its deep LIS absorption lines
(\ewlis~$\approx-3$\AA). The galaxy cB58 thus obeys the same
association as the KBSS galaxies between \lya\ emission and position
in the O3-\ewlis\ parameter space.

The structure of Fig.~\ref{fig:o3_vs_lis} therefore suggests that a linear
combination of O3 and \ewlis\ would better predict \ewlya\ than either
quantity alone. That is, we could in principle define a single
parameter which is maximized 
when both O3 and \ewlis\ predict strong \ewlya, is minimized when both
O3 and \ewlis\ predict weak \ewlya, and which takes intermediate
values when O3 and \ewlya\ have contradictory implications for the
value of \ewlya. We develop such a model in
Sec.~\ref{sec:xlya} below.

\section{Combined Model} \label{sec:xlya}

Motivated by the arguments above, we construct a new parameter
$X_\mathrm{LIS}^\mathrm{O3}$ with the following definition:

\begin{align}
X_\mathrm{LIS}^\mathrm{O3}&=\alpha\, \left(\ewlis/\mathrm{\AA}\right)+(1-\alpha)\, \mathrm{O3}\quad.\label{eq:xlya}
\end{align}

This parameter has the behavior described at the end of
Sec.~\ref{sec:correlations}: $X_\mathrm{LIS}^\mathrm{O3}$ is maximized
when both O3 and \ewlis\ are maximized (i.e., when our proxies for both
\lya\ production and \lya\ escape suggest that \ewlya\ should be
strong). Likewise, $X_\mathrm{LIS}^\mathrm{O3}$ will take smaller values
when either or both of O3 and \ewlis\ are small (i.e., when \ewlya\ is
expected to be small according to Fig.~\ref{fig:o3_vs_lis}). We
therefore may expect any equation with the form of Eq.~\ref{eq:xlya}
to predict a strong, monotonically increasing relationship between
\ewlya\ and $X_\mathrm{LIS}^\mathrm{O3}$.

\begin{figure*}
\center
\includegraphics[width=0.8\linewidth]{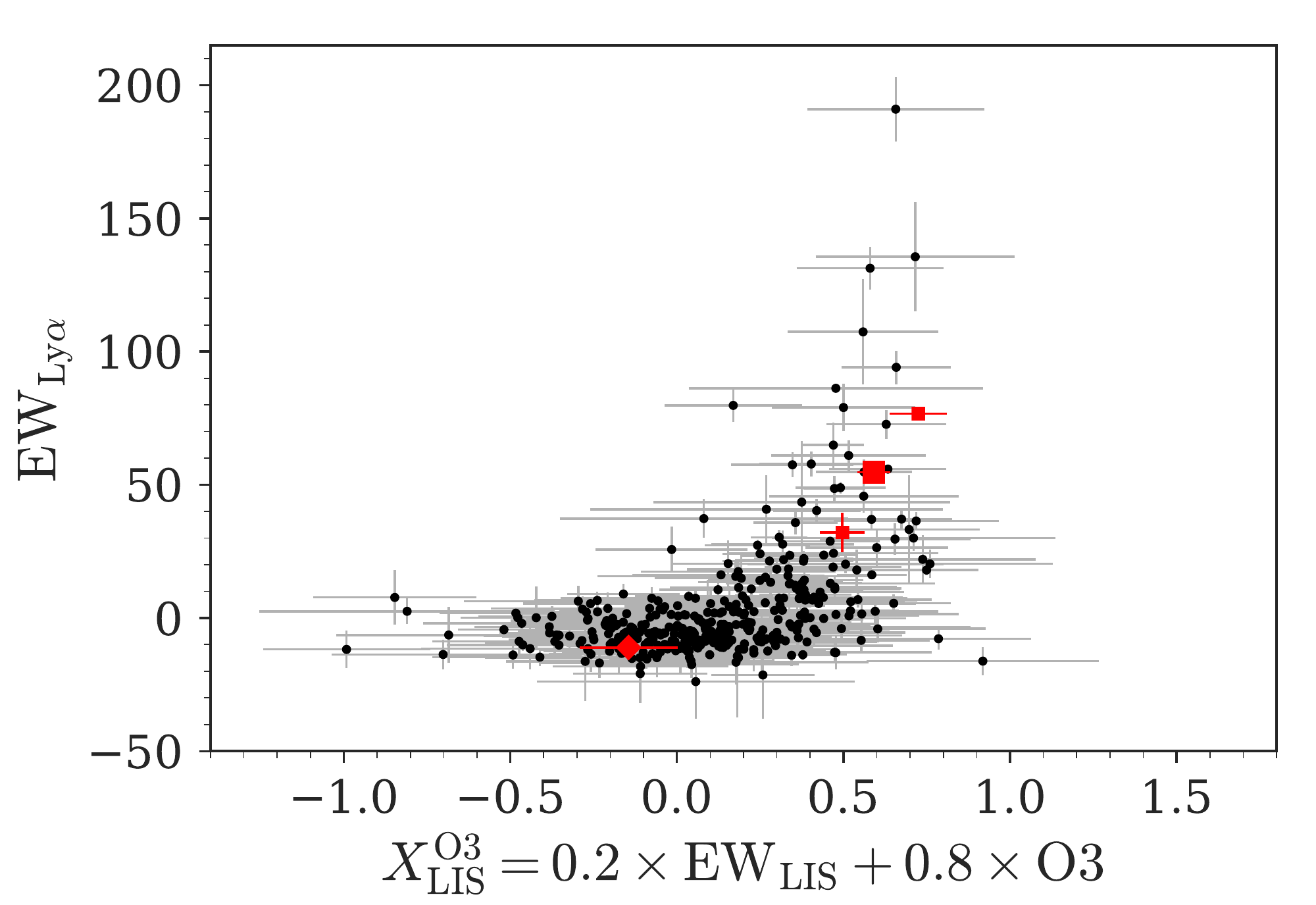}
\caption{\lya\ equivalent width vs. $X_{\mathrm{Ly}\alpha}$, a linear
  combination of O3 and EW$_\mathrm{LIS}$ that maximizes the Spearman
  rank correlation with EW$_{\mathrm{Ly}\alpha}$: $r_{\rm Sp}=0.49$,
  $p=1.5\times10^{-24}$. Approximately half of the
  ordering in observed EW$_{\mathrm{Ly}\alpha}$ is explained by these two
  variables alone, and 90\% of the variance in \ewlya\ is accounted
  for by the combination of an exponential model and the 2D
  measurement uncertainties (Sec.~\ref{sec:expmodel}). As in
  Fig.~\ref{fig:o3_vs_lis}, red squares
  correspond to stacked measurements of 
  faint $L\sim0.1L_*$ \lya-selected galaxies from
  \citet{trainor2015,trainor2016}, and the red diamond corresponds to
  measurements of MS 1512-cB58 from
  \citet{pettini2002} and \citet{teplitz2000}.}
\label{fig:xlya_vs_lya}
\end{figure*}

We then tune
$\alpha$\footnote{Note that $\alpha$ is a 
dimensionless number that sets the weighting of \ewlis\
(measured in \AA) relative to O3 (measured in dex); this arbitrary
parameterization was chosen so that typical values of
$X_\mathrm{LIS}^\mathrm{O3}$ would be of order unity for the galaxies
in our sample.} to maximize the 
predictive power of this relationship. Specifically, we choose the
value of $\alpha$ that maximizes the  rank correlation coefficient
between $X_\mathrm{LIS}^\mathrm{O3}$ and \ewlya, yielding a maximum
correlation of $r_\mathrm{Sp}=0.49$ for
$\alpha\approx0.2$.\footnote{This calculation is performed for the 377
galaxies with robust measurements of \ewlya, \ewlis, and O3; see
Table~\ref{table:correlations}.} Repeating 
this procedure on 1000 bootstrap samples, we find that the optimum
value of $\alpha$ is constrained to $0.19\pm0.06$, and the bootstrap
samples are correlated with \ewlya\ with $r_\mathrm{Sp}=0.49\pm0.04$
for fixed $\alpha=0.2$. The relationship between \ewlya\
and $X_\mathrm{LIS}^\mathrm{O3}$ is displayed in
Fig.~\ref{fig:xlya_vs_lya}. As expected from Fig.~\ref{fig:o3_vs_lis},
strong \lya\ emission is closely associated with large
$X_\mathrm{LIS}^\mathrm{O3}$.

\begin{figure}
\center
\includegraphics[width=\linewidth]{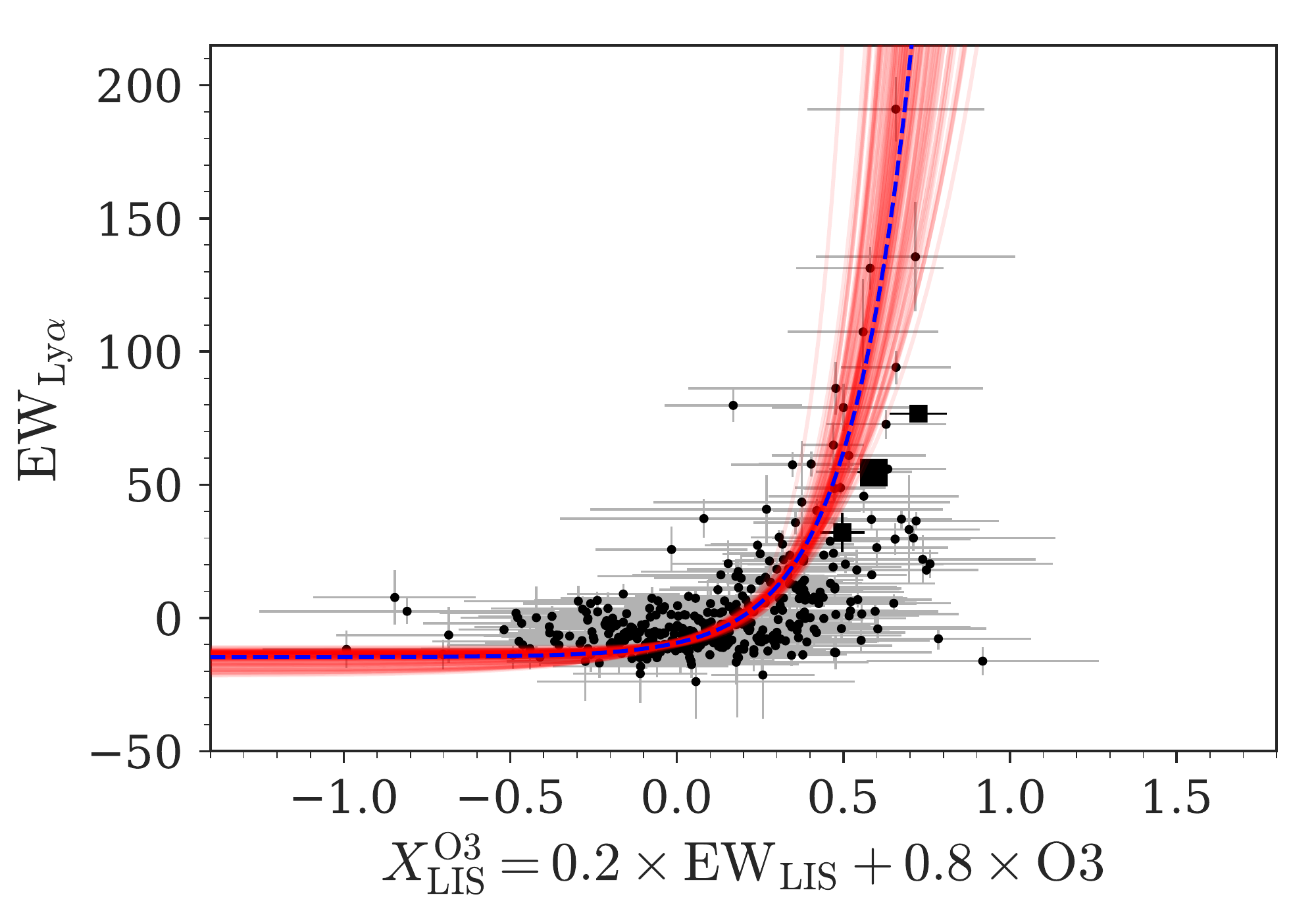}
\caption{Data are the same as in Fig.~\ref{fig:xlya_vs_lya}, with a
  best-fit exponential relationship (blue dashed curve) and 100 fits
  to bootstap realizations of the data (faint red curves). The faint
  galaxy stacks and cB58 spectrum (square and diamond symbols) are not
  used in calculating the best-fit model parameters, but their
  positions are well-described by our exponential model.}
\label{fig:xlya_vs_lya_fit}
\end{figure}
 
\subsection{Exponential Model and Variance} \label{sec:expmodel}

Motivated by the distribution of points in Fig.~\ref{fig:xlya_vs_lya},
we fit an exponential model to the data. Because there are large
uncertainties in both axes, we choose model parameters to minimize the
2D distance between the model and data, scaled by the corresponding
uncertainties in each dimension. In effect, we define a 2D analog of the traditional
$\chi^2$ parameter: 
\begin{align}
\chi_\mathrm{2D}^2&=\sum_i\left(\left(\frac{x_i-x_{c,i}}{\sigma_{x,i}}\right)^2+\left(\frac{y_i-y_{c,i}}{\sigma_{y,i}}\right)^2\right)
\end{align}

\noindent which our model-fitting seeks to minimize. In the above
equation, $x_i$ and $y_i$ represent the values of \ewlya\ and
$X_\mathrm{LIS}^\mathrm{O3}$ for the $i^\mathrm{th}$ object in our
sample; $\sigma_{x,i}$ and $\sigma_{y,i}$ are the associated
uncertainties for that object; and $(x_{c,i},\, y_{c,i})$ is the
closest point on the model curve to the observed values $x_i$ and
$y_i$, scaled by their corresponding uncertainties.  Our
model takes the following form:
\begin{align}
\ewlya&=\mathrm{EW}_0+A\,\mathrm{e}^{X_\mathrm{LIS}^\mathrm{O3}/\beta}
\end{align}
\noindent where the best fit coefficients and 1D marginalized
uncertainties are found to be: 
\begin{align}
\mathrm{EW}_0&=-15\pm2\label{eq:y0fit}\\
A&=5\pm2\label{eq:afit}\\
\beta&=0.19\pm0.04\label{eq:x0fit}
\end{align}

The uncertainties in the model parameters are calculated by repeating the
fit on 500 bootstrap 
samples of the data, where each bootstrap sample contains 377 points
and their corresponding uncertainties selected at random (with
replacement) from the true set of 377 points in the
sample. Fig.~\ref{fig:xlya_vs_lya_fit} displays the best fit curve
along with 100 representative fits to the bootstrap samples to
demonstrate the uncertainty in the fit model.

The best-fit curve corresponds to $\chi_\mathrm{2D}^2=515$ for 377
observations, or $\chi_\mathrm{2D}^2/N_\mathrm{dof}=1.36$. Being above
unity, this value indicates that the typical observations differ from
the best-fit model by more than their formal uncertainties, such that
there is intrinsic scatter in the relationship between \ewlya\ and
$X_\mathrm{LIS}^\mathrm{O3}$ that is neither described by our model
nor by our estimated observational uncertainties.

In order to determine the fraction of the total variance in \ewlya\
captured by the combination of our model and our measurment
uncertainties, we perform the following exercise. We begin by
assigning each object a fiducial value of \ewlya\ and
$X_\mathrm{LIS}^\mathrm{O3}$ according to the nearest point to the
observed values on the best-fit curve (where proximity to the curve is
calculated in 2D, weighted by the uncertainty in each dimension). Each
point is then perturbed in 
both dimensions, with the perturbation drawn from a gaussian
distribution $\phi(\mu,\sigma)$ with $\mu=0$ and $\sigma$ equal to the estimated
uncertainty in \ewlya\ or $X_\mathrm{LIS}^\mathrm{O3}$ for that
object. The resulting simulated data thus represents a hypothetical
sample consistent with the model, with scatter given only by the
observational uncertainties.

A new fit to
the resulting simulated data set is calculated (i.e., new coefficients
are calculated for Eqs.~\ref{eq:y0fit}$-$\ref{eq:x0fit}), and the
following statistics are calculated to assess the variance in the
simulated data: (1) the Spearman rank correlation coefficient $r_s$ of the
simulated data, and (2) the $\chi_\mathrm{2D}^2$ coefficient assessing
the goodness of fit of the simulated data to its own best-fit
model. Repeating this process 100 times, we find that the resulting
distribution of statistics have $\langle r_s\rangle=0.59\pm0.02$ and
$\langle\chi_\mathrm{2D}^2/N_\text{dof}\rangle=0.98\pm0.04$. Comparing
these values to the same statistics for the real data ($r_s=0.49$,
$\chi_\mathrm{2D}^2/N_\text{dof}=1.36$), we see that the simulated
data have a tighter correlation and closer agreement with the fitted
relation than do our real data;  again, this indicates that the real
data have additional sources of intrinsic scatter not described by our
model or our estimated measurement errors. 

We therefore model the intrinsic scatter in the relationship of \ewlya\ and
$X_\mathrm{LIS}^\mathrm{O3}$ by assuming an underlying equation of the
following form:
\begin{align}
\ewlya&=f(X_\mathrm{LIS}^\mathrm{O3})+\phi(0,\sigma_\mathrm{int})
\end{align}

\noindent where $\phi(\mu,\sigma)$ is a number drawn from the gaussian
distribution with $\mu=0$, and $\sigma_\mathrm{int}$ describes the intrinsic
scatter in \ewlya\ at fixed $X_\mathrm{LIS}^\mathrm{O3}$.\footnote{Note
  that $\sigma_\mathrm{int}$ is assumed not to vary with
  $X_\mathrm{LIS}^\mathrm{O3}$ for simplicity. While the observed
  distribution of \ewlya\ shows significantly more scatter at large
  $X_\mathrm{LIS}^\mathrm{O3}$  than at smaller values, we find that
  this effect is entirely consistent with the trend of increasing
  measurement uncertainties on both axes as
  $X_\mathrm{LIS}^\mathrm{O3}$ and \ewlya\ increase.} In this manner, we can
simulate values of \ewlya\ and   $X_\mathrm{LIS}^\mathrm{O3}$ drawn
from the model distribution (including random perturbations
corresponding to the estimated measurements uncertainties, as above),
but with an additional term corresponding to the assumed intrinsic
scatter that can be increased until 
the simulated data have similar total scatter (parameterized by $r_s$ and
$\chi_\mathrm{2D}^2$) to the observed data.

In practice, we find that a value $\sigma_\mathrm{int}=7\pm1\u{\AA}$
produces the best match to the statistical properties of the observed
data, with $r_s=0.50\pm0.03$ and
$\chi_\mathrm{2D}^2/N_\text{dof}=1.33\pm0.06$. Adopting this value for
$\sigma_\mathrm{int}$ implies that the intrinsic variance in \ewlya\ not
described by our model is $\sigma_\mathrm{int}^2=50\u[2]{\AA}$,
while the total variance in \ewlya\ in our data set is
$\sigma_{\mathrm{obs}}^2=512\u[2]{\AA}$. Assuming we have
$\sigma_{\mathrm{obs}}^2=\sigma_{\mathrm{mod}}^2+\sigma_{\mathrm{int}}^2+\sigma_{\mathrm{meas}}^2$,
we find that $\sim$90\% of the total variance in \ewlya\ is accounted for by
our exponential model and the estimated measurement errors.

The apparent success of our two-parameter model for predicting
\ewlya\ deserves some inspection, particularly in light of the
well-known tendency (described in Sec.~\ref{sec:intro} and below) for
\lya\ emission to display substantial scatter with respect
to galaxy properties. We address this topic in Sec.~\ref{sec:discussion}.

\section{Conditional Probabilities for \lya\ Detection} \label{sec:pdist}

Despite the apparent success of the model above in self-consistently describing the
behavior of \ewlya, it has several
shortcomings. Specifically, the model described above allows us to predict the net
\lya\ emission 
of a given galaxy based on measurements of \ewlis\ and O3, but
Figs.~\ref{fig:xlya_vs_lya}$-$\ref{fig:xlya_vs_lya_fit} reveal
substantial observational scatter in this relation that is not
described by our exponential model. Furthermore, it is not obvious that an
exponential model is physically
meaningful for describing the dependence of \lya\ emission on these
properties.

An alternative method of describing the dependence of \lya\ emission
on galaxy properties would be to relinquish analytical functions for
\ewlya\ in favor of a non-parametric model for the
conditional probability of detecting \lya, given a value for one or
more other galaxy parameters. While this method does not provide an expected
numerical value for \ewlya, it allows us to explicitly
describe how the detection fraction (as well as the stochasiticity in
observed \lya\ emission) varies as a function of galaxy properties.

\subsection{Methodology}\label{sec:pdist_method}

For the majority\footnote{We do not present conditional probability
  functions for E$(B-V)_\mathrm{neb}$; the conditional PDF is similar to
  that of E$(B-V)_\mathrm{SED}$ but weaker.} of the empirical parameters listed in
Table~\ref{table:correlations}, we construct conditional probability
functions in two ways. First, we bin the full set of galaxies for
which each indicator is measured into bins with widths that are
allowed to vary in order to contain at least 30
galaxies per bin.\footnote{Note that the number of bins therefore depends on the
  total number of galaxies for which a given parameter is measured;
  see Table~\ref{table:correlations}.} Within each bin, the fraction of
galaxies with detected \lya\ in net emission ($\ewlya>0$) is plotted
as a yellow bar in the corresponding panel of Figs.~\ref{fig:pdist1}$-$\ref{fig:pdist2};
the fraction of galaxies that are net \lya\ absorbers ($\ewlya\le
0$) is plotted as a blue bar in the negative direction. This empirical
\lya-emitter fraction as a function of an observed
parameter $X$ can be interpreted as the conditional probability
distribution $P($\ewlya~$>0\,|\,X)$, hereafter $P_\mathrm{em}^{\mathrm{Ly}\alpha}(X)$. 

Displaying the empirical \lya-emitter fraction in this way has the useful
property that every galaxy contributes to the number of absorbers or
emitters for a single bin, which means that each bin is independent. However,
assigning each galaxy to a specific bin based on the observed value of
a given \lya-predicting parameter neglects the fact that the parameter values
that define the horizontal axes of
Figs.~\ref{fig:pdist1}$-$\ref{fig:pdist2} have their own 
observational uncertainties, which inhibits their assignment to a single
specific bin. Likewise, the observational uncertainty on \ewlya\
prevents a clean separation between observed \lya-emitters and
absorbers. For this reason, we construct a second, unbinned estimator of
$P_\mathrm{em}^{\mathrm{Ly}\alpha}(X)$ that explicitly incorporates both of these uncertainties.

Our unbinned estimator of $P_\mathrm{em}^{\mathrm{Ly}\alpha}(X)$ represents each
galaxy observation as a pair of 1D gaussian probability distributions
of the form $\phi(X\,|\,\mu=X_i,\sigma=\sigma_{X,i})$, where $X_i$ and
$\sigma_{X,i}$ are the observed value and observational uncertainty on
parameter $X$ for galaxy $i$. Two distributions of this form are
generated for each galaxy, with one distribution being normalized by
the probability of the observed galaxy being a \lya\ emitter, and the
second normalized by the probability of being an absorber. Formally,
we calculate the probability of an observed galaxy being a \lya\ emitter from the
measured value of \ewlya\ and its corresponding uncertainty
$\sigma_{\mathrm{EW,Ly}\alpha}$:
\begin{align}
P_{\mathrm{em},i}&=\frac{1}{2}\mathrm{erf}\left(\frac{\mathrm{EW}_{\mathrm{Ly}\alpha,i}}{\sqrt{2}\,\sigma_{\mathrm{EW,Ly}\alpha,i}}\right)+\frac{1}{2}\\
P_{\mathrm{abs},i}&=1-P_{\mathrm{em},i}
\end{align}
Thus, a galaxy with \ewlya~$>0$ and \ewlya~$\gg
\sigma_{\mathrm{EW,Ly}\alpha}$ will have $P_\mathrm{em}\approx 1$ and
$P_\mathrm{abs}\approx 0$; the ``emitter'' and ``absorber'' gaussian
probability distributions are then normalized such that their
integrals equal $P_\mathrm{em}$ and 
$P_\mathrm{abs}$, respectively. The inferred incidence $\eta$ of \lya\
emitters (absorbers) in our sample is then the sum of all the emitter
(absorber) distributions:
\begin{align}
\eta_\mathrm{em}(X)&=\sum_i P_{\mathrm{em},i}\,\phi(X\,|\,X_i,\sigma_{X,i})\\
\eta_\mathrm{abs}(X)&=\sum_iP_{\mathrm{abs},i}\,\phi(X\,|\,X_i,\sigma_{X,i})
\end{align}

\begin{figure*}
\center
\includegraphics[width=0.48\linewidth]{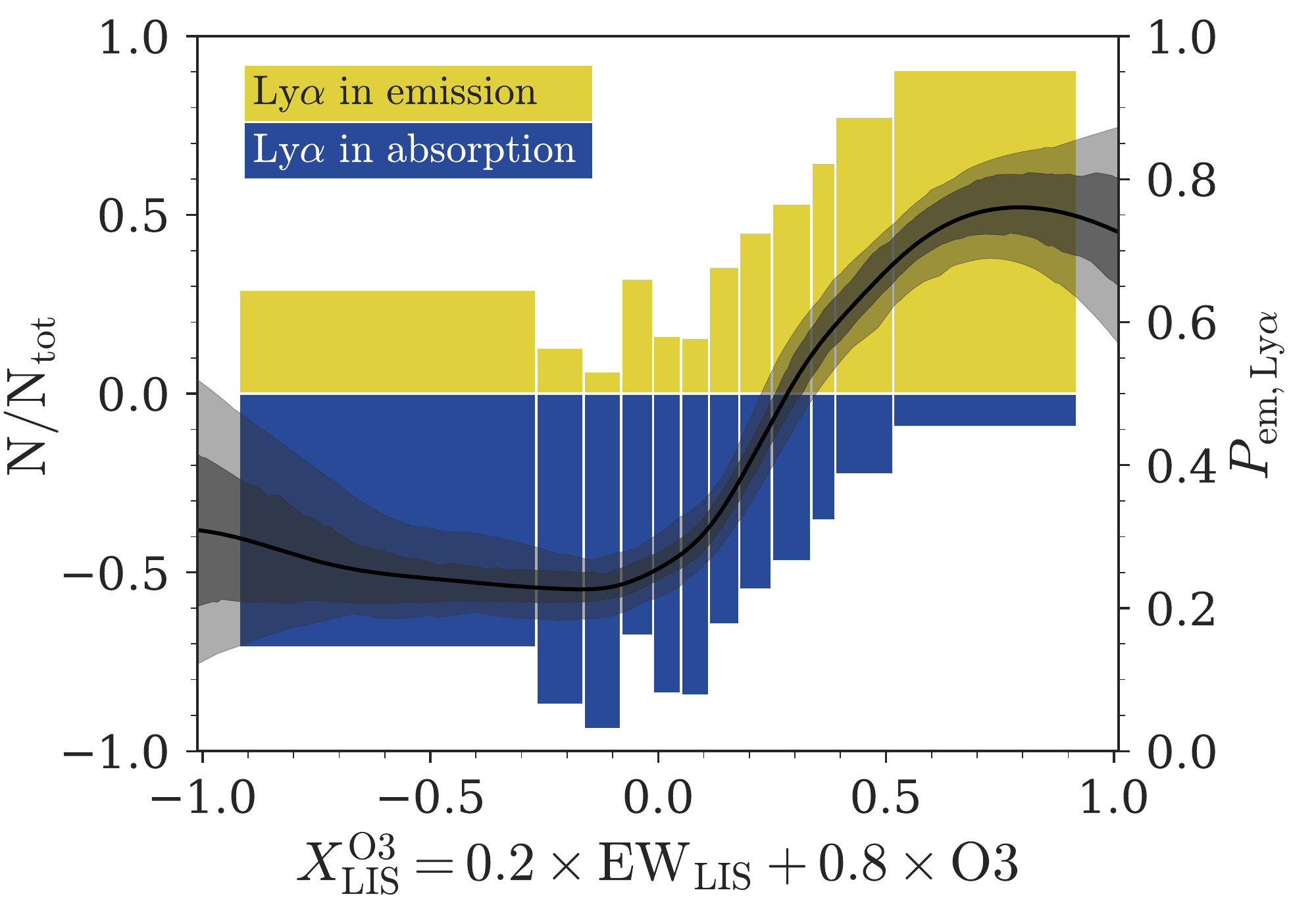}\hfill \includegraphics[width=0.48\linewidth]{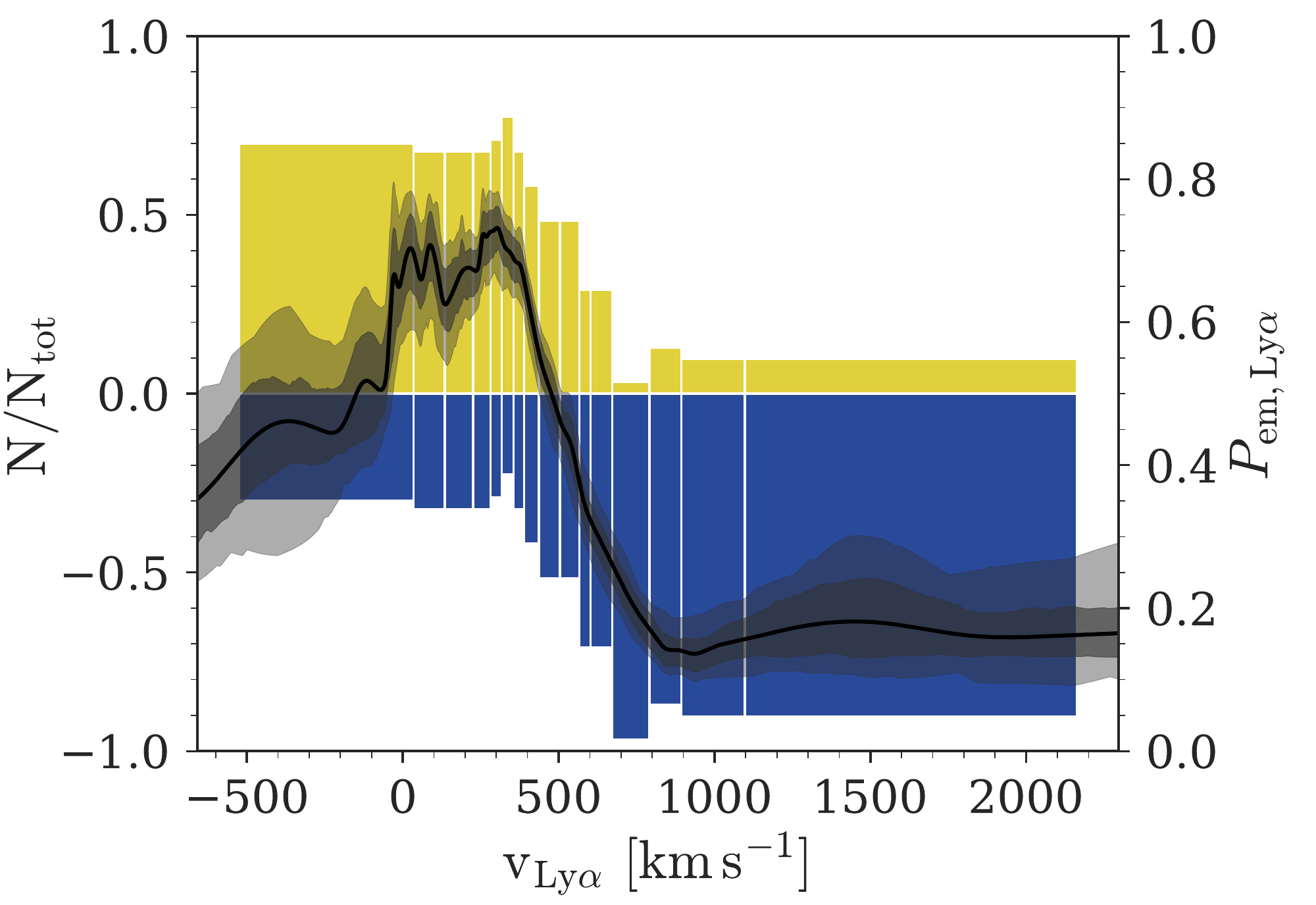}
\vspace{5mm}

\includegraphics[width=0.48\linewidth]{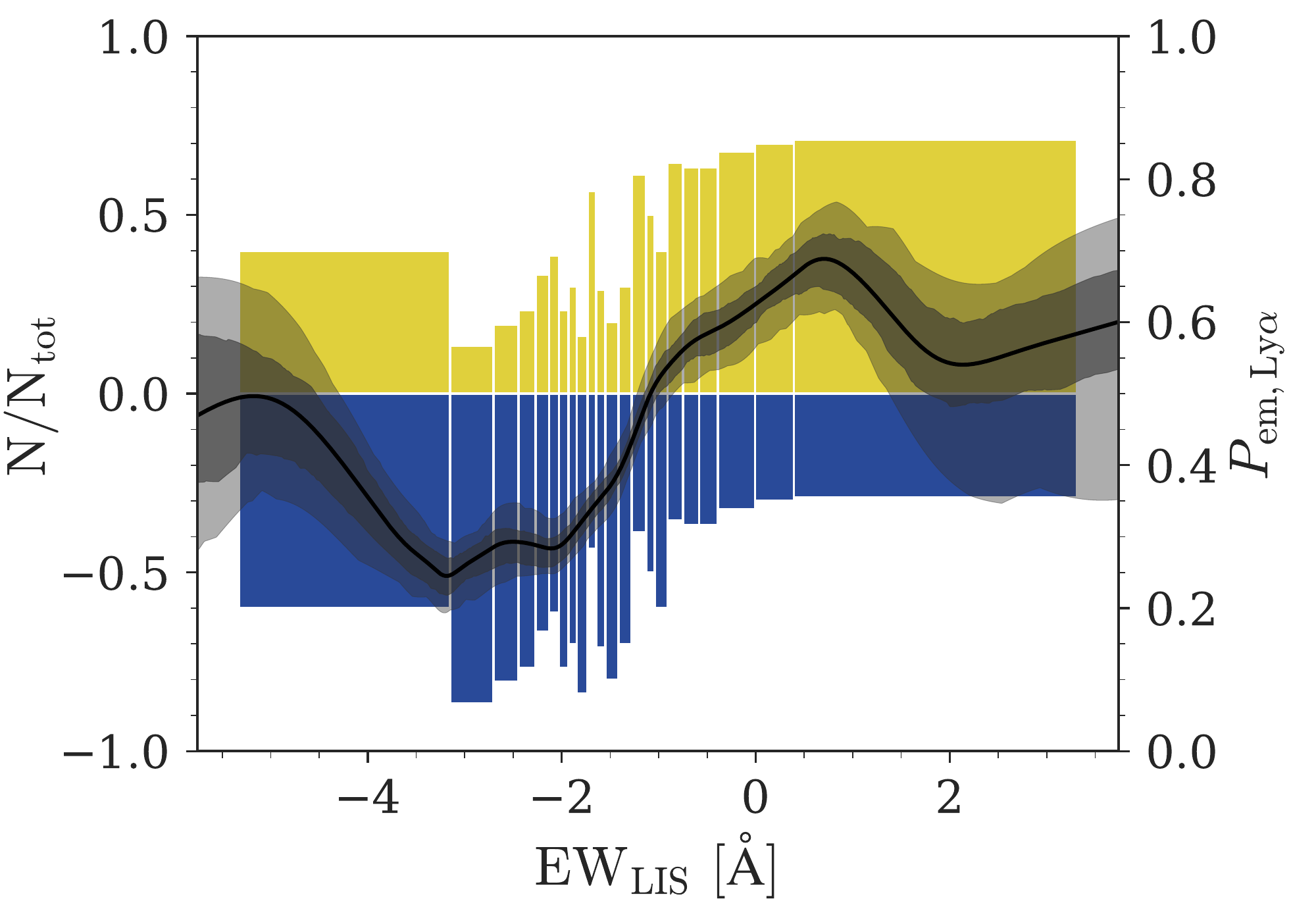}\hfill \includegraphics[width=0.48\linewidth]{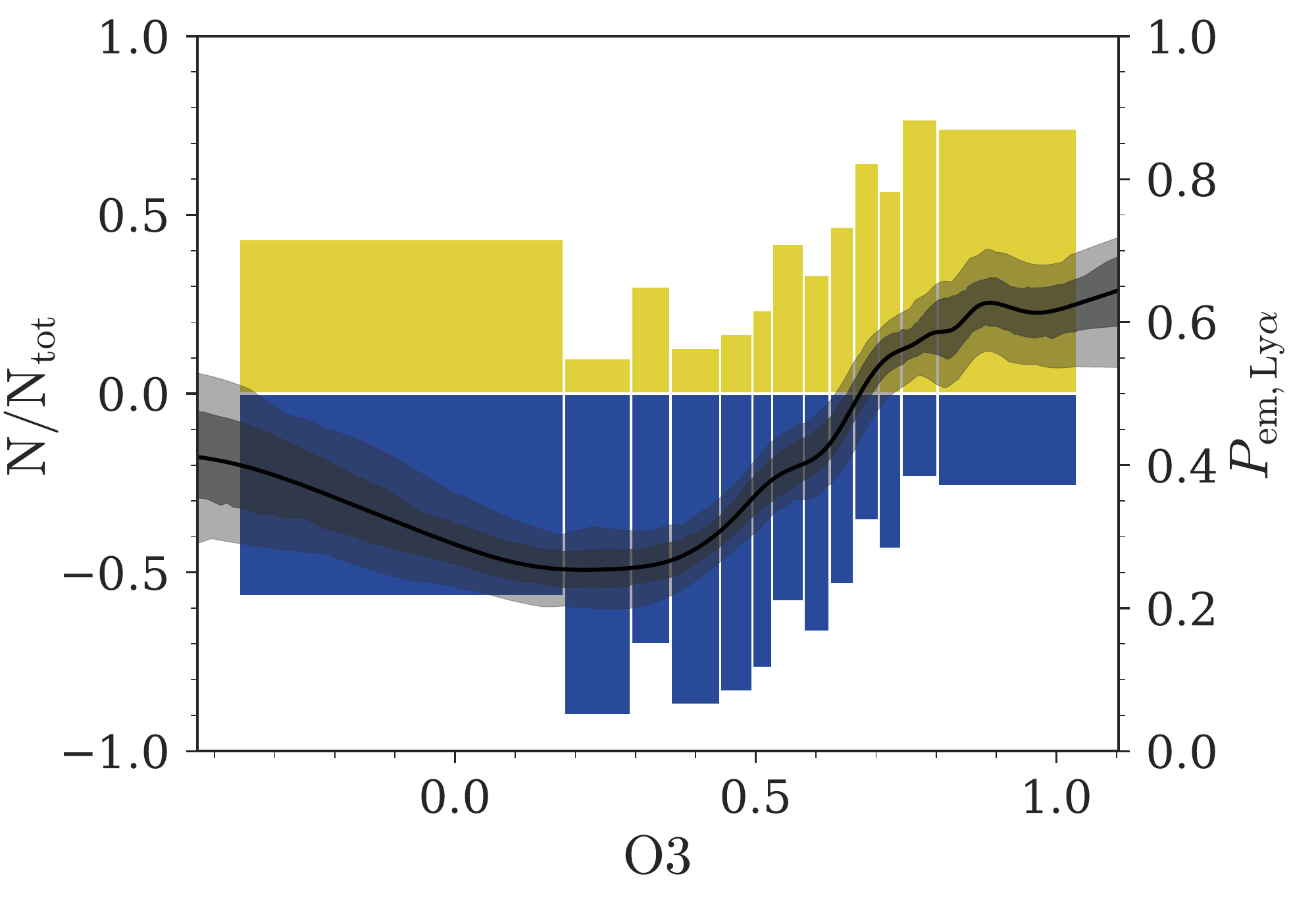}
\vspace{5mm}

\includegraphics[width=0.48\linewidth]{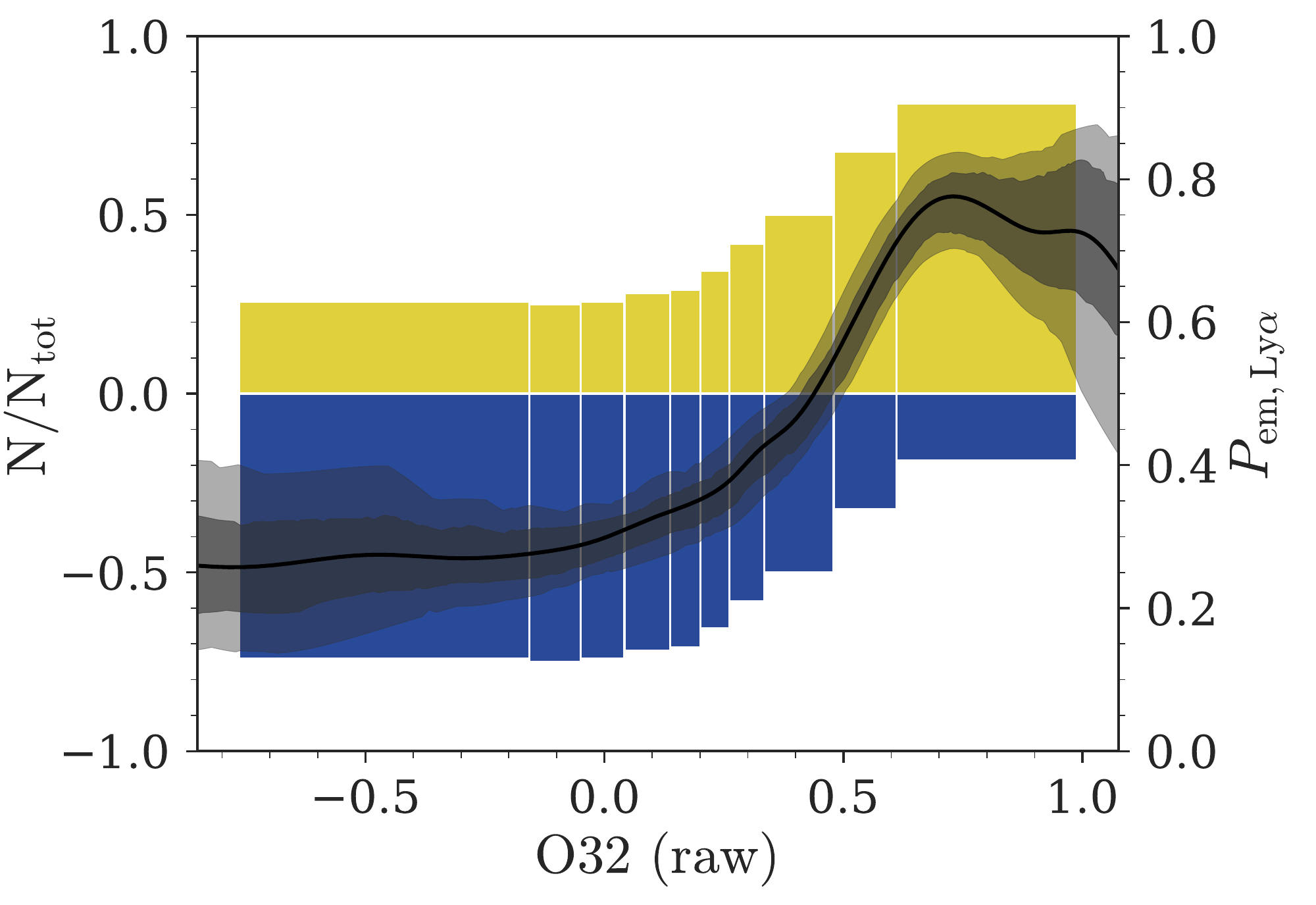}\hfill\includegraphics[width=0.48\linewidth]{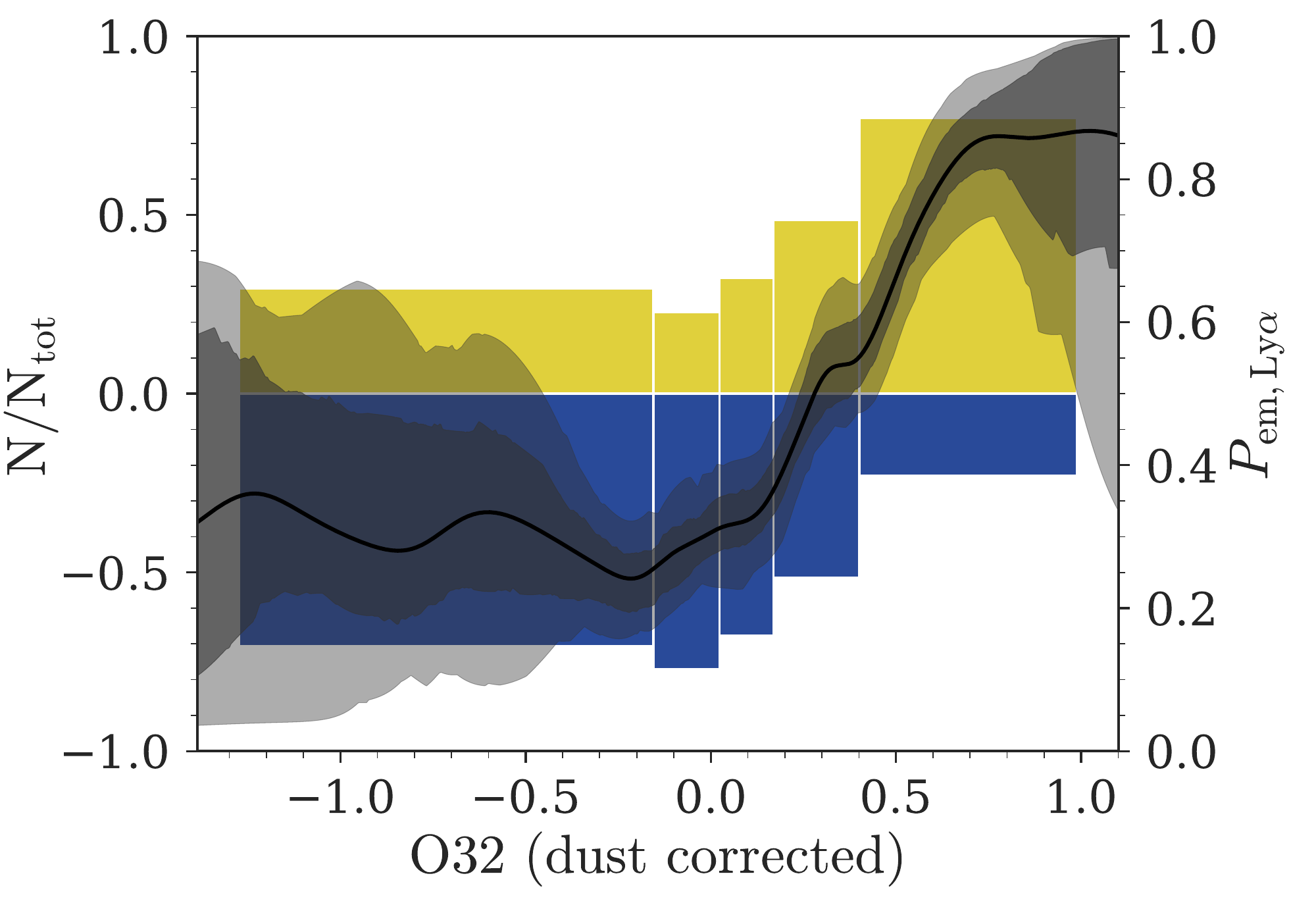}
\caption{Conditional probability distributions for detecting \lya\ in
  net emission (i.e., with $\ewlya>0$) as a
  function of other galaxy parameters. Yellow (blue) bars represent
  the fraction of \lya\ emitters (absorbers) found within a given bin
  as shown by the left-hand vertical axis.
Bins sizes and boundaries are determined in order to ensure at least
30 objects are included per bin while allowing the number of bins to vary. Black lines indicate the conditional
probability of \lya\ detection (according to the right-hand vertical
axis) at a given value of the parameter shown on the horizontal
axis. Dark (light) gray shaded
regions indicate the 68\% (95\%) confidence intervals on this
conditional probability. Curves are determined based on the unbinned,
non-parametric model described in
Sec.~\ref{sec:pdist_method}, which depends on the measured parameter
values and their uncertainties. $P_\mathrm{em}^{\mathrm{Ly}\alpha}$ represents
$P($\ewlya~$>0\,|\,X)$, where $X$ is the parameter given on the
horizontal axis. The parameters displayed here are all relatively
effective predictors of \lya\ emission, with
$X_\mathrm{LIS}^\mathrm{O3}$ and O32 being particularly effective.}
\label{fig:pdist1}
\end{figure*}

\begin{figure*}
\center
\includegraphics[width=0.48\linewidth]{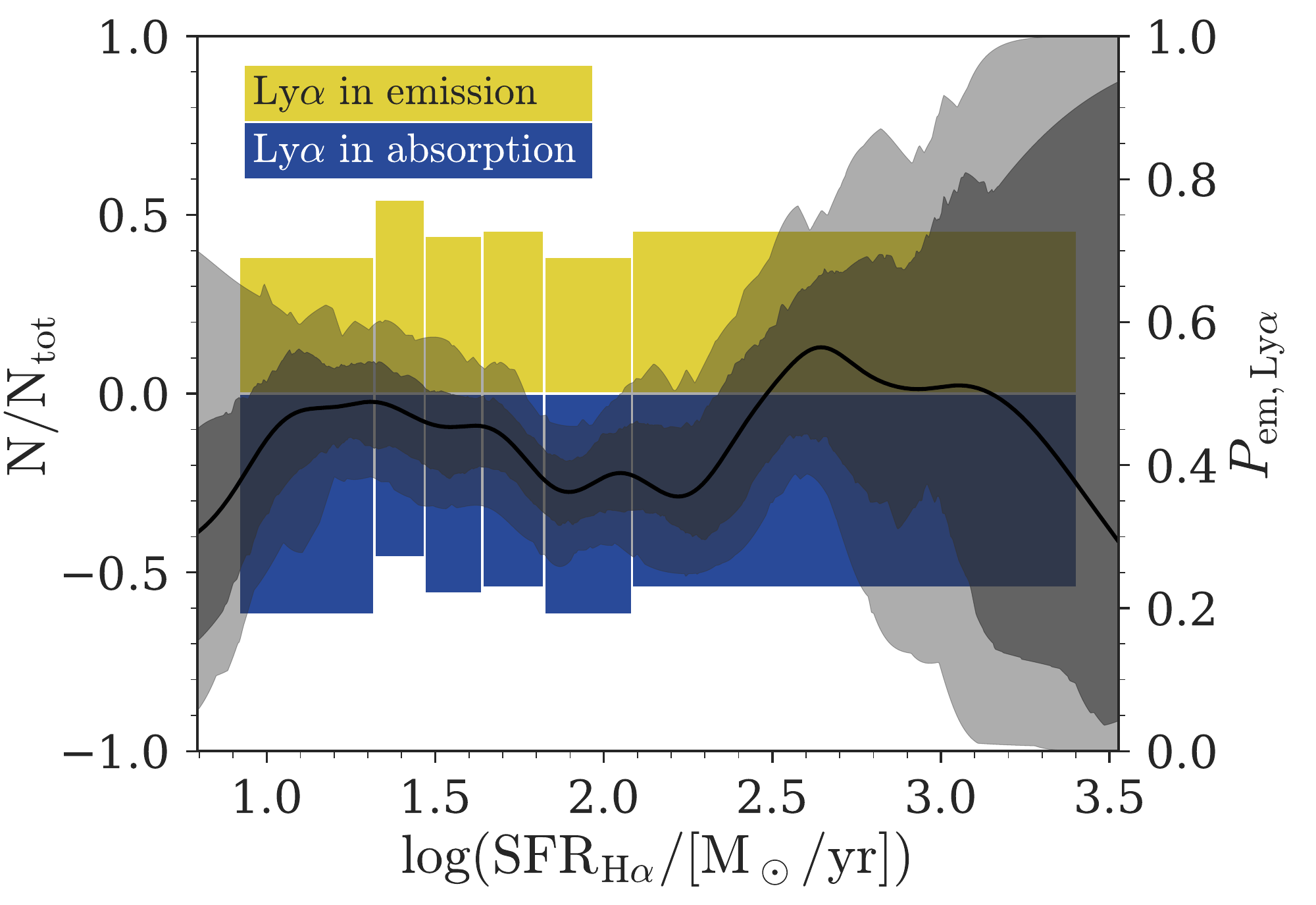}\hfill\includegraphics[width=0.48\linewidth]{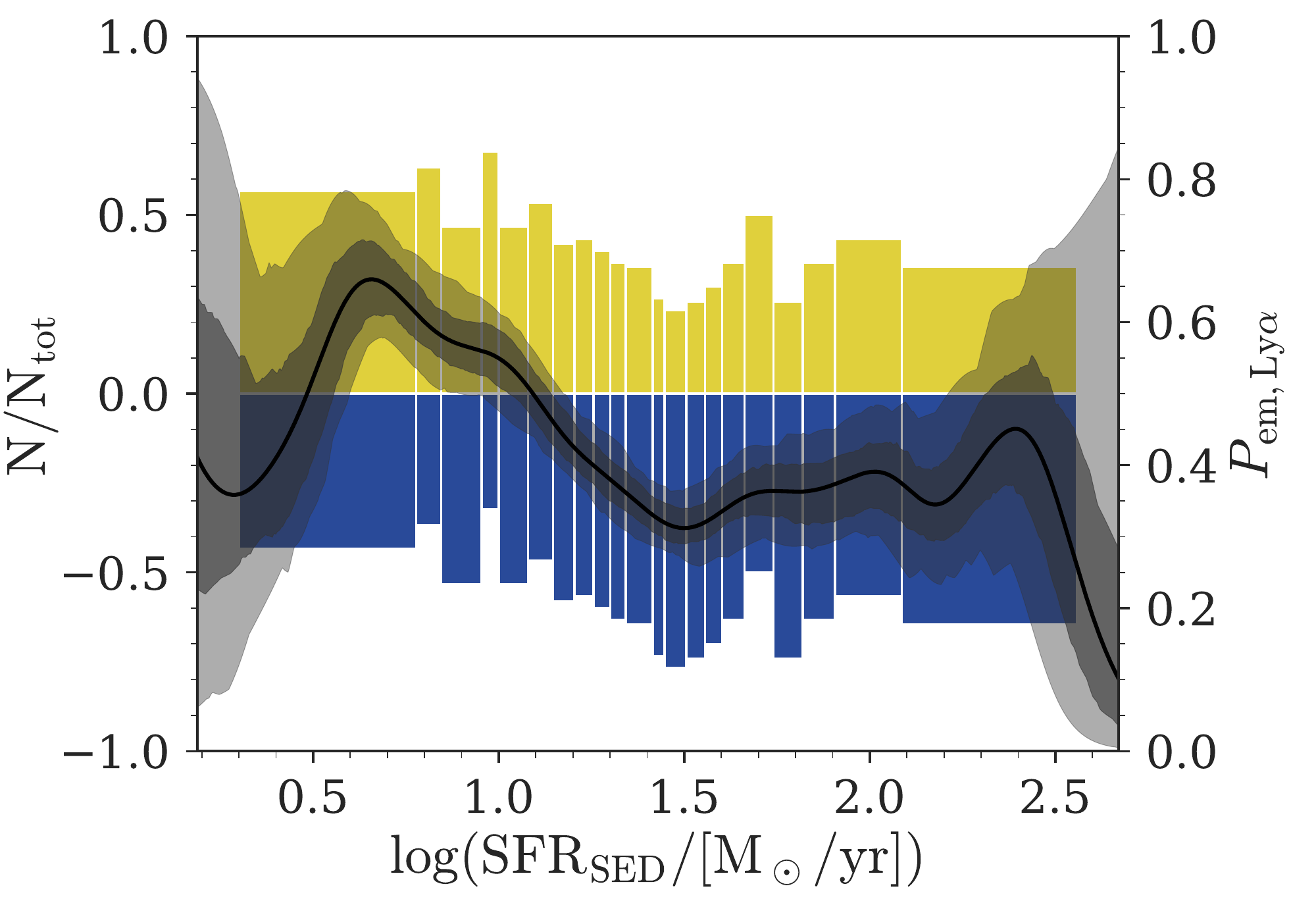}
\vspace{5mm}

\includegraphics[width=0.48\linewidth]{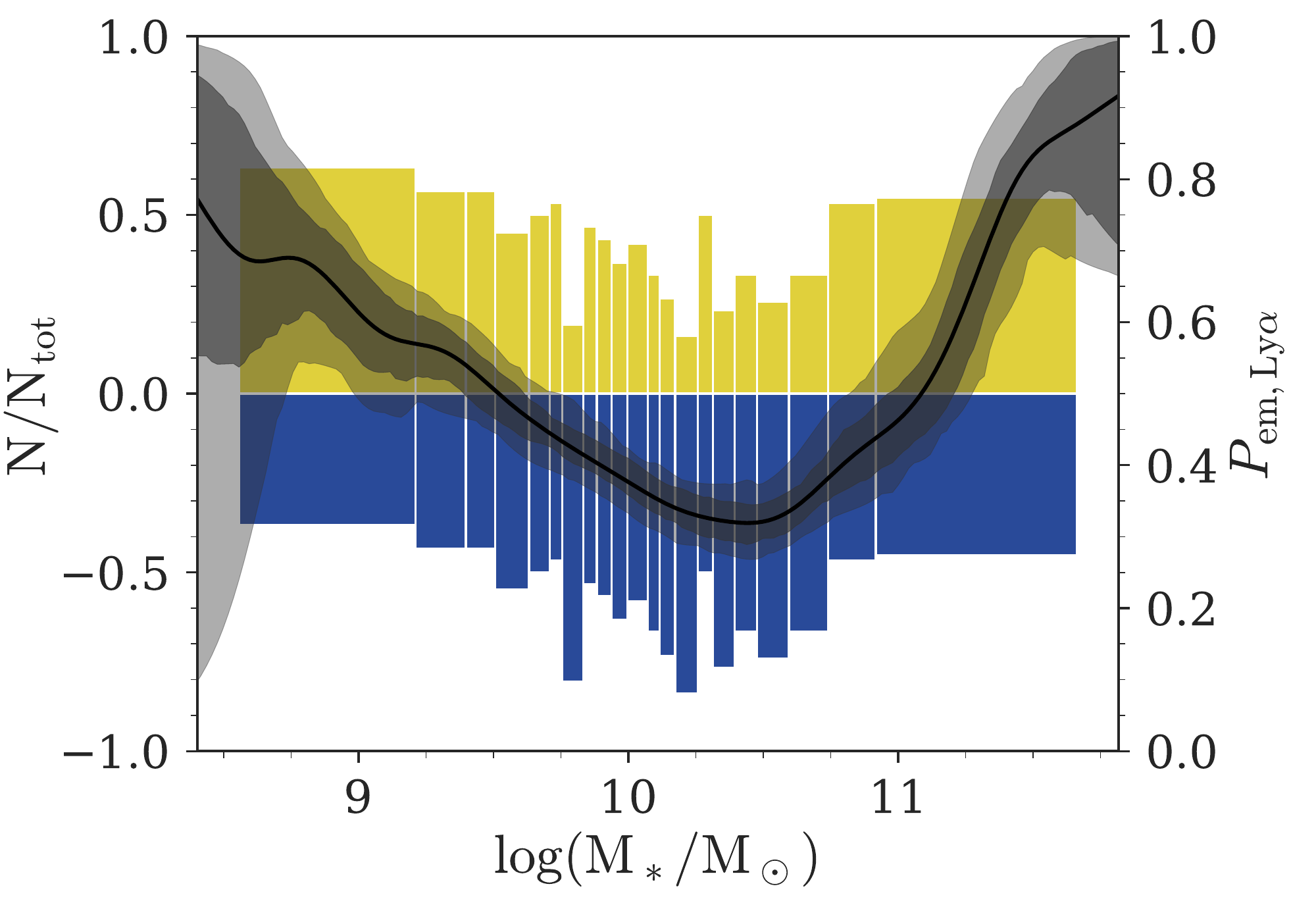}\hfill\includegraphics[width=0.48\linewidth]{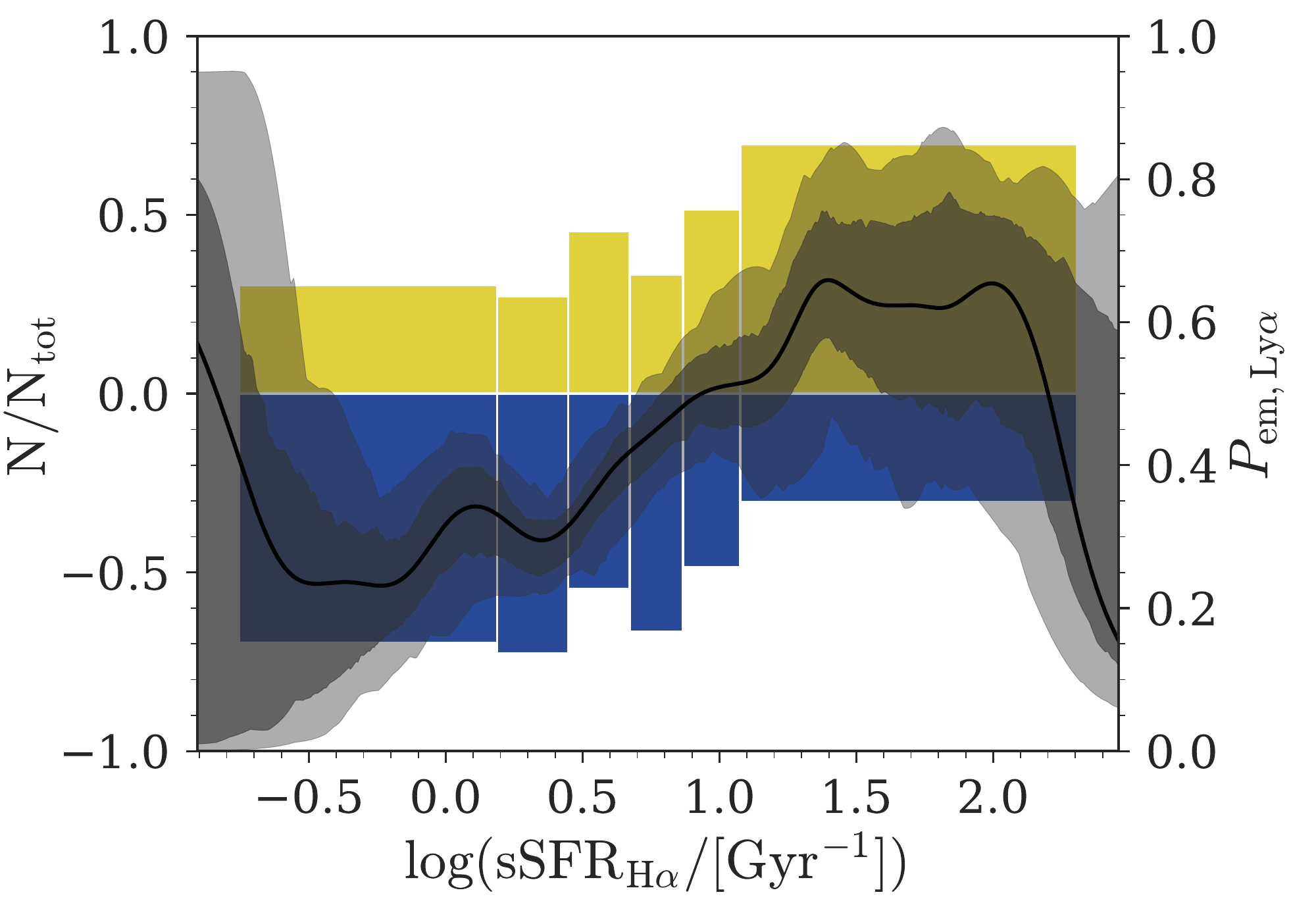}
\vspace{5mm}

\includegraphics[width=0.48\linewidth]{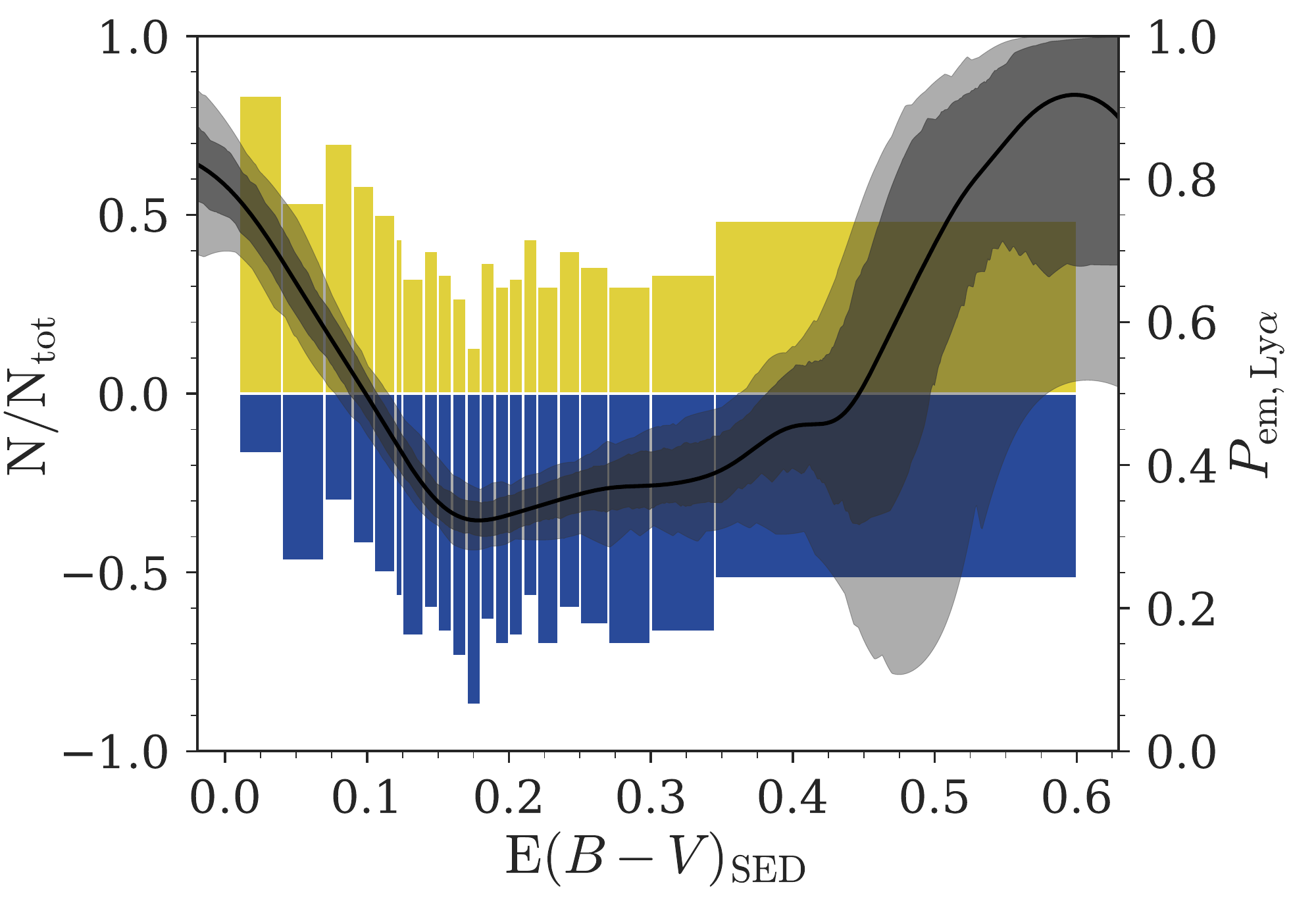}\hfill\includegraphics[width=0.48\linewidth]{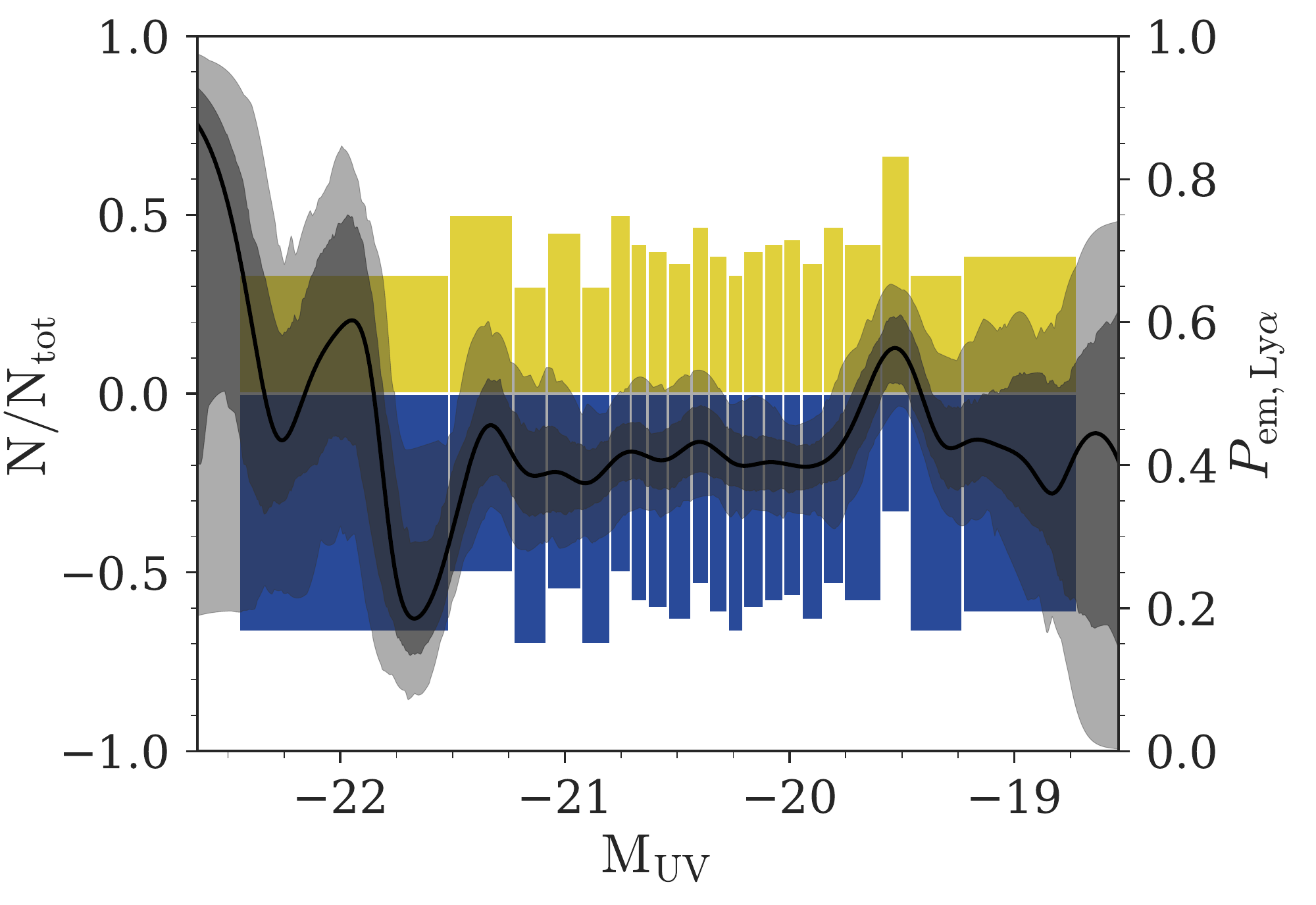}
\caption{Continuation of Fig.~\ref{fig:pdist1}. Conditional probability distributions for the probability of \lya\
  detection as a function of SFR, SED-fit parameters, and other
  photometric galaxy parameters. The parameters in this figure are all relatively
  weak predictors of \ewlya\ compared to the parameters displayed in
  Fig.~\ref{fig:pdist1}; measuring the value of one of these parameters
  does not generally provide a strong prior on the probability of detecting
  \lya\ in emission.} 
\label{fig:pdist2}
\end{figure*}
\renewcommand{\thefigure}{\arabic{figure}}

A galaxy that is a clear \lya\ emitter ($P_\mathrm{em}\approx 1$)
will therefore increase the integral of $\eta_\mathrm{em}$ by one over
the distribution of $X$. This
contribution to the incidence will be localized to a specific value \footnote{To reduce the sample variance of
  this estimator, we replace $\sigma_{X,i}$ with the distance $\Delta
  X_i$ to its nearest neighbor in the observed distribution of $X$ in
  cases where $\Delta X_i>\sigma_{X,i}$. This occurs for $\sim$2\% of
  objects, typically in the extrema of the distribution.} of
$X$ if $\sigma_{X,i}$ is small, whereas the contribution to the total
incidence will be spread across a large fraction of the distribution
if $\sigma_{X,i}$ is large. Finally, the inferred probability of \lya\ emission given the
observation of a galaxy with measured parameter value $X$ is taken to be
the inferred incidence of \lya\ emitters divided by the total
incidence of emitters:
\begin{align}
P_\mathrm{em}^{\mathrm{Ly}\alpha}(X)&=\frac{\eta_\mathrm{em}(X)}{\eta_\mathrm{em}(X)+\eta_\mathrm{abs}(X)}\,\,.
\end{align}

This conditional probability is displayed as a black curve in each
panel of Figs.~\ref{fig:pdist1}$-$\ref{fig:pdist2}. Confidence intervals for this curve
are calculated by repeating the process above for 100 bootstrap
realizations of the data and identifying the central 68\% and 95\%
intervals; these are shown as gray shaded regions in Figs.~\ref{fig:pdist1}$-$\ref{fig:pdist2}.

\subsection{Analysis of conditional probability distributions}\label{sec:pdist_analysis}

The conditional probability distributions displayed in
Figs.~\ref{fig:pdist1}$-$\ref{fig:pdist2} reinforce the relationships between \ewlya\ and
other galaxy properties shown in Table~\ref{table:correlations}. Net
\lya\ emission is strongly associated with weak LIS absorption
(\ewlis~$\gtrsim-1$) and high ionization/excitation
($\mathrm{O3}\gtrsim0.7$; $\mathrm{O32}\gtrsim 0.5$). In particular,
our $X_\mathrm{LIS}^\mathrm{O3}$ metric is able to more clearly discriminate
between \lya\ emitters and absorbers than either \ewlis\ or O3 alone;
even the most extreme values of these latter metrics only predict
\ewlya~$>0$ with $\sim$60\% probability, whereas the highest values of
$X_\mathrm{LIS}^\mathrm{O3}$ predict \lya\ in net emission in $\sim$80\% of
cases. Likewise, $\lesssim$25\% of galaxies with
$X_\mathrm{LIS}^\mathrm{O3}<0$ display \lya\ in net emission.

The distribution for \vlya\ is also shown in Fig.~\ref{fig:pdist1},
which demonstrates the strong dependence of
$P_\mathrm{em}^{\mathrm{Ly}\alpha}$ on \vlya\ over the range
$0\lesssim v_{\mathrm{Ly}\alpha}\lesssim 800\u{km}\u[-1]{s}$ and very
low probability of \lya\ emission for galaxies with
$v_{\mathrm{Ly}\alpha}\gtrsim 800\u{km}\u[-1]{s}$
($P_\mathrm{em}^{\mathrm{Ly}\alpha}\approx0.1-0.2$). $P_\mathrm{em}^{\mathrm{Ly}\alpha}$
peaks for $v_{\mathrm{Ly}\alpha}\approx 0$, consistent with the model
described in Secs.~\ref{sec:intro}~\&~\ref{sec:vlya}.

Other than \vlya, O32 is again the most direct predictor of net \lya\ emission or
absorption: even the dust-uncorrected (i.e., raw) values are
approximately as effective at predicting \ewlya~$>0$ as
$X_\mathrm{LIS}^\mathrm{O3}$, and the dust-corrected measurements predict \lya\
emission with $>$80\% probability at the highest O32 values and
$\lesssim$20\% at the lowest values. As discussed in
Secs.~\ref{sec:o32}$-$\ref{sec:o3}, this strong correlation may be
due in part to the fact that dust-corrected O32 is a quite direct
measure of the ionization parameter in ionization-bounded nebulae, but
it may also be due to the fact that elevated O32 may indicate
density-bounded nebulae that facilitate \lya\ escape as well as
production. Nonetheless, an observer who merely wishes to predict the
net \lya\ emission of a galaxy (remaining agnostic to the
circumstances that facilitate this emission) will find O32 to be an
effective indicator of this emission. However, the caveat to the
effectiveness of this indicator remains the observational difficulty
of obtaining high-S/N measurements of the [\ion{O}{3}] and
[\ion{O}{2}] emission lines (the latter of which can be extremely faint in
the high-ionization galaxies typical at high-$z$) as well as the
Balmer lines necessary to correct for differential attenuation by
dust. This effect is seen in the small number of bins (which must each
contain 30$+$ galaxies) in each of the O32 plots, as well as in the
broad confidence intervals for the corresponding unbinned relations.

Note that in
many of the panels of Figs.~\ref{fig:pdist1}$-$\ref{fig:pdist2}, the
trends apparently reverse toward the extrema 
of a given parameter value. Given that these regions of parameter space
are sparsely populated (as can be seen based on the width of the blue
and yellow bars) and the confidence intervals on
$P_\mathrm{em}^{\mathrm{Ly}\alpha}$ diverge, we interpret majority of this
apparent aberrant behavior as a regression toward the overall average
rate of \lya\ emission
($P_\mathrm{em}^{\mathrm{Ly}\alpha}(X)\approx\langle
P_\mathrm{em}^{\mathrm{Ly}\alpha}\rangle \approx 0.5$) when the parameter
uncertainties are large. This effect is particularly noticeable in the
panel for \ewlis, where the highest and lowest bins appear
particularly dominated by observational error -- in this case, we
expect that the trend between \ewlis\ and \ewlya\ is monotonically 
positive over the range in which both parameters are
well-measured. Conversely, the relatively tight confidence intervals
for $X_\mathrm{LIS}^\mathrm{O3}$ and O32 appear to indicate real flattening
in their relationships with $P_\mathrm{em}^{\mathrm{Ly}\alpha}$; some
non-negligible stochasticity in \ewlya\ appears to be present that is not
accounted for by these factors even when they are well-measured, as
demonstrated by the $\sim$20\% of \lya\ emitters (absorbers) that are
present even at the lowest (highest) values of these parameters.

Fig.~\ref{fig:pdist2} displays the conditional probability
distributions for several parameters that are only weakly correlated
with \ewlya, again reinforcing the results shown in
Table~\ref{table:correlations}. In particular,
SFR$_{\mathrm{H}\alpha}$ and M$_\mathrm{UV}$ display negligible
predictive power related to \lya\ emission over the parameter space
sampled by our galaxies. E$(B-V)_\mathrm{SED}$ is an effective
predictor of $P_\mathrm{em}^{\mathrm{Ly}\alpha}$
at the lowest reddenings (E$(B-V)_\mathrm{SED}\lesssim 0.2$),\footnote{See
  Sec.~\ref{sec:discussion} for a description of other recent studies
  of E$(B-V)$ vs. \ewlya.} but \lya\ emitters
and absorbers appear almost equally common among galaxies with higher
reddening values. Surprisingly, the incidence of \lya\ emitters appears to
grow slightly with increasing E$(B-V)_\mathrm{SED}$ for E$(B-V)_\mathrm{SED}> 0.2$.
This effect may be dominated by the relatively uncertain values of
E$(B-V)_\mathrm{SED}$, which is somewhat degenerate with stellar
population age, causing a regression toward the population mean.
$P_\mathrm{em}^{\mathrm{Ly}\alpha}$ shows a moderate increase
as SFR$_\mathrm{SED}$ decreases or sSFR increases.

Stellar mass
($M_*$) displays perhaps the most interesting relationship with
\ewlya\ that is not obvious from the Spearman correlation alone: \lya\
emission probability increases significantly among both the lowest-
and highest-mass galaxies. The low-mass relationship (along with the trends
in SFR and sSFR) may reflect the tendency for low-mass, low-SFR galaxies
to have relatively porous interstellar media and high nebular excitation
(see, e.g., \citealt{trainor2015,trainor2016}). Although galaxies with
clear signatures of AGN activity were removed from our sample, the
excess of \lya\ emitters at high galaxy masses may reflect residual
AGN in our sample. It is also possible that some galaxy masses in our
sample are over-estimated due to strong line emission
that contaminates the rest-frame optical photometry; SED-fit
masses are especially sensitive to this effect at high redshift (see e.g.,
\citealt{schenker13a}). In this case, the increase in 
\lya\ emission among galaxies with large inferred stellar masses could
be caused by the underlying association between \lya\ and optical 
emission line strength (i.e., nebular excitation). The \lya-emitting
behavior of high-mass galaxies will be investigated in future work.

In general, these conditional probability distributions may be used
to inform analyses of the \lya-detection fraction of galaxies at the
highest redshifts, where the opacity of the neutral IGM may suppress
the observed \lya\ emission from intrinsic \lya\ emitters. By using
other observed properties of galaxies as priors input to the
distributions above, it will be possible to more accurately
characterize the degree to which evolution in both IGM and galaxy
properties shape the distribution of observed \lya\ emission.

\section{Comparison to Recent Work} \label{sec:discussion}

A few recent studies in the low-redshift Universe have
measured correlations between \lya\ emission and other galaxy
properties with the goal of predicting the \lya\
emission. \citet{hayes2014} present data from the Lyman-Alpha
Reference Sample (LARS; \citealt{ostlin2014}), comparing \lya\
emission with 12 different global galaxy properties derived from
imaging and spectroscopy. They find significant correlations in which
normalized\footnote{\citet{hayes2014} consider several metrics for
  \lya\ emission, including
  the total \lya\ luminosity ($L_{\mathrm{Ly}\alpha}$), \ewlya,
  $L_{\mathrm{Ly}\alpha}$/$L_{\mathrm{H}\alpha}$, and $f_{\rm esc,abs}$;
  only the latter three quantities show strong correlations with
  galaxy properties.}  \lya\ emission is highest among galaxies with
low SFR, low dust content (inferred by nebular line ratios or the UV
slope), low mass, and nebular properties indicative of high excitation
and low metallicity. The \citet{hayes2014} study differs from the work
presented here in that the their individual measurements have much
higher signal-to-noise ratios (S/N) but are much fewer in number (12
galaxies in LARS vs. the 703 galaxies in this paper). Furthermore, the
original LARS sample did not include rest-UV continuum spectroscopy
covering the interstellar absorption lines; while these data were
later collected via HST/COS spectroscopy and presented by
\citet{rivera2015}, there is currently no simultaneous analysis of the
predictive power of combined rest-UV and rest-optical spectroscopic
diagnostics of \lya\ emission in LARS.

Recent results by \citet{yang2017} are more directly comparable to
those presented here: \citet{yang2017} analyze HST/COS spectra of 43
``green pea'' galaxies at $z\sim0.1-0.3$ with SDSS optical
spectroscopy. The authors find that the escape fraction of \lya\ is
anticorrelated with the velocity width of the \lya\ line profile, 
the nebular dust extinction, and the stellar mass, as well as
positively correlated with the O32 ratio. Each of these relationships
have Spearman rank-correlation coefficients of $r\sim0.5-0.6$; while
the contributions of observational uncertainties to this scatter are
not explicitly calculated, the quoted uncertainties suggest that these
contributions are neglible. Furthermore,
\citet{yang2017} fit a linear combination of the nebular extinction
E($B-V$) and the velocity offset of the red \lya\ peak, finding that
the resulting relation fits the observed \lya\ escape fraction with a
1$\sigma$ scatter of 0.3 dex. Notably, this multi-parameter
relationship is similar to our own work in
Sec.~\ref{sec:xlya}, but it differs in that both included parameters (the
\lya\ velocity offset and inferred dust extinction) fall into the
category of empirical parameters we have associated with \lya\
\textit{escape} (Sec.~\ref{sec:escape}), rather than including a proxy for the efficiency of
\lya\ production (Sec.~\ref{sec:production}). As with the \citet{hayes2014} study, 
\citet{yang2017} have the advantage over our own work of measuring
individual spectroscopic parameters at high S/N, but they include a much smaller
sample size. The \citet{yang2017} sample also only includes galaxies
selected to be spatially compact, low-mass, and high-excitation,
whereas the KBSS sample includes 15$\times$ more galaxies over a much
broader range of 
properties. Nonetheless, the different selection biases and relative
advantages of low- and high-redshift galaxy samples make these
$z\sim0-0.3$ surveys (including continued HST/COS spectroscopy) an
effective complement to the work presented here.

While not directly focused on predicting \lya\ emission, recent work
from the HiZELS survey \citep{geach2008,sobral2009a} has pointed to
the complex relationships 
between \lya\ and other recombination line emission. In particular,
\citet{oteo2015} demonstrate that galaxies selected on the basis of
\ha\ emission display only weak \lya\ emission on average. This
result is consistent with the first panel of Fig.~\ref{fig:pdist2}
 and the discussion in Sec.~\ref{sec:photparams} of this paper, as we
 also find $L_{\mathrm{H}\alpha}$ to be a poor predictor of \lya\
 emission. Although the HiZELS galaxies do not have deep rest-UV
 spectra, we expect that their low \lya\ escape fractions would
 be associated with strong LIS absorption, such that they may lie
 near cB58 in the O3-\ewlis\ plane displayed in
 Fig.~\ref{fig:o3_vs_lis} (i.e., the parameter space associated with
 high rates of intrinsic \lya\ production, but low rates of \lya\ escape). 

In more recent work, \citet{du2018} present a systematic
study of the redshift evolution of 
rest-UV spectroscopic properties of galaxies over $z\approx2-4$,
including the variation of \lya\ with other galaxy properties in this
epoch. The \citet{du2018} spectroscopic sample is also drawn from the
KBSS and has substantial overlap with the galaxies presented
here. Broadly, the authors find that the relationship between \ewlya\
and \ewlis\ is invariant with redshift for $2\lesssim z\lesssim 4$,
with a similarly invariant (but weaker) relationship between  \ewlya\
and E$(B-V)_\mathrm{SED}$. In addition to these indicators of \lya\ escape,
\citet{du2018} argue that the association between the equivalent width
of \ion{C}{3}] and \ewlya\ (which they find to be similar at $z\sim 2$
and $z\sim 3$) represents the dependence of \ewlya\ on the instrinsic
production rate of \lya\ emission. One interesting point of comparison
between their results  and those presented here regards the
relationship between \ewlya\ and  M$_\mathrm{UV}$; in their $z\sim 2$
galaxy bin (which is most similar  to the galaxies presented here),
\citet{du2018} find no relationship between \ewlya\ and
M$_\mathrm{UV}$, but they find an increasingly strong relationship
with increasing redshift. Likewise, \citet{oyarzun2017} find a
positive correlation between M$_\mathrm{UV}$ and \ewlya, but the
relationship appears to rely on the inclusion of lower-luminosity
galaxies than are included in this sample (although similar to the
galaxies described in \citealt{trainor2015,trainor2016}; see
Figs.~\ref{fig:o3_vs_lis}~\&~\ref{fig:xlya_vs_lya} here) and the extension to higher redshifts. This variation may help 
explain the high \ewlya\ values seen generically in the
lowest-luminosity galaxies at $z\sim2$ (e.g., \citealt{stark2013}),
which are perhaps better analogs for typical galaxies at the highest
redshifts than the more luminous $z\sim 2$ galaxies described in this
paper.

The results of \citet{du2018} are in general agreement with those presented here,
with the exception that \citet{du2018} limit their analysis to composite
rest-UV spectra (i.e., they include no rest-optical data, nor do they
analyze the spectra of individual galaxies), and they consider the
redshift evolution of these trends. The 
\citeauthor{du2018} composite spectra achieve higher S/N measurements of
individual features than the measurements we consider here, but they
also smooth over the intrinsic 
object-to-object variation among galaxies -- variation that we
highlight in this paper, particularly in Sec.~\ref{sec:pdist}. Together, therefore, these two studies provide a comprehensive view of the average trends among net \lya\ emission
and the processes of production and escape, while also demonstrating
the substantial stochasticity that accompanies these broader trends.

 Another
complementary aspect of these works is that \citet{du2018} demonstrate
that individual parameters related to \lya\ production and
escape (i.e., \ewlis, E$(B-V)$, and proxies for nebular excitation) show similar
relationships with \ewlya\ across $2<z<4$, 
despite the fact that the ubiquity of \ewlya\ emission itself (and its
dependence on M$_\mathrm{UV}$, SFR, and $M_*$) evolves significantly
over this period. This invariance suggests that the models for \lya\
emission developed here -- particularly those shown in
Figs.~\ref{fig:xlya_vs_lya}~\&~\ref{fig:pdist1} -- may be expected to
remain useful even at higher redshifts where the intrinsic \lya\
emission of galaxies is more difficult to directly measure.

\section{Conclusions} \label{sec:conclusions}

We have presented an empirical analysis of factors affecting \lya\
production and escape in a sample of 703 star-forming galaxies from the Keck
Baryonic Structure Survey at $z\approx
1.5-3.5$. Our primary indicators of \lya\ escape efficiency include the velocity
offset of the \lya\ line and the equivalent width in absorption of
low-ionization interstellar lines, \ewlis, and we find that these
proxies for \lya\ escape are strongly associated with the directly-measured \lya\ escape
fraction, $f_\mathrm{esc}$. Our indicators of \lya\ production
include the O3 and O32 ratios, which we argue are effective
diagnostics of the ionization conditions within the \ion{H}{2} regions from
which \lya\ photons originate. Several other galaxy parameters
including stellar mass, star-formation rate, luminosity, and
reddenning are shown to have much weaker relationships with observed
\lya\ emission.

We propose that \ewlis\ and O3 are the most useful predictors of
\ewlya\ because of their potential for observability in cases when the \lya\
line is not directly detectable (or may be strongly affected by IGM
absorption) and because of their strong individual correlations with
\ewlya\ and lack of correlation with each other. We then construct a
new quantity, $X_\mathrm{LIS}^\mathrm{O3}$, which is a linear
combination of \ewlis\ and O3 with coefficients chosen to maximize the
association between $X_\mathrm{LIS}^\mathrm{O3}$ and \ewlya.

We find that the combination of O3 and \ewlis\ predicts net \ewlya\
with less scatter than any single variable captured by
our survey that does not require measurement of the \lya\ line;
$\sim$50\% of the ordering in observed \ewlya\ 
is captured by $X_\mathrm{LIS}^\mathrm{O3}$. After accounting for
measurement uncertainties and fitting an exponential model for
\ewlya\ as a function of $X_\mathrm{LIS}^\mathrm{O3}$, we estimate
that the combination of our model and observational error account for
90\% of the total variance in \ewlya\ at fixed $X_\mathrm{LIS}^\mathrm{O3}$.

We also estimate the conditional probability of detecting net \lya\
emission or absorption in slit spectroscopy of a galaxy as a function of various galaxy
parameters. We find that galaxies with $X_\mathrm{LIS}^\mathrm{O3}>0.6$
have an 80\% probability of being net \lya\ emitters, while those with
$X_\mathrm{LIS}^\mathrm{O3}<0$ have less than a 25\%
probability of exhibiting net emission. Similarly strong variation in
the probability of net \lya\ emission is seen when adopting a prior
based on O32, while constraints on photometric or SED-fit parameters
or \ha-based SFR have negligible utility as priors over the parameter
space probed by our sample.

Given the many factors affecting net \lya\ emission, our
two-parameter model for $X_\mathrm{LIS}^\mathrm{O3}$ is remarkably
successful at describing the 
variation in \lya\ emission across a large, heterogenous set of
star-forming galaxies. We suggest that this success indicates that the
wide variety of processes affecting \lya\ emission can be broadly
categorized as relating to \lya\ production or escape, and capturing
these two different ``meta-parameters'' is an essential component of
any model for \lya\ emission from galaxies.


\acknowledgments

\noindent We are indebted to the staff of the W.M. Keck Observatory who keep the
instruments and telescopes running effectively. We also wish to extend
thanks to those of Hawaiian ancestry on whose sacred mountain we are
privileged to be guests. This work has been supported in part by the US
National Science Foundation through grants AST-0908805 and
AST-1313472. We also acknowledge support from the JPL/Caltech
President's and Director's Program.

\software{Astropy \citep{astropy2013},
Matplotlib \citep{matplotlib2007}}

\bibliographystyle{apj}
\bibliography{lya_escape}

\begin{thebibliography}{94}
\expandafter\ifx\csname natexlab\endcsname\relax\def\natexlab#1{#1}\fi

\bibitem[{{Astropy Collaboration} {et~al.}(2013){Astropy Collaboration},
  {Robitaille}, {Tollerud}, {Greenfield}, {Droettboom}, {Bray}, {Aldcroft},
  {Davis}, {Ginsburg}, \& {Price-Whelan}}]{astropy2013}
{Astropy Collaboration}, {Robitaille}, T.~P., {Tollerud}, E.~J., {Greenfield},
  P., {Droettboom}, M., {Bray}, E., {Aldcroft}, T., {Davis}, M., {Ginsburg},
  A., \& {Price-Whelan}, A.~M. 2013, \aap, 558, A33

\bibitem[{{Atek} {et~al.}(2009){Atek}, {Kunth}, {Schaerer}, {Hayes},
  {Deharveng}, {{\"O}stlin}, \& {Mas-Hesse}}]{atek2009}
{Atek}, H., {Kunth}, D., {Schaerer}, D., {Hayes}, M., {Deharveng}, J.~M.,
  {{\"O}stlin}, G., \& {Mas-Hesse}, J.~M. 2009, \aap, 506, L1

\bibitem[{{Atek} {et~al.}(2010){Atek}, {Malkan}, {McCarthy}, {Teplitz},
  {Scarlata}, {Siana}, {Henry}, {Colbert}, {Ross}, {Bridge}, {Bunker},
  {Dressler}, {Fosbury}, {Martin}, \& {Shim}}]{atek2010}
{Atek}, H., {Malkan}, M., {McCarthy}, P., {Teplitz}, H.~I., {Scarlata}, C.,
  {Siana}, B., {Henry}, A., {Colbert}, J.~W., {Ross}, N.~R., {Bridge}, C.,
  {Bunker}, A.~J., {Dressler}, A., {Fosbury}, R.~A.~E., {Martin}, C., \&
  {Shim}, H. 2010, \apj, 723, 104

\bibitem[{{Bacon} {et~al.}(2015){Bacon}, {Brinchmann}, {Richard}, {Contini},
  {Drake}, {Franx}, {Tacchella}, {Vernet}, {Wisotzki}, {Blaizot}, {Bouch{\'e}},
  {Bouwens}, {Cantalupo}, {Carollo}, {Carton}, {Caruana}, {Cl{\'e}ment},
  {Dreizler}, {Epinat}, {Guiderdoni}, {Herenz}, {Husser}, {Kamann}, {Kerutt},
  {Kollatschny}, {Krajnovic}, {Lilly}, {Martinsson}, {Michel-Dansac},
  {Patricio}, {Schaye}, {Shirazi}, {Soto}, {Soucail}, {Steinmetz}, {Urrutia},
  {Weilbacher}, \& {de Zeeuw}}]{bacon2015}
{Bacon}, R., {Brinchmann}, J., {Richard}, J., {Contini}, T., {Drake}, A.,
  {Franx}, M., {Tacchella}, S., {Vernet}, J., {Wisotzki}, L., {Blaizot}, J.,
  {Bouch{\'e}}, N., {Bouwens}, R., {Cantalupo}, S., {Carollo}, C.~M., {Carton},
  D., {Caruana}, J., {Cl{\'e}ment}, B., {Dreizler}, S., {Epinat}, B.,
  {Guiderdoni}, B., {Herenz}, C., {Husser}, T.-O., {Kamann}, S., {Kerutt}, J.,
  {Kollatschny}, W., {Krajnovic}, D., {Lilly}, S., {Martinsson}, T.,
  {Michel-Dansac}, L., {Patricio}, V., {Schaye}, J., {Shirazi}, M., {Soto}, K.,
  {Soucail}, G., {Steinmetz}, M., {Urrutia}, T., {Weilbacher}, P., \& {de
  Zeeuw}, T. 2015, \aap, 575, A75

\bibitem[{{Baldwin} {et~al.}(1981){Baldwin}, {Phillips}, \&
  {Terlevich}}]{baldwin1981}
{Baldwin}, J.~A., {Phillips}, M.~M., \& {Terlevich}, R. 1981, \pasp, 93, 5

\bibitem[{{Borthakur} {et~al.}(2014){Borthakur}, {Heckman}, {Leitherer}, \&
  {Overzier}}]{borthakur2014}
{Borthakur}, S., {Heckman}, T.~M., {Leitherer}, C., \& {Overzier}, R.~A. 2014,
  Science, 346, 216

\bibitem[{{Brammer} {et~al.}(2012){Brammer}, {van Dokkum}, {Franx},
  {Fumagalli}, {Patel}, {Rix}, {Skelton}, {Kriek}, {Nelson}, {Schmidt},
  {Bezanson}, {da Cunha}, {Erb}, {Fan}, {F{\"o}rster Schreiber}, {Illingworth},
  {Labb{\'e}}, {Leja}, {Lundgren}, {Magee}, {Marchesini}, {McCarthy},
  {Momcheva}, {Muzzin}, {Quadri}, {Steidel}, {Tal}, {Wake}, {Whitaker}, \&
  {Williams}}]{brammer2012}
{Brammer}, G.~B., {van Dokkum}, P.~G., {Franx}, M., {Fumagalli}, M., {Patel},
  S., {Rix}, H.-W., {Skelton}, R.~E., {Kriek}, M., {Nelson}, E., {Schmidt},
  K.~B., {Bezanson}, R., {da Cunha}, E., {Erb}, D.~K., {Fan}, X., {F{\"o}rster
  Schreiber}, N., {Illingworth}, G.~D., {Labb{\'e}}, I., {Leja}, J.,
  {Lundgren}, B., {Magee}, D., {Marchesini}, D., {McCarthy}, P., {Momcheva},
  I., {Muzzin}, A., {Quadri}, R., {Steidel}, C.~C., {Tal}, T., {Wake}, D.,
  {Whitaker}, K.~E., \& {Williams}, A. 2012, \apjs, 200, 13

\bibitem[{{Bruzual} \& {Charlot}(2003)}]{bruzual2003}
{Bruzual}, G. \& {Charlot}, S. 2003, \mnras, 344, 1000

\bibitem[{{Calzetti} {et~al.}(2000){Calzetti}, {Armus}, {Bohlin}, {Kinney},
  {Koornneef}, \& {Storchi-Bergmann}}]{calzetti2000}
{Calzetti}, D., {Armus}, L., {Bohlin}, R.~C., {Kinney}, A.~L., {Koornneef}, J.,
  \& {Storchi-Bergmann}, T. 2000, \apj, 533, 682

\bibitem[{{Cantalupo} {et~al.}(2014){Cantalupo}, {Arrigoni-Battaia},
  {Prochaska}, {Hennawi}, \& {Madau}}]{cantalupo2014}
{Cantalupo}, S., {Arrigoni-Battaia}, F., {Prochaska}, J.~X., {Hennawi}, J.~F.,
  \& {Madau}, P. 2014, \nat, 506, 63

\bibitem[{{Cantalupo} {et~al.}(2012){Cantalupo}, {Lilly}, \&
  {Haehnelt}}]{can12}
{Cantalupo}, S., {Lilly}, S.~J., \& {Haehnelt}, M.~G. 2012, \mnras, 425, 1992

\bibitem[{{Cantalupo} {et~al.}(2005){Cantalupo}, {Porciani}, {Lilly}, \&
  {Miniati}}]{can05}
{Cantalupo}, S., {Porciani}, C., {Lilly}, S.~J., \& {Miniati}, F. 2005, \apj,
  628, 61

\bibitem[{{Cardelli} {et~al.}(1989){Cardelli}, {Clayton}, \&
  {Mathis}}]{cardelli1989}
{Cardelli}, J.~A., {Clayton}, G.~C., \& {Mathis}, J.~S. 1989, \apj, 345, 245

\bibitem[{{Chabrier}(2003)}]{chabrier2003}
{Chabrier}, G. 2003, \pasp, 115, 763

\bibitem[{{Cowie} \& {Hu}(1998)}]{cowie1998}
{Cowie}, L.~L. \& {Hu}, E.~M. 1998, \aj, 115, 1319

\bibitem[{{de Barros} {et~al.}(2016){de Barros}, {Vanzella}, {Amor{\'{\i}}n},
  {Castellano}, {Siana}, {Grazian}, {Suh}, {Balestra}, {Vignali}, {Verhamme},
  {Zamorani}, {Mignoli}, {Hasinger}, {Comastri}, {Pentericci},
  {P{\'e}rez-Montero}, {Fontana}, {Giavalisco}, \& {Gilli}}]{debarros2016}
{de Barros}, S., {Vanzella}, E., {Amor{\'{\i}}n}, R., {Castellano}, M.,
  {Siana}, B., {Grazian}, A., {Suh}, H., {Balestra}, I., {Vignali}, C.,
  {Verhamme}, A., {Zamorani}, G., {Mignoli}, M., {Hasinger}, G., {Comastri},
  A., {Pentericci}, L., {P{\'e}rez-Montero}, E., {Fontana}, A., {Giavalisco},
  M., \& {Gilli}, R. 2016, \aap, 585, A51

\bibitem[{{Dopita} \& {Sutherland}(2003)}]{dop03}
{Dopita}, M.~A. \& {Sutherland}, R.~S. 2003, Astrophysics of the diffuse
  universe, Berlin, New York: Springer, 2003.~Astronomy and astrophysics
  library, ISBN 3540433627

\bibitem[{{Du} {et~al.}(2018){Du}, {Shapley}, {Reddy}, {Jones}, {Stark},
  {Steidel}, {Strom}, {Rudie}, {Erb}, {Ellis}, \& {Pettini}}]{du2018}
{Du}, X., {Shapley}, A.~E., {Reddy}, N.~A., {Jones}, T., {Stark}, D.~P.,
  {Steidel}, C.~C., {Strom}, A.~L., {Rudie}, G.~C., {Erb}, D.~K., {Ellis},
  R.~S., \& {Pettini}, M. 2018, \apj, 860, 75

\bibitem[{{Erb} {et~al.}(2016){Erb}, {Pettini}, {Steidel}, {Strom}, {Rudie},
  {Trainor}, {Shapley}, \& {Reddy}}]{erb2016}
{Erb}, D.~K., {Pettini}, M., {Steidel}, C.~C., {Strom}, A.~L., {Rudie}, G.~C.,
  {Trainor}, R.~F., {Shapley}, A.~E., \& {Reddy}, N.~A. 2016, \apj, 830, 52

\bibitem[{{Erb} {et~al.}(2014){Erb}, {Steidel}, {Trainor}, {Bogosavljevi{\'c}},
  {Shapley}, {Nestor}, {Kulas}, {Law}, {Strom}, {Rudie}, {Reddy}, {Pettini},
  {Konidaris}, {Mace}, {Matthews}, \& {McLean}}]{erb2014}
{Erb}, D.~K., {Steidel}, C.~C., {Trainor}, R.~F., {Bogosavljevi{\'c}}, M.,
  {Shapley}, A.~E., {Nestor}, D.~B., {Kulas}, K.~R., {Law}, D.~R., {Strom},
  A.~L., {Rudie}, G.~C., {Reddy}, N.~A., {Pettini}, M., {Konidaris}, N.~P.,
  {Mace}, G., {Matthews}, K., \& {McLean}, I.~S. 2014, \apj, 795, 33

\bibitem[{{Finkelstein} {et~al.}(2011){Finkelstein}, {Hill}, {Gebhardt},
  {Adams}, {Blanc}, {Papovich}, {Ciardullo}, {Drory}, {Gawiser}, {Gronwall},
  {Schneider}, \& {Tran}}]{finkelstein2011}
{Finkelstein}, S.~L., {Hill}, G.~J., {Gebhardt}, K., {Adams}, J., {Blanc},
  G.~A., {Papovich}, C., {Ciardullo}, R., {Drory}, N., {Gawiser}, E.,
  {Gronwall}, C., {Schneider}, D.~P., \& {Tran}, K.-V. 2011, \apj, 729, 140

\bibitem[{{Fletcher} {et~al.}(2018){Fletcher}, {Robertson}, {Nakajima},
  {Ellis}, {Stark}, \& {Inoue}}]{fletcher2018}
{Fletcher}, T.~J., {Robertson}, B.~E., {Nakajima}, K., {Ellis}, R.~S., {Stark},
  D.~P., \& {Inoue}, A. 2018, ArXiv e-prints

\bibitem[{{Geach} {et~al.}(2008){Geach}, {Smail}, {Best}, {Kurk}, {Casali},
  {Ivison}, \& {Coppin}}]{geach2008}
{Geach}, J.~E., {Smail}, I., {Best}, P.~N., {Kurk}, J., {Casali}, M., {Ivison},
  R.~J., \& {Coppin}, K. 2008, \mnras, 388, 1473

\bibitem[{{Hagen} {et~al.}(2016){Hagen}, {Zeimann}, {Behrens}, {Ciardullo},
  {Grasshorn Gebhardt}, {Gronwall}, {Bridge}, {Fox}, {Schneider}, {Trump},
  {Blanc}, {Chiang}, {Chonis}, {Finkelstein}, {Hill}, {Jogee}, \&
  {Gawiser}}]{hagen2016}
{Hagen}, A., {Zeimann}, G.~R., {Behrens}, C., {Ciardullo}, R., {Grasshorn
  Gebhardt}, H.~S., {Gronwall}, C., {Bridge}, J.~S., {Fox}, D.~B., {Schneider},
  D.~P., {Trump}, J.~R., {Blanc}, G.~A., {Chiang}, Y.-K., {Chonis}, T.~S.,
  {Finkelstein}, S.~L., {Hill}, G.~J., {Jogee}, S., \& {Gawiser}, E. 2016,
  \apj, 817, 79

\bibitem[{{Hayes} {et~al.}(2014){Hayes}, {{\"O}stlin}, {Duval}, {Sandberg},
  {Guaita}, {Melinder}, {Adamo}, {Schaerer}, {Verhamme}, {Orlitov{\'a}},
  {Mas-Hesse}, {Cannon}, {Atek}, {Kunth}, {Laursen}, {Ot{\'{\i}}-Floranes},
  {Pardy}, {Rivera-Thorsen}, \& {Herenz}}]{hayes2014}
{Hayes}, M., {{\"O}stlin}, G., {Duval}, F., {Sandberg}, A., {Guaita}, L.,
  {Melinder}, J., {Adamo}, A., {Schaerer}, D., {Verhamme}, A., {Orlitov{\'a}},
  I., {Mas-Hesse}, J.~M., {Cannon}, J.~M., {Atek}, H., {Kunth}, D., {Laursen},
  P., {Ot{\'{\i}}-Floranes}, H., {Pardy}, S., {Rivera-Thorsen}, T., \&
  {Herenz}, E.~C. 2014, \apj, 782, 6

\bibitem[{{Hayes} {et~al.}(2010){Hayes}, {{\"O}stlin}, {Schaerer}, {Mas-Hesse},
  {Leitherer}, {Atek}, {Kunth}, {Verhamme}, {de Barros}, \&
  {Melinder}}]{hayes2010}
{Hayes}, M., {{\"O}stlin}, G., {Schaerer}, D., {Mas-Hesse}, J.~M., {Leitherer},
  C., {Atek}, H., {Kunth}, D., {Verhamme}, A., {de Barros}, S., \& {Melinder},
  J. 2010, \nat, 464, 562

\bibitem[{{Henry} {et~al.}(2015){Henry}, {Scarlata}, {Martin}, \&
  {Erb}}]{henry2015}
{Henry}, A., {Scarlata}, C., {Martin}, C.~L., \& {Erb}, D. 2015, \apj, 809, 19

\bibitem[{{Hoag} {et~al.}(2019){Hoag}, {Brada{\v{c}}}, {Huang}, {Mason},
  {Treu}, {Schmidt}, {Trenti}, {Strait}, {Lemaux}, {Finney}, \&
  {Paddock}}]{hoag2019}
{Hoag}, A., {Brada{\v{c}}}, M., {Huang}, K., {Mason}, C., {Treu}, T.,
  {Schmidt}, K.~B., {Trenti}, M., {Strait}, V., {Lemaux}, B.~C., {Finney},
  E.~Q., \& {Paddock}, M. 2019, \apj, 878, 12

\bibitem[{{Hunter}(2007)}]{matplotlib2007}
{Hunter}, J.~D. 2007, Computing in Science and Engineering, 9, 90

\bibitem[{{Izotov} {et~al.}(2016){Izotov}, {Schaerer}, {Thuan}, {Worseck},
  {Guseva}, {Orlitov{\'a}}, \& {Verhamme}}]{izotov2016}
{Izotov}, Y.~I., {Schaerer}, D., {Thuan}, T.~X., {Worseck}, G., {Guseva},
  N.~G., {Orlitov{\'a}}, I., \& {Verhamme}, A. 2016, \mnras, 461, 3683

\bibitem[{{Izotov} {et~al.}(2018){Izotov}, {Worseck}, {Schaerer}, {Guseva},
  {Thuan}, {Fricke}, \& {Orlitov{\'a}}}]{izotov2018}
{Izotov}, Y.~I., {Worseck}, G., {Schaerer}, D., {Guseva}, N.~G., {Thuan},
  T.~X., {Fricke}, Verhamme, A., \& {Orlitov{\'a}}, I. 2018, \mnras, 478, 4851

\bibitem[{{Jones} {et~al.}(2012){Jones}, {Stark}, \& {Ellis}}]{jones2012}
{Jones}, T., {Stark}, D.~P., \& {Ellis}, R.~S. 2012, \apj, 751, 51

\bibitem[{{Kennicutt}(1998)}]{kennicutt1998}
{Kennicutt}, Jr., R.~C. 1998, \araa, 36, 189

\bibitem[{{Kollmeier} {et~al.}(2010){Kollmeier}, {Zheng}, {Dav{\'e}}, {Gould},
  {Katz}, {Miralda-Escud{\'e}}, \& {Weinberg}}]{kol10}
{Kollmeier}, J.~A., {Zheng}, Z., {Dav{\'e}}, R., {Gould}, A., {Katz}, N.,
  {Miralda-Escud{\'e}}, J., \& {Weinberg}, D.~H. 2010, \apj, 708, 1048

\bibitem[{{Kunth} {et~al.}(1998){Kunth}, {Mas-Hesse}, {Terlevich}, {Terlevich},
  {Lequeux}, \& {Fall}}]{kunth1998}
{Kunth}, D., {Mas-Hesse}, J.~M., {Terlevich}, E., {Terlevich}, R., {Lequeux},
  J., \& {Fall}, S.~M. 1998, \aap, 334, 11

\bibitem[{{Martin} {et~al.}(2015){Martin}, {Dijkstra}, {Henry}, {Soto},
  {Danforth}, \& {Wong}}]{martin2015}
{Martin}, C.~L., {Dijkstra}, M., {Henry}, A., {Soto}, K.~T., {Danforth}, C.~W.,
  \& {Wong}, J. 2015, \apj, 803, 6

\bibitem[{{Mason} {et~al.}(2018){Mason}, {Treu}, {Dijkstra}, {Mesinger},
  {Trenti}, {Pentericci}, {de Barros}, \& {Vanzella}}]{mason2018}
{Mason}, C.~A., {Treu}, T., {Dijkstra}, M., {Mesinger}, A., {Trenti}, M.,
  {Pentericci}, L., {de Barros}, S., \& {Vanzella}, E. 2018, \apj, 856, 2

\bibitem[{{McLean} {et~al.}(2012){McLean}, {Steidel}, {Epps}, {Konidaris},
  {Matthews}, {Adkins}, {Aliado}, {Brims}, {Canfield}, {Cromer}, {Fucik},
  {Kulas}, {Mace}, {Magnone}, {Rodriguez}, {Rudie}, {Trainor}, {Wang}, {Weber},
  \& {Weiss}}]{mclean2012}
{McLean}, I.~S., {Steidel}, C.~C., {Epps}, H.~W., {Konidaris}, N., {Matthews},
  K.~Y., {Adkins}, S., {Aliado}, T., {Brims}, G., {Canfield}, J.~M., {Cromer},
  J.~L., {Fucik}, J., {Kulas}, K., {Mace}, G., {Magnone}, K., {Rodriguez}, H.,
  {Rudie}, G., {Trainor}, R., {Wang}, E., {Weber}, B., \& {Weiss}, J. 2012, in
  Society of Photo-Optical Instrumentation Engineers (SPIE) Conference Series,
  Vol. 8446, Society of Photo-Optical Instrumentation Engineers (SPIE)
  Conference Series

\bibitem[{{McLinden} {et~al.}(2011){McLinden}, {Finkelstein}, {Rhoads},
  {Malhotra}, {Hibon}, {Richardson}, {Cresci}, {Quirrenbach}, {Pasquali},
  {Bian}, {Fan}, \& {Woodward}}]{mclinden2011}
{McLinden}, E.~M., {Finkelstein}, S.~L., {Rhoads}, J.~E., {Malhotra}, S.,
  {Hibon}, P., {Richardson}, M.~L.~A., {Cresci}, G., {Quirrenbach}, A.,
  {Pasquali}, A., {Bian}, F., {Fan}, X., \& {Woodward}, C.~E. 2011, \apj, 730,
  136

\bibitem[{{Momose} {et~al.}(2014){Momose}, {Ouchi}, {Nakajima}, {Ono},
  {Shibuya}, {Shimasaku}, {Yuma}, {Mori}, \& {Umemura}}]{momose14}
{Momose}, R., {Ouchi}, M., {Nakajima}, K., {Ono}, Y., {Shibuya}, T.,
  {Shimasaku}, K., {Yuma}, S., {Mori}, M., \& {Umemura}, M. 2014, \mnras, 442,
  110

\bibitem[{{Momose} {et~al.}(2016){Momose}, {Ouchi}, {Nakajima}, {Ono},
  {Shibuya}, {Shimasaku}, {Yuma}, {Mori}, \& {Umemura}}]{momose2016}
---. 2016, \mnras, 457, 2318

\bibitem[{{Nakajima} {et~al.}(2016){Nakajima}, {Ellis}, {Iwata}, {Inoue},
  {Kusakabe}, {Ouchi}, \& {Robertson}}]{nakajima2016}
{Nakajima}, K., {Ellis}, R.~S., {Iwata}, I., {Inoue}, A.~K., {Kusakabe}, H.,
  {Ouchi}, M., \& {Robertson}, B.~E. 2016, \apjl, 831, L9

\bibitem[{{Nakajima} {et~al.}(2013){Nakajima}, {Ouchi}, {Shimasaku},
  {Hashimoto}, {Ono}, \& {Lee}}]{nakajima2013}
{Nakajima}, K., {Ouchi}, M., {Shimasaku}, K., {Hashimoto}, T., {Ono}, Y., \&
  {Lee}, J.~C. 2013, \apj, 769, 3

\bibitem[{{Oke} {et~al.}(1995){Oke}, {Cohen}, {Carr}, {Cromer}, {Dingizian},
  {Harris}, {Labrecque}, {Lucinio}, {Schaal}, {Epps}, \& {Miller}}]{oke1995}
{Oke}, J.~B., {Cohen}, J.~G., {Carr}, M., {Cromer}, J., {Dingizian}, A.,
  {Harris}, F.~H., {Labrecque}, S., {Lucinio}, R., {Schaal}, W., {Epps}, H., \&
  {Miller}, J. 1995, \pasp, 107, 375

\bibitem[{{Osterbrock}(1989)}]{osterbrock1989}
{Osterbrock}, D.~E. 1989, {Astrophysics of gaseous nebulae and active galactic
  nuclei}

\bibitem[{{{\"O}stlin} {et~al.}(2014){{\"O}stlin}, {Hayes}, {Duval},
  {Sandberg}, {Rivera-Thorsen}, {Marquart}, {Orlitov{\'a}}, {Adamo},
  {Melinder}, {Guaita}, {Atek}, {Cannon}, {Gruyters}, {Herenz}, {Kunth},
  {Laursen}, {Mas-Hesse}, {Micheva}, {Ot{\'{\i}}-Floranes}, {Pardy}, {Roth},
  {Schaerer}, \& {Verhamme}}]{ostlin2014}
{{\"O}stlin}, G., {Hayes}, M., {Duval}, F., {Sandberg}, A., {Rivera-Thorsen},
  T., {Marquart}, T., {Orlitov{\'a}}, I., {Adamo}, A., {Melinder}, J.,
  {Guaita}, L., {Atek}, H., {Cannon}, J.~M., {Gruyters}, P., {Herenz}, E.~C.,
  {Kunth}, D., {Laursen}, P., {Mas-Hesse}, J.~M., {Micheva}, G.,
  {Ot{\'{\i}}-Floranes}, H., {Pardy}, S.~A., {Roth}, M.~M., {Schaerer}, D., \&
  {Verhamme}, A. 2014, \apj, 797, 11

\bibitem[{{Oteo} {et~al.}(2015){Oteo}, {Sobral}, {Ivison}, {Smail}, {Best},
  {Cepa}, \& {P{\'e}rez-Garc{\'\i}a}}]{oteo2015}
{Oteo}, I., {Sobral}, D., {Ivison}, R.~J., {Smail}, I., {Best}, P.~N., {Cepa},
  J., \& {P{\'e}rez-Garc{\'\i}a}, A.~M. 2015, \mnras, 452, 2018

\bibitem[{{Ouchi} {et~al.}(2018){Ouchi}, {Harikane}, {Shibuya}, {Shimasaku},
  {Taniguchi}, {Konno}, {Kobayashi}, {Kajisawa}, {Nagao}, {Ono}, {Inoue},
  {Umemura}, {Mori}, {Hasegawa}, {Higuchi}, {Komiyama}, {Matsuda}, {Nakajima},
  {Saito}, \& {Wang}}]{ouchi2018}
{Ouchi}, M., {Harikane}, Y., {Shibuya}, T., {Shimasaku}, K., {Taniguchi}, Y.,
  {Konno}, A., {Kobayashi}, M., {Kajisawa}, M., {Nagao}, T., {Ono}, Y.,
  {Inoue}, A.~K., {Umemura}, M., {Mori}, M., {Hasegawa}, K., {Higuchi}, R.,
  {Komiyama}, Y., {Matsuda}, Y., {Nakajima}, K., {Saito}, T., \& {Wang}, S.-Y.
  2018, \pasj, 70, S13

\bibitem[{{Oyarz{\'u}n} {et~al.}(2017){Oyarz{\'u}n}, {Blanc}, {Gonz{\'a}lez},
  {Mateo}, \& {Bailey}}]{oyarzun2017}
{Oyarz{\'u}n}, G.~A., {Blanc}, G.~A., {Gonz{\'a}lez}, V., {Mateo}, M., \&
  {Bailey}, John~I., I. 2017, \apj, 843, 133

\bibitem[{{Oyarz{\'u}n} {et~al.}(2016){Oyarz{\'u}n}, {Blanc}, {Gonz{\'a}lez},
  {Mateo}, {Bailey}, {Finkelstein}, {Lira}, {Crane}, \&
  {Olszewski}}]{oyarzun16}
{Oyarz{\'u}n}, G.~A., {Blanc}, G.~A., {Gonz{\'a}lez}, V., {Mateo}, M.,
  {Bailey}, III, J.~I., {Finkelstein}, S.~L., {Lira}, P., {Crane}, J.~D., \&
  {Olszewski}, E.~W. 2016, \apjl, 821, L14

\bibitem[{{Pellegrini} {et~al.}(2012){Pellegrini}, {Oey}, {Winkler}, {Points},
  {Smith}, {Jaskot}, \& {Zastrow}}]{pellegrini2012}
{Pellegrini}, E.~W., {Oey}, M.~S., {Winkler}, P.~F., {Points}, S.~D., {Smith},
  R.~C., {Jaskot}, A.~E., \& {Zastrow}, J. 2012, \apj, 755, 40

\bibitem[{{Pentericci} {et~al.}(2011){Pentericci}, {Fontana}, {Vanzella},
  {Castellano}, {Grazian}, {Dijkstra}, {Boutsia}, {Cristiani}, {Dickinson},
  {Giallongo}, {Giavalisco}, {Maiolino}, {Moorwood}, {Paris}, \&
  {Santini}}]{pentericci2011}
{Pentericci}, L., {Fontana}, A., {Vanzella}, E., {Castellano}, M., {Grazian},
  A., {Dijkstra}, M., {Boutsia}, K., {Cristiani}, S., {Dickinson}, M.,
  {Giallongo}, E., {Giavalisco}, M., {Maiolino}, R., {Moorwood}, A., {Paris},
  D., \& {Santini}, P. 2011, \apj, 743, 132

\bibitem[{{Pentericci} {et~al.}(2018){Pentericci}, {Vanzella}, {Castellano},
  {Fontana}, {De Barros}, {Grazian}, {Marchi}, {Bradac}, {Conselice},
  {Cristiani}, {Dickinson}, {Finkelstein}, {Giallongo}, {Guaita}, {Koekemoer},
  {Maiolino}, {Santini}, \& {Tilvi}}]{pentericci2018}
{Pentericci}, L., {Vanzella}, E., {Castellano}, M., {Fontana}, A., {De Barros},
  S., {Grazian}, A., {Marchi}, F., {Bradac}, M., {Conselice}, C.~J.,
  {Cristiani}, S., {Dickinson}, M., {Finkelstein}, S.~L., {Giallongo}, E.,
  {Guaita}, L., {Koekemoer}, A.~M., {Maiolino}, R., {Santini}, P., \& {Tilvi},
  V. 2018, \aap, 619, A147

\bibitem[{{Pettini} {et~al.}(2002){Pettini}, {Rix}, {Steidel}, {Adelberger},
  {Hunt}, \& {Shapley}}]{pettini2002}
{Pettini}, M., {Rix}, S.~A., {Steidel}, C.~C., {Adelberger}, K.~L., {Hunt},
  M.~P., \& {Shapley}, A.~E. 2002, \apj, 569, 742

\bibitem[{{Pettini} {et~al.}(2000){Pettini}, {Steidel}, {Adelberger},
  {Dickinson}, \& {Giavalisco}}]{pettini2000}
{Pettini}, M., {Steidel}, C.~C., {Adelberger}, K.~L., {Dickinson}, M., \&
  {Giavalisco}, M. 2000, \apj, 528, 96

\bibitem[{{Rakic} {et~al.}(2012){Rakic}, {Schaye}, {Steidel}, \&
  {Rudie}}]{rakic2012}
{Rakic}, O., {Schaye}, J., {Steidel}, C.~C., \& {Rudie}, G.~C. 2012, \apj, 751,
  94

\bibitem[{{Reddy} {et~al.}(2012){Reddy}, {Pettini}, {Steidel}, {Shapley},
  {Erb}, \& {Law}}]{reddy2012}
{Reddy}, N.~A., {Pettini}, M., {Steidel}, C.~C., {Shapley}, A.~E., {Erb},
  D.~K., \& {Law}, D.~R. 2012, \apj, 754, 25

\bibitem[{{Reddy} {et~al.}(2008){Reddy}, {Steidel}, {Pettini}, {Adelberger},
  {Shapley}, {Erb}, \& {Dickinson}}]{reddy2008}
{Reddy}, N.~A., {Steidel}, C.~C., {Pettini}, M., {Adelberger}, K.~L.,
  {Shapley}, A.~E., {Erb}, D.~K., \& {Dickinson}, M. 2008, \apjs, 175, 48

\bibitem[{{Rhoads} {et~al.}(2000){Rhoads}, {Malhotra}, {Dey}, {Stern},
  {Spinrad}, \& {Jannuzi}}]{rhoads2000}
{Rhoads}, J.~E., {Malhotra}, S., {Dey}, A., {Stern}, D., {Spinrad}, H., \&
  {Jannuzi}, B.~T. 2000, \apjl, 545, L85

\bibitem[{{Rivera-Thorsen} {et~al.}(2015){Rivera-Thorsen}, {Hayes},
  {{\"O}stlin}, {Duval}, {Orlitov{\'a}}, {Verhamme}, {Mas-Hesse}, {Schaerer},
  {Cannon}, {Ot{\'{\i}}-Floranes}, {Sandberg}, {Guaita}, {Adamo}, {Atek},
  {Herenz}, {Kunth}, {Laursen}, \& {Melinder}}]{rivera2015}
{Rivera-Thorsen}, T.~E., {Hayes}, M., {{\"O}stlin}, G., {Duval}, F.,
  {Orlitov{\'a}}, I., {Verhamme}, A., {Mas-Hesse}, J.~M., {Schaerer}, D.,
  {Cannon}, J.~M., {Ot{\'{\i}}-Floranes}, H., {Sandberg}, A., {Guaita}, L.,
  {Adamo}, A., {Atek}, H., {Herenz}, E.~C., {Kunth}, D., {Laursen}, P., \&
  {Melinder}, J. 2015, \apj, 805, 14

\bibitem[{{Roberts-Borsani} {et~al.}(2016){Roberts-Borsani}, {Bouwens},
  {Oesch}, {Labbe}, {Smit}, {Illingworth}, {van Dokkum}, {Holden}, {Gonzalez},
  {Stefanon}, {Holwerda}, \& {Wilkins}}]{roberts-borsani2016}
{Roberts-Borsani}, G.~W., {Bouwens}, R.~J., {Oesch}, P.~A., {Labbe}, I.,
  {Smit}, R., {Illingworth}, G.~D., {van Dokkum}, P., {Holden}, B., {Gonzalez},
  V., {Stefanon}, M., {Holwerda}, B., \& {Wilkins}, S. 2016, \apj, 823, 143

\bibitem[{{Robertson} {et~al.}(2015){Robertson}, {Ellis}, {Furlanetto}, \&
  {Dunlop}}]{robertson2015}
{Robertson}, B.~E., {Ellis}, R.~S., {Furlanetto}, S.~R., \& {Dunlop}, J.~S.
  2015, \apjl, 802, L19

\bibitem[{Rousseeuw \& Croux(1993)}]{rousseeuw1993}
Rousseeuw, P.~J. \& Croux, C. 1993, Journal of the American Statistical
  Association, 88, 1273

\bibitem[{{Rudie} {et~al.}(2012){Rudie}, {Steidel}, {Trainor}, {Rakic},
  {Bogosavljevi{\'c}}, {Pettini}, {Reddy}, {Shapley}, {Erb}, \&
  {Law}}]{rudie2012a}
{Rudie}, G.~C., {Steidel}, C.~C., {Trainor}, R.~F., {Rakic}, O.,
  {Bogosavljevi{\'c}}, M., {Pettini}, M., {Reddy}, N., {Shapley}, A.~E., {Erb},
  D.~K., \& {Law}, D.~R. 2012, \apj, 750, 67

\bibitem[{{Sanders} {et~al.}(2016){Sanders}, {Shapley}, {Kriek}, {Reddy},
  {Freeman}, {Coil}, {Siana}, {Mobasher}, {Shivaei}, {Price}, \& {de
  Groot}}]{sanders2016}
{Sanders}, R.~L., {Shapley}, A.~E., {Kriek}, M., {Reddy}, N.~A., {Freeman},
  W.~R., {Coil}, A.~L., {Siana}, B., {Mobasher}, B., {Shivaei}, I., {Price},
  S.~H., \& {de Groot}, L. 2016, \apj, 816, 23

\bibitem[{{Schenker} {et~al.}(2013){Schenker}, {Ellis}, {Konidaris}, \&
  {Stark}}]{schenker13a}
{Schenker}, M.~A., {Ellis}, R.~S., {Konidaris}, N.~P., \& {Stark}, D.~P. 2013,
  \apj, 777, 67

\bibitem[{{Schenker} {et~al.}(2012){Schenker}, {Stark}, {Ellis}, {Robertson},
  {Dunlop}, {McLure}, {Kneib}, \& {Richard}}]{schenker2012}
{Schenker}, M.~A., {Stark}, D.~P., {Ellis}, R.~S., {Robertson}, B.~E.,
  {Dunlop}, J.~S., {McLure}, R.~J., {Kneib}, J.-P., \& {Richard}, J. 2012,
  \apj, 744, 179

\bibitem[{{Shapley} {et~al.}(2003){Shapley}, {Steidel}, {Pettini}, \&
  {Adelberger}}]{sha03}
{Shapley}, A.~E., {Steidel}, C.~C., {Pettini}, M., \& {Adelberger}, K.~L. 2003,
  \apj, 588, 65

\bibitem[{{Shapley} {et~al.}(2016){Shapley}, {Steidel}, {Strom},
  {Bogosavljevi{\'c}}, {Reddy}, {Siana}, {Mostardi}, \& {Rudie}}]{shapley2016}
{Shapley}, A.~E., {Steidel}, C.~C., {Strom}, A.~L., {Bogosavljevi{\'c}}, M.,
  {Reddy}, N.~A., {Siana}, B., {Mostardi}, R.~E., \& {Rudie}, G.~C. 2016, \apj,
  826, L24

\bibitem[{{Siana} {et~al.}(2008){Siana}, {Teplitz}, {Chary}, {Colbert}, \&
  {Frayer}}]{siana2008}
{Siana}, B., {Teplitz}, H.~I., {Chary}, R.-R., {Colbert}, J., \& {Frayer},
  D.~T. 2008, \apj, 689, 59

\bibitem[{{Sobral} {et~al.}(2009){Sobral}, {Best}, {Geach}, {Smail}, {Kurk},
  {Cirasuolo}, {Casali}, {Ivison}, {Coppin}, \& {Dalton}}]{sobral2009a}
{Sobral}, D., {Best}, P.~N., {Geach}, J.~E., {Smail}, I., {Kurk}, J.,
  {Cirasuolo}, M., {Casali}, M., {Ivison}, R.~J., {Coppin}, K., \& {Dalton},
  G.~B. 2009, \mnras, 398, 75

\bibitem[{{Song} {et~al.}(2014){Song}, {Finkelstein}, {Gebhardt}, {Hill},
  {Drory}, {Ashby}, {Blanc}, {Bridge}, {Chonis}, {Ciardullo}, {Fabricius},
  {Fazio}, {Gawiser}, {Gronwall}, {Hagen}, {Huang}, {Jogee}, {Livermore},
  {Salmon}, {Schneider}, {Willner}, \& {Zeimann}}]{song14}
{Song}, M., {Finkelstein}, S.~L., {Gebhardt}, K., {Hill}, G.~J., {Drory}, N.,
  {Ashby}, M.~L.~N., {Blanc}, G.~A., {Bridge}, J., {Chonis}, T., {Ciardullo},
  R., {Fabricius}, M., {Fazio}, G.~G., {Gawiser}, E., {Gronwall}, C., {Hagen},
  A., {Huang}, J.-S., {Jogee}, S., {Livermore}, R., {Salmon}, B., {Schneider},
  D.~P., {Willner}, S.~P., \& {Zeimann}, G.~R. 2014, \apj, 791, 3

\bibitem[{{Stark} {et~al.}(2017){Stark}, {Ellis}, {Charlot}, {Chevallard},
  {Tang}, {Belli}, {Zitrin}, {Mainali}, {Gutkin}, {Vidal-Garc{\'\i}a},
  {Bouwens}, \& {Oesch}}]{stark2017}
{Stark}, D.~P., {Ellis}, R.~S., {Charlot}, S., {Chevallard}, J., {Tang}, M.,
  {Belli}, S., {Zitrin}, A., {Mainali}, R., {Gutkin}, J., {Vidal-Garc{\'\i}a},
  A., {Bouwens}, R., \& {Oesch}, P. 2017, \mnras, 464, 469

\bibitem[{{Stark} {et~al.}(2010){Stark}, {Ellis}, {Chiu}, {Ouchi}, \&
  {Bunker}}]{stark2010}
{Stark}, D.~P., {Ellis}, R.~S., {Chiu}, K., {Ouchi}, M., \& {Bunker}, A. 2010,
  \mnras, 408, 1628

\bibitem[{{Stark} {et~al.}(2013){Stark}, {Schenker}, {Ellis}, {Robertson},
  {McLure}, \& {Dunlop}}]{stark2013}
{Stark}, D.~P., {Schenker}, M.~A., {Ellis}, R., {Robertson}, B., {McLure}, R.,
  \& {Dunlop}, J. 2013, \apj, 763, 129

\bibitem[{{Steidel} {et~al.}(2000){Steidel}, {Adelberger}, {Shapley},
  {Pettini}, {Dickinson}, \& {Giavalisco}}]{steidel2000}
{Steidel}, C.~C., {Adelberger}, K.~L., {Shapley}, A.~E., {Pettini}, M.,
  {Dickinson}, M., \& {Giavalisco}, M. 2000, \apj, 532, 170

\bibitem[{{Steidel} {et~al.}(2003){Steidel}, {Adelberger}, {Shapley},
  {Pettini}, {Dickinson}, \& {Giavalisco}}]{steidel2003}
---. 2003, \apj, 592, 728

\bibitem[{{Steidel} {et~al.}(2011){Steidel}, {Bogosavljevi{\'c}}, {Shapley},
  {Kollmeier}, {Reddy}, {Erb}, \& {Pettini}}]{steidel2011}
{Steidel}, C.~C., {Bogosavljevi{\'c}}, M., {Shapley}, A.~E., {Kollmeier},
  J.~A., {Reddy}, N.~A., {Erb}, D.~K., \& {Pettini}, M. 2011, \apj, 736, 160

\bibitem[{{Steidel} {et~al.}(2018){Steidel}, {Bogosavljevi{\'c}}, {Shapley},
  {Reddy}, {Rudie}, {Pettini}, {Trainor}, \& {Strom}}]{steidel2018}
{Steidel}, C.~C., {Bogosavljevi{\'c}}, M., {Shapley}, A.~E., {Reddy}, N.~A.,
  {Rudie}, G.~C., {Pettini}, M., {Trainor}, R.~F., \& {Strom}, A.~L. 2018,
  \apj, 869, 123

\bibitem[{{Steidel} {et~al.}(2010){Steidel}, {Erb}, {Shapley}, {Pettini},
  {Reddy}, {Bogosavljevi{\'c}}, {Rudie}, \& {Rakic}}]{steidel2010}
{Steidel}, C.~C., {Erb}, D.~K., {Shapley}, A.~E., {Pettini}, M., {Reddy}, N.,
  {Bogosavljevi{\'c}}, M., {Rudie}, G.~C., \& {Rakic}, O. 2010, \apj, 717, 289

\bibitem[{{Steidel} {et~al.}(2014){Steidel}, {Rudie}, {Strom}, {Pettini},
  {Reddy}, {Shapley}, {Trainor}, {Erb}, {Turner}, {Konidaris}, {Kulas}, {Mace},
  {Matthews}, \& {McLean}}]{steidel2014}
{Steidel}, C.~C., {Rudie}, G.~C., {Strom}, A.~L., {Pettini}, M., {Reddy},
  N.~A., {Shapley}, A.~E., {Trainor}, R.~F., {Erb}, D.~K., {Turner}, M.~L.,
  {Konidaris}, N.~P., {Kulas}, K.~R., {Mace}, G., {Matthews}, K., \& {McLean},
  I.~S. 2014, \apj, 795, 165

\bibitem[{{Steidel} {et~al.}(2004){Steidel}, {Shapley}, {Pettini},
  {Adelberger}, {Erb}, {Reddy}, \& {Hunt}}]{steidel2004}
{Steidel}, C.~C., {Shapley}, A.~E., {Pettini}, M., {Adelberger}, K.~L., {Erb},
  D.~K., {Reddy}, N.~A., \& {Hunt}, M.~P. 2004, \apj, 604, 534

\bibitem[{{Steidel} {et~al.}(2016){Steidel}, {Strom}, {Pettini}, {Rudie},
  {Reddy}, \& {Trainor}}]{steidel2016}
{Steidel}, C.~C., {Strom}, A.~L., {Pettini}, M., {Rudie}, G.~C., {Reddy},
  N.~A., \& {Trainor}, R.~F. 2016, \apj, 826, 159

\bibitem[{{Strom} {et~al.}(2018){Strom}, {Steidel}, {Rudie}, {Trainor}, \&
  {Pettini}}]{strom2018}
{Strom}, A.~L., {Steidel}, C.~C., {Rudie}, G.~C., {Trainor}, R.~F., \&
  {Pettini}, M. 2018, \apj, 868, 117

\bibitem[{{Strom} {et~al.}(2017){Strom}, {Steidel}, {Rudie}, {Trainor},
  {Pettini}, \& {Reddy}}]{strom2017}
{Strom}, A.~L., {Steidel}, C.~C., {Rudie}, G.~C., {Trainor}, R.~F., {Pettini},
  M., \& {Reddy}, N.~A. 2017, \apj, 836, 164

\bibitem[{{Teplitz} {et~al.}(2000){Teplitz}, {McLean}, {Becklin}, {Figer},
  {Gilbert}, {Graham}, {Larkin}, {Levenson}, \& {Wilcox}}]{teplitz2000}
{Teplitz}, H.~I., {McLean}, I.~S., {Becklin}, E.~E., {Figer}, D.~F., {Gilbert},
  A.~M., {Graham}, J.~R., {Larkin}, J.~E., {Levenson}, N.~A., \& {Wilcox},
  M.~K. 2000, \apjl, 533, L65

\bibitem[{{Theios} {et~al.}(2019){Theios}, {Steidel}, {Strom}, {Rudie},
  {Trainor}, \& {Reddy}}]{theios2019}
{Theios}, R.~L., {Steidel}, C.~C., {Strom}, A.~L., {Rudie}, G.~C., {Trainor},
  R.~F., \& {Reddy}, N.~A. 2019, \apj, 871, 128

\bibitem[{{Trainor} \& {Steidel}(2013)}]{trainor2013}
{Trainor}, R. \& {Steidel}, C.~C. 2013, \apjl, 775, L3

\bibitem[{{Trainor} \& {Steidel}(2012)}]{trainor2012}
{Trainor}, R.~F. \& {Steidel}, C.~C. 2012, \apj, 752, 39

\bibitem[{{Trainor} {et~al.}(2015){Trainor}, {Steidel}, {Strom}, \&
  {Rudie}}]{trainor2015}
{Trainor}, R.~F., {Steidel}, C.~C., {Strom}, A.~L., \& {Rudie}, G.~C. 2015,
  \apj, 809, 89

\bibitem[{{Trainor} {et~al.}(2016){Trainor}, {Strom}, {Steidel}, \&
  {Rudie}}]{trainor2016}
{Trainor}, R.~F., {Strom}, A.~L., {Steidel}, C.~C., \& {Rudie}, G.~C. 2016,
  \apj, 832, 171

\bibitem[{{Veilleux} \& {Osterbrock}(1987)}]{veilleux1987}
{Veilleux}, S. \& {Osterbrock}, D.~E. 1987, The Astrophysical Journal
  Supplement Series, 63, 295

\bibitem[{{Wisotzki} {et~al.}(2016){Wisotzki}, {Bacon}, {Blaizot},
  {Brinchmann}, {Herenz}, {Schaye}, {Bouch{\'e}}, {Cantalupo}, {Contini},
  {Carollo}, {Caruana}, {Courbot}, {Emsellem}, {Kamann}, {Kerutt}, {Leclercq},
  {Lilly}, {Patr{\'\i}cio}, {Sandin}, {Steinmetz}, {Straka}, {Urrutia},
  {Verhamme}, {Weilbacher}, \& {Wendt}}]{wisotzki2016}
{Wisotzki}, L., {Bacon}, R., {Blaizot}, J., {Brinchmann}, J., {Herenz}, E.~C.,
  {Schaye}, J., {Bouch{\'e}}, N., {Cantalupo}, S., {Contini}, T., {Carollo},
  C.~M., {Caruana}, J., {Courbot}, J.~B., {Emsellem}, E., {Kamann}, S.,
  {Kerutt}, J., {Leclercq}, F., {Lilly}, S.~J., {Patr{\'\i}cio}, V., {Sandin},
  C., {Steinmetz}, M., {Straka}, L.~A., {Urrutia}, T., {Verhamme}, A.,
  {Weilbacher}, P.~M., \& {Wendt}, M. 2016, \aap, 587, A98

\bibitem[{{Yang} {et~al.}(2017){Yang}, {Malhotra}, {Gronke}, {Rhoads},
  {Leitherer}, {Wofford}, {Jiang}, {Dijkstra}, {Tilvi}, \& {Wang}}]{yang2017}
{Yang}, H., {Malhotra}, S., {Gronke}, M., {Rhoads}, J.~E., {Leitherer}, C.,
  {Wofford}, A., {Jiang}, T., {Dijkstra}, M., {Tilvi}, V., \& {Wang}, J. 2017,
  \apj, 844, 171

\end{thebibliography}

\end{document}